\documentclass[superscriptaddress,11pt]{article}%
\usepackage{authblk}
\usepackage{array}
\usepackage{type1cm}
\usepackage{lettrine}
\usepackage{graphicx}
\usepackage{geometry}
\usepackage{psfrag}
\usepackage{amsmath}
\usepackage{amsfonts}
\usepackage{appendix}
\usepackage[margin=10pt,font=footnotesize,labelfont=sc,format=hang,labelformat=parens]{caption}
\usepackage{amssymb}%
\providecommand{\U}[1]{\protect\rule{.1in}{.1in}}
\numberwithin{equation}{section}
\begin{document}

\title{Towards an Anomaly-Free Quantum Dynamics for a Weak Coupling Limit of
Euclidean Gravity: Diffeomorphism Covariance}
\author{Madhavan Varadarajan}
\affil{Raman Research Institute\\Bangalore-560 080, India}

\maketitle

\begin{abstract}
The $G_{\mathrm{Newton}}\rightarrow0$ limit of Euclidean gravity introduced by
Smolin is described by a generally covariant $\mathrm{U}(1)^{3}$ gauge theory.
In an earlier paper, Tomlin and Varadarajan constructed
the quantum Hamiltonian constraint of density weight
$4/3$ for this $\mathrm{U}(1)^{3}$ theory so as to produce a
non-trivial anomaly free LQG-type representation of the Poisson bracket between
a pair of Hamiltonian constraints.
These constructions involved a choice of regulating coordinate patches.
The use of these coordinate patches is in apparent conflict with 
spatial diffeomorphism covariance. In this work we show how
an appropriate choice of coordinate patches together with suitable
modifications of these constructions results in the diffeomorphism covariance
of the continuum limit action of the Hamiltonian constraint operator, while preserving the 
anomaly free property of the continuum limit action of its commutator.

\end{abstract}

\affil{Raman Research Institute\\Bangalore-560 080, India}

\thispagestyle{empty}
\let\oldthefootnote\thefootnote\renewcommand{\thefootnote}{\fnsymbol{footnote}}
\footnotetext{Email:madhavan@rri.res.in}
\let\thefootnote\oldthefootnote

\section{Introduction}
A key open problem in LQG is that of the construction of the Hamiltonian constraint operator.
The Hamiltonian constraint  operators currently used in LQG \cite{qsd,habitat1,aajurekreview} 
do not display a {\em non- trivial} anomaly free representation of the classical constraint algebra.
In this regard, studies of
Parameterized Field Theory \cite{ppftham} and the Husain- Kucha{\v r} model \cite{hk}
have yielded valuable lessons. Among other things these studies suggest
that  a nontrivial representation requires
a Hamiltonian constraint of density weight greater than unity,  in contrast to 
the widely used density weight one Hamiltonian constraint. 
While the two toy models above share certain structural features with gravity by virtue
of their general covariance, there is one key structural aspect of gravity which they do not share. 
The structure of their constraint algebras is that of a Lie algebra. In contrast
the constraint algebra of gravity is not a Lie algebra, the reason being that the Poisson bracket between
a pair of Hamiltonian constraints contains structure functions.

Hence, Laddha initiated the study of a $U(1)^3$ generally covariant theory 
with a constraint algebra similar to that of gravity. The model is obtained by replacing the internal triad 
rotation group $SU(2)$ of Euclidean
gravity by $U(1)^3$ and was introduced earlier by Smolin \cite{lee} who obtained it as a novel weak coupling 
limit of Euclidean gravity.
The constraint algebra is, as in gravity, an open algebra
with structure functions and the structure functions appear in  
the Poisson bracket between a pair of Hamiltonian constraints

Laddha, Henderson and Tomlin \cite{aloku131} studied an LQG type  `polymer' quantization of this model in 2+1 dimensions and
Tomlin and Varadarajan \cite{p1} analysed its polymer quantization in 3+1 dimensions.
These studies were aimed at the construction of the Hamiltonian constraint
operator of the  theory in such a way as to result in a {\em non trivial} anomaly free representation of the Poisson brackets between a pair of Hamiltonian constraints. As mentioned above non- triviality seems to require the use
of higher density constraints.
The general arguments of 
\cite{ppftham,diffeo} indicate a
density weight $4/3$ in 3+1 dimensions \cite{p1} and $3/2$ in 2+1 dimensions 
\cite{aloku131}. 
As in the case of the standard density weight one Hamiltonian constraint 
\cite{qsd,aajurekreview,habitat1,habitat2}, the higher 
density weight constraints act non- trivially only on vertices of the spin network being acted upon. However, in contrast to 
the unit density weight case, it turns out that
the action of the constraint operator depends on the choice
of a regulating coordinate patch at the vertex \cite{aloku131,p1}. 

An immediate concern is if the choice of these coordinate patches 
is in conflict with {\em diffeomorphism covariance}. By diffeomorphism covariance, we mean the condition:
\begin{equation}
{\hat U}(\phi ) {\hat C}(N) {\hat U}^{\dagger} (\phi ) = {\hat C} (\phi_* N) 
\label{dc}
\end{equation}
where ${\hat U}(\phi )$ is the unitary operator which implements the diffeomorphism $\phi$ and $\phi_* N$ is the 
pull back of the lapse $N$ by  the diffeomorphism $\phi$.
The infinitesmal version of this condition in classical theory is expressed through the Poisson bracket:
\begin{equation}
\{C(N), D({\vec M})\} = C({\cal L}_{\vec M}N)
\end{equation}
where $D({\vec M})$ is the diffeomorphism constraint smeared with the shift ${\vec M}$ and 
${\cal L}_{\vec M}$ is the Lie derivative with respect to this shift. 
In this work we show how the constructions of Reference \cite{p1} (henceforth referred to as Paper 1 or P1)
can be improved so as to incorporate condition (\ref{dc}) while still retaining a nontrivial anomaly
free representation of the commutator between a pair of Hamiltonian constraints.

In P1, we first constructed the Hamiltonian constraint at finite triangulation and then took its continuum 
limit. The condition (\ref{dc}) refers to the continuum limit  and is not necessarily a property of the 
Hamiltonian constraint at finite triangulation. 
The continuum limit is taken with respect to a certain
operator topology whose definition depends on certain `vertex smooth' distributions. 
These distributions lie in the algebraic dual to the finite span of charge network
states
\footnote{Charge networks are the $U(1)^3$ analog of the  spin networks of LQG. The algebraic dual is 
the space of linear mappings from the finite span of charge networks into the complex numbers.}
and are called
vertex smooth algebraic states or VSA states in P1. 
The corresponding topology is called the VSA topology.
Our detailed technical claim is then as follows.
We show that the continuum limit of the action of a single Hamiltonian constraint 
is diffeomorphism covariant i.e. that 
\begin{equation}
\lim_{\delta \rightarrow 0} {\hat U}(\phi ) {\hat C}_{\delta}(N) {\hat U}^{\dagger} (\phi ) = 
\lim_{\delta \rightarrow 0}{\hat C}_{\delta} (\phi_* N) 
\label{dc1}
\end{equation}
where the subscript $\delta$ refers to an operator at  finite triangulation
and the $\delta \rightarrow 0$ limit is taken in the VSA topology.

In order that condition (\ref{dc1}) holds,
we require that coordinate patches associated with vertices of diffeomorphic
charge networks be themselves related by diffeomorphisms. This is in contrast to P1 wherein the choice
of coordinate patches was not constrained by such relations. While the choice of patches made here
results in the satisfaction of condition (\ref{dc1}),  it turns out that this choice necessarily
renders the continuum limit of the commutator between two Hamiltonian constraints ill defined. 
\footnote{Deformations generated by the Hamiltonian constraint at different values of $\delta >0$ are
related by diffeomorphisms. Diffeomorphism covariance implicates the use of coordinate patches 
related by such diffeomorphisms. In the commutator the second constraint acts on deformations
created by the first. As $\delta\rightarrow 0$, coordinate patches associated with these deformations
 become sick. This leads to a divergence of the commutator.} 
In order to obtain a well defined continuum limit, we slightly modify the operator topology.
Once this is done, a well defined continuum limit results. The resulting operator topology,
together with the choice of coordinate patches which ensure condition (\ref{dc1}), 
probes the structure of the deformations produced by the finite triangulation constraints, ${\hat C}_{\delta}(N)$,
in a more detailed manner than in P1. As a result, an anomaly free commutator between a pair of Hamiltonian
constraints is obtained only if the deformations produced by the finite triangulation constraints 
are specified in more detail than in P1. The appropriate detailed specification is that of a `conical' deformation.
With this final element, we have a single Hamiltonian constraint action which is  diffeomorphism covariant, as well
as a commutator which is well defined and anomaly free. 

To summarise, this work contains three improvements over the basic framework of P1:\\
\noindent (i) A diffeomorphism covariant choice of coordinate patches and a proof of condition (\ref{dc1}).\\
\noindent (ii)A slightly modified operator topology which renders the commutator between a pair of Hamiltonian
constraints well defined.\\
\noindent (iii)A more detailed choice of the deformations generated by the finite triangulation Hamiltonian 
constraint so as to obtain an anomaly free commutator.

Before we describe the layout of this paper we note the following.
From P1, the finite triangulation Hamiltonian constraint acts only at the vertices of a chargenetwork. 
Its continuum limit is nontrivial only if these vertices 
are characterised by certain diffeomorphism invariant 
properties of the coloured edges which emanate from them. The VSA states which define this limit
are expressible as linear combinations of charge network bra states. Since, in P1 and in this paper, we
restrict attention to bra states with a single such nontrivial vertex, it turns out that the continuum
limit action can be nontrivial only on chargenets with a single nontrivial vertex.
\footnote{We expect that our considerations in P1 admit a straightforward generalization to the case
wherein the bras have multiple nontrivial vertices. We leave the details of such a generaliztion for future work.} 
As a result (and as we shall show in section 2 ),  in our analyis of diffeomorphism covariance it  suffices to 
consider states with a single nontrivial vertex and to restrict attention to a coordinate 
patch around each such vertex. 

The layout of this paper is as follows.
In section 2 we fix a reference charge network in each diffeomorphism class of charge networks and define
coordinate patches on nontrivial vertices of diffeomorphic charge networks as 
diffeomorphic images of the coordinate patch
associated with the nontrivial vertex of the  reference charge network. 
Next, we show that condition (\ref{dc}) translates to a condition involving 
diffeomorphisms which preserve the reference charge network, and the graph structure around 
its nontrivial vertex. This condition 
is fulfilled if any diffeomorphism which preserves
the reference  charge network being acted upon is such that its associated $GL(3,R)$ transformation
at the tangent space of the nontrivial vertex is proportional to a rotation.
In section 3 we show that a reference coordinate system can be chosen at the nontrivial vertex
of each reference charge network in such a way that the $GL(3,R)$ transformations of section 2 are
indeed proportional to rotations. 

In section 4 we 
evaluate the commutator between a pair of Hamiltonian constraints and show that its continuum limit
is ill defined. In sections 5 and 6 we suitably modify our constructions so as to obtain a well defined
anomaly free continuum commutator while preserving the diffeomorphism covariance of the continuum limit of the 
single Hamiltonian  constraint action. In section 5 we  modify
the deformations at finite triangulation so that they acquire a certain ``conical'' structure.
In section 6,  
we modify the VSA topology by modifying the VSA states of P1. Every such state is a weighted sum over 
a certain set of bras.
We modify this set of bras so as to incorporate the modifications of section 5, and, more importantly,
change the weights in the 
weighted sum over these bras. In P1 each bra in such a  sum  was weighted with the evaluation
of a smooth (more precisely, $C^k$ semianalytic) complex function on the spatial manifold at the nontrivial vertex of the bra. Here we extend
the functional dependence to multiple copies of the spatial manifold and the evaluation of the function
to certain  additional vertices of the bra it multiplies. The continuum limit of the commutator involves the
coincidence of some of these additional vertices together with the nontrivial vertex.
A well defined commutator is then obtained by suitably restricting
the `short distance'  behaviour of the function in this coincidence limit.

In section 7 we compute the continuum limit of the single action of the Hamiltonian constraint, demonstrate
its diffeomorphism covariance and compute the continuum limit of the commutator between a pair of Hamiltonian
constraints. We show that the modifications of sections 5 and 6 result in a well defined, anomaly free
continuum commutator.
Section 8 contains a discussion of our results and of open questions, the key open questions being 
the generalization of our results in this paper to the case of gauge group $SU(2)$ and the improvement of 
our $U(1)^3$ considerations from an operator topology based continuum limit to a genuine habitat.

Our work here is based entirely on the contents of P1. In the interests of brevity, we shall assume familiarity with
the notation and contents of P1. The reader interested in following the exposition of this paper is
invited to study P1. The reader unfamiliar with the material in P1 may skip directly to section 8 and peruse sections 8.1, 8.3 and 8.4.

\section{The diffeomorphism covariance condition.}

In section 2.1 we define the notion of nontrivial vertices  and nontrivial states 
and show that it suffices to restrict our attention to nontrivial states.
In section 2.2 we describe our choice of coordinate patches around nontrivial vertices.
In section 2.3 we modify some of the notation of P1 so as to better suit our purposes here.
In section 2.4 we specify the deformations produced by the Hamiltonian constraint at finite triangulation
in such a way that they are consistent with the considerations of P1 and such that they interact 
covariantly with the choice of coordinate patches made in section 2.2.
In section 2.5 we emphasize that the diffeomorphism invariance  of the inverse metric eigenvalue
is more natural in the diffeomorphism covariant context of this work than in P1. 
In section 2.6 we show that condition (\ref{dc1}) is equivalent to a condition on the $GL(3,R)$
transformations induced by $\phi$ on the tangent spaces at nontrivial vertices.

As in P1, we use the semianalytic category \cite{lost} in all our considerations. Thus, the Cauchy slice, 
coordinate patches, diffeomorphisms and (edges of) graphs  are semianalytic and $C^k, k>>1$. 

\subsection{Restriction to nontrivial vertices.}
Let the space of VSA states of section 6, P1 be ${\cal S}_{VSA}$. Each such state is specified by 
a vertex smooth function $f$ and the set of bras $B_{VSA}$ and denoted as $\Psi^f_{B_{VSA}}$. 
From Step 1, section 3.1 of P1 the VSA continuum limit of the Hamiltonian constraint exists iff 
$\lim_{\delta\rightarrow 0}\Psi^f_{B_{VSA}} ({\hat C}_{\delta}(N) |c\rangle)$ exists for every 
$\Psi^f_{B_{VSA}} \in {\cal S}_{VSA}$ and 
every chargenet $c$.
 From section 6, P1, this limit is nontrivial only if  $c$ has a single vertex (which we 
termed nontrivial in section 1) with the following diffeomorphism invariant structure:\\
\noindent (i) The inverse volume operator of P1 has nonvanishing eigenvalue at the vertex.
Such vertices are called nondegenerate. 
\footnote{See, however, section 8.2 of this paper and Footnote 12 of  P1.}
From P1, gauge invariance of $c$ together with nondegeneracy
implies that the vertex must have valence greater than 3.\\
\noindent (ii) No triplet of edge tangents at the vertex is coplanar (such vertices are called GR or Grot- Rovelli
in P1).\\
\noindent (iii) The vertex has a preferred edge which cannot be mapped to any other edge at the vertex by 
any diffeomorphism.
\footnote{In P1 such an edge ended in a trivalent $C^1$ kink whereas the others ended in trivalent
$C^0$ kinks. For our considerations here it is enough that there exists a single edge which is preferred by virtue of
some distinguishing 
diffeomorphism invariant property; the
detailed nature of this property is not important.}

In detail, condition (\ref{dc1}) is defined as the conditions:
\begin{equation}
\lim_{\delta\rightarrow 0}\Psi^f_{B_{VSA}} ({\hat U}^{\dagger}(\phi ){\hat C}_{\delta}(N) {\hat U}(\phi)|c\rangle)
=\lim_{\delta\rightarrow 0}\Psi^f_{B_{VSA}} ({\hat C}_{\delta}(\phi_*N) |c\rangle) \;\;\forall c, \Psi^f_{B_{VSA}}.
\label{dc2}
\end{equation}
For all chargenets $c$ which do not have one ( i.e. have  none or more than one) nontrivial vertex it follows
from P1 that both sides of equation (\ref{dc2}) vanish 
by virtue of 
the diffeomorphism invariant 
nature of  (i)- (iii) above. Hence we need only show that the condition (\ref{dc2}) holds for 
$c$ with a single nontrivial vertex. Let us call such $c$ as {\em nontrivial} and restrict our attention 
to nontrivial $c$ from now on.

\subsection{Diffeomorphism covariant choice of coordinate patches}

We choose a `reference' charge network $c_0$ in each diffeomorphism class of nontrivial charge networks
$[c_0]$, and choose a reference coordinate patch $\{x\}_{c_0}$ 
around the nontrivial vertex $v$ of the reference charge network. 
For each distinct diffeomorphic charge network $c \in [c_0], c\neq c_0$, we choose a diffeomorphism $\beta$
such that 
\begin{equation}
|c>={\hat U}(\beta )|c_0\rangle  =:|c_{0}^{\beta}\rangle. 
\end{equation}
Let the set of these diffeomorphisms
be ${\cal D}_{[c_0]}$. We choose the coordinate patch at the vertex $\beta(v)$ of the charge network $c^{\beta}_0$
to be obtained by the push forward of the reference coordinate patch by $\beta$. We denote this by
\begin{equation}
\{x \}_{c^{\beta}_0}:=  \beta^* \{x \}_{c_0}
\label{cpatch1}
\end{equation}
In what follows, if the context is clear, we shall condense our notation and define:
\begin{equation}
\{x\}_{c_0}=:\{x\} \;\;\;\;\;,
 \{x \}_{c^{\beta}_0}=: \{x_{\beta} \}
\label{cpatch2}
\end{equation}

In order that the commutator be anomaly free, it is necessary to restrict  the choice of reference chargenetworks
and reference coordinate patches as follows.
\footnote{A key feature which ensures the anomaly free action of the commutator in the non- covariant setting of P1 
is that the coordinate patches at the nontrivial vertex $v_{I_{v},\delta }^{\prime}$ of 
the deformed charge nets 
$c(i,v_{I_{v},\delta}^{\prime}), i=1,2,3$ generated by ${\hat C}_{\delta}(N)$ and the deformed charge net
 $c(v_{I_{v},\delta}^{\prime})$ generated by the electric diffeomorphism operator, ${\hat D}_{\delta} ({\vec N}_i)$,
are identical. The restrictions below are a reflection of this feature  in the diffeomorphsim covariant context of
this work.} 
First we develop some nomenclature to phrase the nature of the restrictions. In what follows, recall
that the graph underlying the charge net $c$ is denoted by $\gamma (c)$.
\\

\noindent{\em Definition 1. The $(i, I_v)$ child ${\tilde c}$  of $c$:}
Let ${\tilde c}, c$ be nontrivial chargenets 
with nontrivial $M$ valent vertices ${\tilde v},v, v\neq {\tilde v}$. 
Let the $M$ distinct edges in $c$ emanating from $v$ be denoted as $\{e_{J_v}, J_v=1,.., M\}$. 
The charge net ${\tilde c}$
is an $(i, I_v)$ child of $c$ if the following hold:
\\
\noindent (i) $\gamma ({\tilde c}) \supset \gamma (c)$
\\
\noindent (ii) On the interior of every $C^k$ semianalytic edge $e_{J_v}\subset \gamma (c) $ emanating from $v$
there exists a point ${\tilde v}_{J_v}$ which is connected to ${\tilde v}$ by a unique $C^k$ semianalytic edge
${\tilde e}_{J_v}\subset \gamma ({\tilde c})$ such that 
\begin{eqnarray}
&&\gamma ({\tilde c})= \gamma (c) {{\cup}} (\cup_{J_v} {\tilde e}_{J_v}),  \\
&&{\tilde e}_{J_v}\cap \gamma (c) = {\tilde v}_{J_v},\\
&& {\tilde e}_{J_v}\cap{\tilde e}_{K_v\neq J_v} = {\tilde v}.
\end{eqnarray}
In addition ${\tilde v}_{J_v\neq I_v}$ are $C^0$ kink vertices of $\gamma ({\tilde c}) $ and ${\tilde v}_{I_v}$ is
a $C^1$ kink vertex of $\gamma ({\tilde c}) $.
\\
\noindent
(iii) The charge labels of $c,{\tilde c}$ agree  on 
$\gamma (c) - \cup_{J_v} e_{v, {\tilde v}_{J_v}}$ where $e_{v, {\tilde v}_{J_v}}$ is that part of the edge $e_{J_v}$
which connects $v$ to ${\tilde v}_{J_v}$. The charges ${\tilde q}^k_{J_v}$ on ${\tilde e}_{J_v}$ in $\tilde c$ 
are related by an ``$i$- flip''
to the charges $q^k_{J_v}$ on $e_{J_v}$ in $c$ where the $i$ flipping is defined as in the second equation of 
section 4.4.3, P1 so that 
\begin{equation}
{\tilde q}^j_{J_v}= \delta^{ij} q^j_{J_v} - \sum_{k}\epsilon^{ijk}q^k_{J_v}.
\end{equation}
Note that the charges in $\tilde c$ on $e_{v, {\tilde v}_{J_v}}$ are then uniquely determined by 
gauge invariance and that the vertex $v$ in $\tilde c$ becomes degenerate since the edges
$e_{v, {\tilde v}_{J_v}}$ have vanishing $i$th charge.
\\

\noindent{\em Definition 2. The unflipped sibling of an ${(i,I_v})$ child:}
Let ${\tilde c}$ be  an $(i, I_v)$ child of $c$. Then ${\hat c}$ is the (unique) unflipped 
sibling of ${\tilde c}$ iff, in the notation used in Definition 1, the following hold:\\
\noindent (i) $\gamma ({\hat c}) = \gamma ({\tilde c}) - \cup_{J_v} e_{v, {\tilde v}_{J_v}}$
\\
\noindent (ii) The charges in ${\hat c}$ on $\gamma ({\hat c}) \cap \gamma (c) $ agree with those
on $\gamma ({\hat c}) \cap \gamma (c) $ in $c$. 

Note that ${\tilde v}_{J_v\neq I_v}$ are bivalent $C^0$ kink vertices of ${\hat c}$ and ${\tilde v}_{ I_v}$
is a bivalent $C^1$ kink vertex of ${\hat c}$ and  that by virtue of (ii), Definition 2
and the bivalent nature of these vertices, 
gauge invariance implies that the charges on ${\tilde e}_{J_v}$  in ${\hat c}$ agree with those
on $e_{J_v}$ in $c$. Note also that the vertex $v$ of ${\tilde c}$
has disappeared in ${\hat c}$.\\

\noindent{\em Definition 3. Flipped $I_v$ Siblings:} Fix $I_v$ in Definition 1. 
Let ${\tilde c}_i, i=1,2,3$ be $(i, I_v)$ children of $c$. The 3 chargenets 
${\tilde c}_i, i=1,2,3$ are siblings of each other iff $\gamma ({\tilde c}_1)=\gamma ({\tilde c}_2)= 
\gamma ({\tilde c}_3)$.
\\

An example of an $(i, I_v)$ child of $c$ is the charge net $c(i, v_{I_{v},\delta }^{\prime})$ of P1.
The unflipped sibling of $c(i, v_{I_{v},\delta }^{\prime})$ is the charge net $c(v_{I_{v},\delta }^{\prime})$ of
P1. The chargenets $c(j\neq i, v_{I_{v},\delta }^{\prime})$ of P1 are the flipped siblings of 
$c(i, v_{I_{v},\delta }^{\prime})$.
Recall from P1 that $c(k, v_{I_{v},\delta }^{\prime}), k=1,2,3$ are generated by the action of the finite triangulation
Hamiltonian constraint on $c$ and that $c(v_{I_{v},\delta }^{\prime})$ is generated by the action of the 
finite triangulation electric diffeomorphsim operator on $c$.

We are now ready to phrase the restrictions
 on the reference chargenets and reference coordinate patches alluded to in the beginning of this section as follows.
Let $[{\bar c}^1], [{\bar c}^2]$  be such that there exist $c^1\in [{\bar c}^1]$, $c^2\in [{\bar c}^2]$
such that either 
(a) $c^1, c^2$ are flipped $I_v$ siblings, or,  (b)$c^1$ is the unflipped $I_v$ sibling of $c^2$ (or vice versa).
Then we require that the reference charge nets $c^1_0, c^2_0$  for $[{\bar c}^1], [{\bar c}^2]$
be, correspondingly either (a) flipped $I_v$ siblings of each other or (b) be related in such a way that
one is the unflipped $I_v$ sibling of the other. 
Further,
we require that the reference coordinate charts for $c^1_0, c^2_0$ be identical(
Note that $c^1_0, c^2_0$ which satisfy these requirements necessarily have the same nontrivial vertex).
\\

\noindent{\em Note:} Due to our proof of diffeomorphism covariance in section 3 (see especially section 3.5)
it is {\em not} necessary to correlate the sets of reference diffeomorphisms for siblings.

Next, we fine 
tune some of the notation of section 4, P1 for our purposes here.

\subsection{Notation for deformed charge networks}

The last equation of section 4.4,P1, evaluated on a nontrivial charge network state with the nontrivial vertex $v$ is:
\begin{equation}
\hat{C}_{\delta}[N] c (A)=\frac{\hbar}{2\mathrm{i}}\frac{3}%
{4\pi}N(x(v))\nu_{v}^{-2/3}\sum_{I_v, i}q_{I_{v}}^{i}\frac{1}{\delta
}c(i,v_{I_{v},\delta}^{\prime}) 
\label{cdeltancfinal}
\end{equation}
Here $c(i,v_{I_{v},\delta}^{\prime})$ denotes the nontrivial charge network state obtained by deforming the 
nontrivial charge network state $c(A)$ along the edge $I_v$ emanating from the nontrivial vertex $v$ so that the
position of the nontrivial vertex is deformed from $v$ to $v_{I_{v},\delta}^{\prime}$. The distance between
$v$ and $v_{I_{v},\delta}^{\prime}$ is of $O(\delta )$ as measured by the coordinate patch at $v$. The 
edges emanataing from the deformed vertex $v_{I_{v},\delta}^{\prime}$ are colored by charges which are
determined by `flipping'  the charges at $v$ in $c$, the flipping determined by the value of the $U(1)^3$
index $i$.  

For our purposes here it is important to keep track of {\em which} coordinate system is being used to specify
the deformation. We denote the coordinate patch associated with the vertex $v$ of $c$ by $\{x\}_c$ and 
write the original and deformed charge net states as:
\begin{equation}
|c\rangle \equiv c(A),  \;\;\;\;\;\;\; 
|c,\{x\}_c, i,v_{I_{v},\delta}^{\prime}\rangle 
\equiv c(i,v_{I_{v},\delta}^{\prime}) .
\end{equation}
In order to display the role of the coordinate patch in the evaluation of the density weighted lapse, we use the
notation:
\begin{equation}
N(v, \{x \}_c ) \equiv N(x(v)).
\end{equation}
With this, equation (\ref{cdeltancfinal}) reads:
\begin{equation}
\hat{C}_{\delta}[N]|c\rangle=
\frac{\hbar}{2\mathrm{i}}\frac{3}%
{4\pi}N(v, \{x\}_c )\nu_{v}^{-2/3}\sum_{I_{v},i}q_{I_{v}}^{i}\frac{1}{\delta
}|c, \{x\}_c, i,v_{I_{v},\delta^{\prime}} \rangle
\label{cdeltaket}
\end{equation}
We shall often drop the subscript $c$ of $\{x\}_c$ and simply write $\{x\}$ when the context is clear .

Next, let $\chi$ be a diffeomorphism which maps $c$ to $ \chi (c)=:c^{\chi}$. Let the nontrivial vertex $v$ of  $c$ be
$M$ valent. Then $\chi$ maps the set of $M$ edges at $v$ in $c$ to the set of $M$ edges at
$\chi (v)$ in $c^{\chi}$. By virtue of the invertibility of $\chi$, this mapping between edge sets
is bijective. Hence  if $I_v$ indexes the edge set at $v$ and $I_{\chi (v)}$ indexes the edge set at 
$\chi (v)$, then $\chi$ maps each distinct value of $I_v \in \{1,2,., M\}$ to a distinct value of 
$I_{\chi (v)} \in \{1,2,., M\}$. Thus $\chi$ defines a permutation $\pi_{\chi}$ on the set
$\{1,2,., M\}$ so that the $I_v^{\;\rm th}$ 
edge at $v$ is mapped to the $(\pi_{\chi}(I_v))^{\rm th}$ edge at 
$\chi (v)$. 
Let the charges on $c^{\chi}$ be $q^{ (\chi ) i}_{J_{\chi (v)}}, J_{\chi (v)}= 1,..,M$. Then we have that 
\begin{equation}
q^{(\chi )i}_{   \pi_{\chi (J_v)} }= q^{i}_{J_{v}}.
\label{2.12a}
\end{equation}
In particular, if the diffeomorphism $\chi$ is a symmetry of $c$ so that $c^{\chi}= c$ then it follows that
$q^{ (\chi ) i }_{\pi_{\chi (J_v)}}= q^{i}_{ \pi_{\chi (J_v)}}$. It then follows from equation (\ref{2.12a}) that 
\begin{equation}
q^{i}_{ \pi_{\chi (J_v)}}= q^{i}_{J_{v}} \;\;{\rm if }\;\;
c^{\chi}= c
\label{2.12b}
\end{equation}
\subsection{Diffeomorphism covariant deformations}

The Hamiltonian constraint generates the deformations (\ref{cdeltancfinal}), (\ref{cdeltaket})
on the charge network $|c\rangle$. In P1 deformations of diffeomorphic charge nets were not
necessarily related in any way. Here we define deformations of diffeomorphic charge networks
to be related by diffeomorphisms as follows.

Let the deformations at parameter $\delta$ be defined as in P1 for the reference charge network $|c_0>$ 
in terms of the reference coordinate system $\{x\}$. 
The deformed charge network (see equation (\ref{cdeltaket}))
obtained by deforming $|c_0\rangle$ along the edge $I_v$ with charge flips determined by $i$, 
is $|c_0, \{x\}, i,v_{I_{v},\delta}^{\prime}) \rangle$. We define the corresponding deformation of the 
diffeomorphic charge net $|c_0^{\alpha}\rangle$, $\alpha \in {\cal D}_{[c_0]}$,
 to be the image of $|c_0, \{x\}, i,v_{I_{v},\delta}^{\prime} \rangle$ by the diffeomorphism $\alpha$.
Thus, the deformed charge network 
obtained by deforming $|c^{\alpha}_0\rangle$ along the  $\pi_{\alpha} (I_v)$th edge
 with charge flips determined by $i$, 
is $ {\hat U}(\alpha )|c_0, \{x\}, i,v_{I_{v},\delta}^{\prime} \rangle$ where, as defined in the previous
section, $\pi_{\alpha} (I_v)$ is the image of $I_v$
by $\alpha$ and ${\hat U}(\alpha )$ is the
unitary operator corresponding to the diffeomorphism $\alpha$.

Recall that the coordinate system $\{x_{\alpha}\}$ associated with the nontrivial vertex $\alpha (v)$ of 
$c^{\alpha}_0$ is the image of the reference coordinate system by $\alpha$. 
Since the deformation of the reference charge network satisfies all the requirements of P1 and is of $O(\delta )$
in the coordinate system $\{x \}$, it immediately follows that 
that the 
deformation of $c_0^{\alpha}$ defined above is of $O(\delta )$  as measured by the coordinate system $\{x_{\alpha}\}$
and that this deformation
also satisfies all the requirements of P1.
Hence, in the notation of section 2.3, it follows that we may set:
\begin{equation}
|c^{\alpha}_0, \{x_{\alpha}\}, i,v_{ \pi_{\alpha}(I_{v})  ,\delta}^{\prime} \rangle
:= {\hat U}(\alpha )|c_0, \{x\}, i,v_{I_{v},\delta}^{\prime} \rangle
\label{dcdeform}
\end{equation}

\subsection{Diffeomorphism invariance of the inverse metric eigenvalue}

The considerations of 
Appendix A, P1  show that the inverse metric eigenvalue ${\nu}^{-\frac{2}{3}}$
derived there is diffeomorphism invariant. There, the expressions are based
on a non- covariant choice of coordinate patches. Diffeomorphism invariance
of the eigenvalue requires that the length of the regulating holonomies
be adjusted so as to cancel out factors which come from the change in
the regulating holonomy edge tangents under diffeomorphisms. More in detail, there we require that 
the combination $B_{123}\lambda ({\vec e})$ be constant over each 
 diffeomorphism
class of charge networks, the parameter $B_{123}$ characterising how long the edge holonomies
are and the parameter $\lambda ({\vec e})$ quantifying the change
in the regulating holonomy edge tangents. In P1, as a result of the non- covariant choice of coordinate patches,
the value of 
$\lambda ({\vec e})$ differs from chargenet to chargenet in a single diffeomorphism class. Consequently
we need to adjust $B_{123}$ accordingly.
In contrast, here, due to the diffeomorphism covariant choice of coordinate patches, it is easy to check that 
the value of $\lambda ({\vec e})$ is constant over the entire diffeomorphism class so that the parameter
$B_{123}$ needs no adjustment. With regard to the analysis of the inverse metric eigenvalue, this is the {\em only} difference between the non-covariant context of P1 and the
covariant setting of this paper.

Clearly, the diffeomorphism invariance of the eigen value is more natural
in the diffeomorphism covariant context of this paper.
It is also easy to verify the constancy of the eigenvalue over the diffeomorphism classes of flipped children
as well as their unflipped sibling. This follows from the fact that the restrictions on the choice of reference 
charge nets in Appendix A,P1 are identical to the `sibling' restrictions on the choice of reference chargenets
in section 2.2.

\subsection{The diffeomorphism covariance condition and $GL(3,R)$.}

Note that there exists a unique $\alpha \in {\cal D}_{[c_0]}$ such that 
\begin{equation}
{\hat U}(\phi )|c_0\rangle = {\hat U}(\alpha )|c_0\rangle =:|c_{0}^{\alpha}\rangle
\label{phi=alpha}
\end{equation}
Next, consider the left hand side of 
equation (\ref{dc2}) when $c$ is a nontrivial reference charge net (i.e.  $c=c_0$) with the single 
nontrivial vertex $v$. 
From equations (\ref{phi=alpha}), (\ref{cdeltaket}), (\ref{dcdeform}) and (\ref{2.12a})
we have that 
\begin{eqnarray}
\hat{C}_{\delta}[N]{\hat U}(\phi )|c_0\rangle &=&
\hat{C}_{\delta}[N]|c^{\alpha}_0\rangle \nonumber\\
&=&\frac{\hbar}{2\mathrm{i}}\frac{3}%
{4\pi}N(\alpha (v), \{x_{\alpha}\})\nu_{\alpha (v)}^{-2/3}\sum_{I_{v},i}
     q_{   \pi_{\alpha}(I_{v})   }^{   (\alpha )i} \frac{1}{\delta
}|c^{\alpha}_0, \{x_{\alpha}\}, i,v_{ \pi_{\alpha}(I_{v})  ,\delta}^{\prime} \rangle 
\\
&=&\frac{\hbar}{2\mathrm{i}}\frac{3}%
{4\pi}N(\alpha (v), \{x_{\alpha}\})\nu_{\alpha (v)}^{-2/3}\sum_{I_{v},i}
     q_{I_{v} }^{i} 
\frac{1}{\delta}|c^{\alpha}_0, \{x_{\alpha}\}, i,v_{ \pi_{\alpha}(I_{v})  ,\delta}^{\prime} \rangle 
\label{cnphic0}
\end{eqnarray}
To reiterate, the notation of section 2.3 reminds us that the deformations produced by 
$\hat{C}_{\delta}[N]$ are defined in terms of the coordinate system $\{x_{\alpha}\}$ at the vertex
$\alpha (v)$ of the charge network $|c^{\alpha}_0\rangle$ where $|c^{\alpha}_0\rangle$, $\{x_{\alpha}\}$ and 
$\alpha (v)$ are the images of the reference charge network $|c_0\rangle$, the reference coordinate patch $\{x\}$ 
and the  nontrivial vertex 
$v$, by the diffeomorphism $\alpha$.

Next, consider the action of ${\hat U}^{\dagger}(\phi )$ on 
$|c^{\alpha}_0, \{x_{\alpha}\}, i,v_{ \pi_{\alpha}(I_{v})  ,\delta}^{\prime}) \rangle$.
From equation (\ref{dcdeform}), we have that 
\begin{equation}
{\hat U}^{\dagger}(\phi )|c^{\alpha}_0, \{x_{\alpha}\}, i,v_{ \pi_{\alpha}(I_{v})  ,\delta}^{\prime} \rangle
= {\hat U}(\phi^{-1}\circ\alpha )|c_0, \{x\}, i,v_{I_{v},\delta}^{\prime} \rangle
\label{phicalpha}
\end{equation}
From equation (\ref{phi=alpha}) it follows that the diffeomorphism $\phi^{-1}\alpha$ leaves the 
reference charge network invariant so that we have that:
\begin{equation}
{\hat U}(\phi^{-1}\circ\alpha )|c_0\rangle = |c_0\rangle
\label{symmref}
\end{equation}
Consider the coordinate system $\{x_{\phi^{-1}\circ\alpha}\}$ which is obtained as the push forward of the
coordinate system $\{x \}$ by $\phi^{-1}\circ\alpha$. 
Since the charge net $|c_0, \{x\}, i,v_{I_{v},\delta}^{\prime} \rangle$
is obtained by deforming the reference charge net $|c_0\rangle$ along the lines of P1 and since the 
deformation is of $O(\delta )$ in the reference coordinate system $\{x \}$, it follows that 
the charge net  ${\hat U}(\phi^{-1}\circ\alpha )|c_0, \{x\}, i,v_{I_{v},\delta}^{\prime} \rangle$
may be viewed as a deformation of the charge net ${\hat U}(\phi^{-1}\circ\alpha )|c_0\rangle =|c_0\rangle$
which satisfies all the requirements of P1 and which is of $O(\delta )$ in the coordinate system
$\{x_{\phi^{-1}\circ\alpha}\}$. Thus we may write:
\begin{equation}
|c_0, \{x_{\phi^{-1}\circ\alpha}\}, i,v_{ \pi_{\phi^{-1}\circ\alpha}(I_{v})  ,\delta}^{\prime} \rangle
= {\hat U}(\phi^{-1}\circ\alpha )|c_0, \{x\}, i,v_{I_{v},\delta}^{\prime} \rangle
\label{symmdef}
\end{equation}

From equations 
(\ref{cnphic0}), (\ref{phicalpha}) and (\ref{symmdef}) 
we have that 
\begin{eqnarray}
{\hat U}(\phi )^{\dagger}\hat{C}_{\delta}[N]{\hat U}(\phi )|c_0\rangle 
&=&\frac{\hbar}{2\mathrm{i}}\frac{3}%
{4\pi}N(\alpha (v), \{x_{\alpha}\})\nu_{\alpha (v)}^{-2/3}
\nonumber\\
& &\sum_{I_{v},i}
q^i_{I_v}
\frac{1}{\delta
}|c_0, \{x_{\phi^{-1}\circ\alpha}\}, i,v_{ \pi_{\phi^{-1}\circ\alpha}(I_{v})  ,\delta}^{\prime} \rangle
\label{2.19}
\end{eqnarray}

Focus on the first line of the right hand side of the above equation.
From the diffeomorphism invariance of the inverse volume operator eigenvalue (see section 2.5), it follows that 
$\nu_{\alpha (v)} = \nu_v$ where $\nu_v$ is the inverse volume eigenvalue for the nontrivial vertex
of $c_0$. Next,
denoting the push forward of the 
reference coordinate system $\{x \}$ by $\phi$ as $\{x_{\phi}\}$, we have, from the density $-\frac{1}{3}$ 
property of the lapse that
\begin{equation}
N(\alpha (v), \{x_{\alpha}\})= N(\phi (v), \{x_{\phi}\}) 
|\frac{\partial x_{\phi}}{\partial x_{\alpha}}|^{-\frac{1}{3}}.
\end{equation}
Here, we have used $\alpha (v)= \phi (v)$ and denoted the determinant of the Jacobian between the 
$\{x_{\phi}\}$ and $\{x_{\alpha}\}$ coordinates at $\phi (v)$ by 
$|\frac{\partial x_{\phi}}{\partial x_{\alpha}}|$
Note that since all diffeomorphisms of interest are connected to identity and hence are orientation preserving, 
this determinant is 
positive. Next note that the Jacobian $\frac{\partial x_{\phi}}{\partial x_{\alpha}}$ at the point $p= \phi (v)$
is related to a Jacobian at the point $p=v$ through the following identities which follow from the 
fact that the coordinate systems involved may be obtained as suitable push forwards of the reference 
coordinate system. In obvious notation, we have that:
\begin{eqnarray}
\frac{\partial x^{\mu}_{\phi}(p)}{\partial x^{\nu}_{\alpha}(p)}|_{p=\alpha (v)}
&=& \frac{\partial x^{\mu}(\phi^{-1}(p))}{\partial x^{\nu}({\alpha}^{-1}(p))}|_{p=\alpha (v)}
= \frac{\partial x^{\mu}(\phi^{-1}\circ \alpha\circ \alpha^{-1}(p))}
{\partial x^{\nu}({\alpha}^{-1}(p))}|_{\alpha^{-1}(p)= v}\nonumber \\
&=&
\frac{\partial x^{\mu}(\phi^{-1}\circ \alpha(p))}
{\partial x^{\nu}(p)}|_{p= v}\nonumber\\
&=&
\frac{\partial x^{\mu}_{\alpha^{-1}\circ\phi}(p)}{\partial x^{\nu}(p)}|_{p=v}:= G^{\mu}_{\;\;\nu}
\label{defG}
\end{eqnarray}
Noting that $\phi_*N(v, \{x\})= N(\phi (v), \{x_{\phi}\})$ and denoting the determinant of $G^{\mu}_{\;\;\nu}$
by $\det G$, it follows that 
\begin{eqnarray}
{\hat U}(\phi )^{\dagger}\hat{C}_{\delta}[N]{\hat U}(\phi )|c_0\rangle 
&=&\frac{\hbar}{2\mathrm{i}}\frac{3}%
{4\pi}\phi_*N(v, \{x\})\nu_{v}^{-2/3}(\det G)^{-\frac{1}{3}}
\nonumber\\
& &\sum_{I_{v},i}q_{\pi_{\phi^{-1}\circ\alpha}(I_{v})}^{i}\frac{1}{\delta
}|c_0, \{x_{\phi^{-1}\circ\alpha}\}, i,v_{ \pi_{\phi^{-1}\circ\alpha}(I_{v})  ,\delta}^{\prime} \rangle ,
\label{cndeltaG}
\end{eqnarray}
where we have used equation (\ref{2.12b}) and the fact that $\phi^{-1}\circ \alpha$ is a symmetry of $c_0$
to replace $q_{I_v}$  in equation (\ref{2.19}) by $q_{\pi_{\phi^{-1}\circ\alpha}(I_{v})}^{i}$.
Next, we evaluate the continuum limit action of the operator 
${\hat U}(\phi )^{\dagger}\hat{C}_{\delta}[N]{\hat U}(\phi )$. Accordingly, we first evaluate 
the action of the VSA state $\Psi_{B_{VSA}}^f$ on equation (\ref{cndeltaG}). From P1, this action, when nontrivial
is given by
\begin{eqnarray}
 & \Psi_{B_{VSA}}^f ({\hat U}(\phi )^{\dagger}\hat{C}_{\delta}[N]{\hat U}(\phi )|c_0\rangle)  
\;\;\;\;\;\;\;\;\;\;\;\;\;\;\;\;\;\;\;\;\;\;\;\;\;\;\;\;\;\;\;
\;\;\;\;\;\;\;\;\;\;\;\;\;\;\;\;\;\;\;\;\;\;\;\;\;\;\;\;\;\;\;
\nonumber\\
 & =
\frac{\hbar}{2\mathrm{i}}\frac{3}%
{4\pi}\phi_*N( v, \{x\})\nu_{v}^{-2/3}(\det G)^{-\frac{1}{3}}  
\Psi_{B_{VSA}}^f (
\sum_{I_{v},i}q_{\pi_{\phi^{-1}\circ\alpha}(I_{v})}^{i}\frac{1}{\delta
}|c_0, \{x_{\phi^{-1}\circ\alpha}\}, i,v_{ \pi_{\phi^{-1}\circ\alpha}(I_{v})  ,\delta}^{\prime} \rangle ) 
\nonumber\\
 & = 
\frac{\hbar}{2\mathrm{i}}\frac{3}%
{4\pi}\phi_*N(v , \{x\})\nu_{v}^{-2/3}(\det G)^{-\frac{1}{3}}  
\sum_{I_{v},i}q_{\pi_{\phi^{-1}\circ\alpha}(I_{v})}^{i}\frac{1}{\delta
} f (v_{ \pi_{\phi^{-1}\circ\alpha}(I_{v})  ,\delta}^{\prime}) 
\nonumber\\
 & =
\frac{\hbar}{2\mathrm{i}}\frac{3}%
{4\pi}\phi_*N(v, \{x\})\nu_{v}^{-2/3}(\det G)^{-\frac{1}{3}} 
\sum_{I_{v},i}q_{\pi_{\phi^{-1}\circ\alpha}(I_{v})}^{i}
\frac{f (v_{ \pi_{\phi^{-1}\circ\alpha}(I_{v})  ,\delta}^{\prime})- f(v)}{\delta}  
\end{eqnarray}
where in the last line we have used gauge invariance to add the $f(v)$ term. Taking the continuum limit, 
and using the properties of the deformation described in P1, we obtain:
\begin{eqnarray}
\lim_{\delta\rightarrow 0} \Psi_{B_{VSA}}^f ({\hat U}(\phi )^{\dagger}\hat{C}_{\delta}[N]{\hat U}(\phi )|c_0\rangle
&=&
\frac{\hbar}{2\mathrm{i}}\frac{3}%
{4\pi}\phi_*N(v, \{x\})\nu_{v}^{-2/3}(\det G)^{-\frac{1}{3}}\nonumber\\
& & \sum_{I_{v},i}q_{\pi_{\phi^{-1}\circ\alpha}(I_{v})}^{i}
{\hat e}^{\prime a}_{ \pi_{\phi^{-1}\circ\alpha}(I_{v}) } \partial_a f(v).
\label{2.22}
\end{eqnarray}
Here ${\hat e}_{ \pi_{\phi^{-1}\circ\alpha}(I_{v}) }^{\prime a}$ is the unit vector along 
the $\pi_{\phi^{-1}\circ\alpha}(I_{v})$th edge. 
The superscript, ${\hat{\;}}^\prime$
indicates that the tangent vector is unit with respect to the 
coordinate metric associated with the $\{x_{\phi^{-1}\circ\alpha}\}$ system. The index $a$ is a vector index
and we may choose to evaluate the components of the vector in any coordinate system we wish, since
the index $a$ is contracted with $\partial_a$.

We now compare the  vector ${\hat e}_{ \pi_{\phi^{-1}\circ\alpha}(I_{v}) }^{\prime a}$
with the tangent vector ${\hat e}_{ \pi_{\phi^{-1}\circ\alpha}(I_{v}) }^a$ which is  
along the same edge but which is of unit norm in the coordinate metric associated with 
the $\{x\}$ system. 
Clearly, the two vectors are proportional so that:
\begin{equation}
{\hat e}_{ \pi_{\phi^{-1}\circ\alpha}(I_{v}) }^{\prime a}
=\lambda {\hat e}_{ \pi_{\phi^{-1}\circ\alpha}(I_{v}) }^a
\label{deflambda}
\end{equation}
where $\lambda$ is some positive real number. We now compute $\lambda$.
Let the components of the vector ${\hat e}_{ \pi_{\phi^{-1}\circ\alpha}(I_{v}) }^{\prime a}$
in the coordinate system $\{x_{\phi^{-1}\circ\alpha}\}$ be 
${\hat e}_{ \pi_{\phi^{-1}\circ\alpha}(I_{v}) }^{\prime {\mu}^{\prime}}$ and let its components in the 
$\{x\}$ system be ${\hat e}_{ \pi_{\phi^{-1}\circ\alpha}(I_{v}) }^{\prime\mu}$.
Then we have that 
\begin{equation}
{\hat e}_{ \pi_{\phi^{-1}\circ\alpha}(I_{v}) }^{\prime\mu}=
{\hat e}_{ \pi_{\phi^{-1}\circ\alpha}(I_{v}) }^{\prime {\nu}^{\prime}}
\frac{\partial x^{\mu}}{\partial x_{\phi^{-1}\circ\alpha}^{\nu^{\prime}}}
\label{normprimenorm}
\end{equation}
Next, we note the following properties of diffeomorphically related vectors.
Let $\chi$ be a diffeomorphism which maps the point $p$, a coordinate patch $\{x\}$ around the 
point $p$ and the vector $v^a$ in the tangent space at $p$ to $\chi (p),\{x_{\chi}\}, v_{\chi}^a$. 
Then it follows that:\\
\noindent (i) The components $v^{\mu}$ of $v^a$ in the $\{x\}$ system are equal to the components
$v_{\chi}^{\mu^{\prime}}$ of $v_{\chi}^a$ in the $\{x_{\chi}\}$ system i.e. 
\begin{equation}
v_{\chi}^{\mu^{\prime}}= v^{\mu} \delta^{\mu^{\prime}}_{\mu }
\end{equation}
where $\delta^{\mu^{\prime}}_{\mu }$ is the Kronecker delta function.\\
\noindent (ii) The coordinate metrics associated with $\{x\}$ and $\{x_{\chi}\}$
are related by $\chi$, so that the norm of $v_{\chi}^a$ in the $\{x_{\chi}\}$ coordinate metric 
is the same as 
the norm of  $v^a$ in the $\{x\}$- coordinate metric i.e. 
\begin{equation}
\sum_{\mu}(v^{\mu})^2  = \sum_{\mu^{\prime}}(v_{\chi}^{\mu^{\prime}})^2  
\end{equation}
Setting  $\chi = \phi^{-1}\circ\alpha$ and noting  that 
the image of the $I_v$th edge by $\phi^{-1}\circ\alpha$ is 
precisely the $\pi_{\phi^{-1}\circ\alpha}(I_{v})$th edge, we use 
(i)-(ii) above in
equation (\ref{normprimenorm}) to obtain 
\begin{equation}
{\hat e}_{ \pi_{\phi^{-1}\circ\alpha}(I_{v}) }^{\prime\mu}=
{\hat e}_{ I_{v} }^{\nu}
\frac{\partial x^{\mu}}{\partial x_{\phi^{-1}\circ\alpha}^{\nu}},
\label{npn}
\end{equation}
where ${\hat e}_{ I_{v} }^{\nu}$ are the components of ${\hat e}_{ I_{v} }^{a}$ in the $\{x\}$ system.
The partial derivatives on the right hand side of the above equation can be rexpressed in terms of the matrix
$G$ (see equation (\ref{defG}) as follows:
\begin{eqnarray}
\frac{\partial x^{\mu}}{\partial x_{\phi^{-1}\circ\alpha}^{\nu}}
&:=& \frac{\partial x^{\mu}(p)}{\partial x^{\nu}_{\phi^{-1}\circ\alpha}(p)}|_{p=v}\nonumber\\
&=& \frac{ \partial x^{\mu} (\phi^{-1}\circ \alpha  \circ  \alpha^{-1}\circ\phi (p)) }
{\partial x^{\nu}({\alpha}^{-1}\circ{\phi} (p))}|_{p=v}
= \frac{\partial x^{\mu}(\phi^{-1}\circ \alpha (p))}
{\partial x^{\nu}(p)}|_{p= \alpha^{-1}\circ \phi (v)= v}\nonumber \\
&=&
 G^{\mu}_{\;\;\nu}
\label{2.28}
\end{eqnarray}

From equation (\ref{deflambda}) it follows that 
\begin{equation}
\lambda = \sqrt{\sum_{\mu}({\hat e}_{ \pi_{\phi^{-1}\circ\alpha}(I_{v}) }^{\prime\mu})^2}
\end{equation}
Equations (\ref{npn}) and (\ref{2.28}) then imply that 
\begin{equation}
\lambda = \sqrt{
\sum_{\mu,\nu,\tau}({\hat e}_{I_v}^{\nu}G^{\mu}_{\;\;\nu}{\hat e}_{I_v}^{\tau}G^{\mu}_{\;\;\tau})}
\label{2.29}
\end{equation}
Using this in equation (\ref{2.22}) leads to:
\begin{eqnarray}
\lim_{\delta\rightarrow 0} \Psi_{B_{VSA}}^f ({\hat U}(\phi )^{\dagger}\hat{C}_{\delta}[N]{\hat U}(\phi )|c_0\rangle
=
\frac{\hbar}{2\mathrm{i}}\frac{3}%
{4\pi}\phi_*N(\alpha (v), \{x\})\nu_{v}^{-2/3}(\det G)^{-\frac{1}{3}}\nonumber\\
  \sum_{I_{v},i} \sqrt{
\sum_{\mu,\nu,\tau}({\hat e}_{I_v}^{\nu}G^{\mu}_{\;\;\nu}{\hat e}_{I_v}^{\tau}G^{\mu}_{\;\;\tau})}
       q_{\pi_{\phi^{-1}\circ\alpha}(I_{v})}^{i}
{\hat e}^{a}_{ \pi_{\phi^{-1}\circ\alpha}(I_{v}) } \partial_a f(v).
\label{2.31}
\end{eqnarray}

Let us now  make the following assumption.\\
\noindent{\em Assumption:} The matrix $G$ is proportional to a rotation $R$ i.e.
\begin{equation}
G^{\mu}_{\;\;\nu}= CR^{\mu}_{\;\;\nu},\;\;\; RR^{T}={\bf 1},\;\;\; \det R =1
\label{assumeG=cR}
\end{equation}
where $R^T$ denotes the transpose of $R$.

Under this assumption it is easy to see that equation (\ref{2.31}) reduces to
\begin{eqnarray}
&&\lim_{\delta\rightarrow 0} \Psi_{B_{VSA}}^f ({\hat U}(\phi )^{\dagger}\hat{C}_{\delta}[N]{\hat U}(\phi )|c_0\rangle
\nonumber\\
&&=\frac{\hbar}{2\mathrm{i}}\frac{3}%
{4\pi}\phi_*N(v, \{x\})\nu_{v}^{-2/3}
 \sum_{I_{v},i} 
       q_{\pi_{\phi^{-1}\circ\alpha}(I_{v})}^{i}
{\hat e}^{a}_{ \pi_{\phi^{-1}\circ\alpha}(I_{v}) } \partial_a f(v)\nonumber\\
&&=\frac{\hbar}{2\mathrm{i}}\frac{3}%
{4\pi}\phi_*N(v, \{x\})\nu_{v}^{-2/3}
 \sum_{I_{v},i} 
       q_{I_{v}}^{i}
{\hat e}^{a}_{I_{v} } \partial_a f(v)
\label{2.33} \\
&&=
\lim_{\delta\rightarrow 0} \Psi_{B_{VSA}}^f (\hat{C}_{\delta}[\phi_*N]|c_0\rangle),
\label{2.34}
\end{eqnarray}
where the verification of the equality between (\ref{2.33}) and (\ref{2.34}) follows
straightforwardly from the considerations of P1 and where we have used the bijective nature of the 
permutation $\pi_{\phi^{-1}\circ\alpha}$ to go from the second line to (\ref{2.33}).

Next, consider equation (\ref{dc2}) when $c=c_0^{\beta}, \;\beta \in {\cal D}_{[c_0]}$. 
The coordinate system at the nontrivial vertex
$\beta (v)$ of $c_0^{\beta}$ is $\{x_{\beta}\}$. 
There exists a unique ${\gamma}\in{\cal D}_{[c_0]}$
such that  
\begin{equation}
{\hat U} (\phi )| c^{\beta}_0\rangle \equiv {\hat U}(\phi \circ\beta )|c_0\rangle
= 
{\hat U} ( {\gamma} )| c_0\rangle ,
\label{alphabeta}
\end{equation}
so that the coordinate patch associated with ${\hat U} (\phi )| c^{\beta}_0\rangle$ is $\{x_{\gamma}\}$.
Setting ${\bar \alpha}:= \gamma \circ \beta^{-1}$. we have that 
\begin{eqnarray}
&&\{x_{\gamma}\}= {\bar \alpha}^*\beta^*\{x\}= {\bar \alpha}^*\{x_{\beta}\} \label{baralphax}\\
&& {\hat U}(\gamma ) | c_0\rangle = {\hat U}({\bar \alpha} ){\hat U}(\beta )| c_0\rangle
= {\hat U}({\bar \alpha} )| c^{\beta}_0\rangle  . \label{baralphaket}
\end{eqnarray}
From (\ref{alphabeta})- (\ref{baralphaket}) it follows that 
an analysis identical to the one leading to equation (\ref{2.33}) in which
$\{x\}, v,c_0, \alpha$ are replaced with $\{x_{\beta}\}, \beta (v), c^{\beta}_0, {\bar \alpha}$
leads to the desired result (i.e. equation (\ref{dc2}) when $c=c_0^{\beta}$) if we make the following 
assumption:
\begin{equation}
{\bar G}^{\mu}_{\;\;\nu}= {\bar C}{\bar R}^{\mu}_{\;\;\nu},\;\;\; {\bar R}{\bar R}^{T}={\bf 1},\;\;\; \det {\bar R} =1,
\label{assumebarG=cR}
\end{equation}
where we have set:
\begin{equation}
{\bar G}^{\mu}_{\;\;\nu}:= 
\frac{\partial x^{\mu}_{{\bar \alpha}^{-1}\circ \phi\circ \beta}(p)}{\partial x^{\nu}_{\beta}(p)}|_{p=\beta (v)},
\label{defGbar}
\end{equation}
where the coordinate patch $\{x_{{\bar \alpha}^{-1}\circ \phi\circ \beta}\}$ is obtained as the 
push forward of the coordinate patch $\{x_{\beta}\}$ by the diffeomorphsim ${\bar \alpha}^{-1} \circ \phi$.
Note however that:
\begin{eqnarray}
\frac{\partial x^{\mu}_{{\bar \alpha}^{-1}\circ \phi\circ \beta}(p)}{\partial x^{\nu}_{\beta}(p)}|_{p=\beta (v)}
&:=& 
\frac{\partial x^{\mu}(\beta^{-1}\circ \phi^{-1}\circ {\bar \alpha}(p))}{\partial x^{\nu}(\beta^{-1}(p))}|_{p=\beta(v)}
\nonumber\\
&=& 
\frac{\partial x^{\mu}_{\beta^{-1}\circ {\bar \alpha}^{-1}\circ \phi \circ \beta} (p) }
{\partial x^{\nu}(p)}|_{p=v}.
\end{eqnarray}
Note also, that the diffeomorphisms $\alpha^{-1}\circ\phi$ as well as 
$\beta^{-1}\circ {\bar \alpha}^{-1}\circ \phi \circ \beta$ are symmetries of the reference charge network $c_0$
and that $\alpha, \phi, \beta, {\bar \alpha}$ are all connected to identity.
It then follows that 
the assumptions (\ref{assumeG=cR}) and (\ref{assumebarG=cR}) hold if the following condition holds
for all orientation preserving diffeomorphisms $\psi$ 
for which 
${\hat U}(\psi )|c_0\rangle = |c_0\rangle$:
\begin{equation}
G^{\mu}_{\;\;\nu} (\psi )
= C_{\psi} R^{\mu}_{\;\;\nu}(\psi) \;\;\;R(\psi)R^T(\psi )= {\bf 1}\;\;\det R(\psi ) =1 .
\label{dcfinal}
\end{equation}
Here $C_{\psi}$ can be any positive constant and  we have defined the $GL(3,R)$ matrix $G(\psi )$ as 
\begin{equation}
G^{\mu}_{\;\;\nu} (\psi )
:= \frac{\partial x^{\mu}( \psi (p) )}{\partial x^{\nu}(p)}|_{p=v}
\label{2.40}
\end{equation}
We shall prove (a slight generalization of) the condition (\ref{dcfinal}) in section 3.

\section{Proof of diffeomorphism covariance}

In section 3.1 we detail a few useful properties of symmetries of nontrivial (reference) charge networks.
These properties have been stated or implicitly used in our considerations till this point. Section 3.1
serves to collect them together at one place.
In section 3.2, we abstract these properties to the general setting of linear maps on a vector space
and provide a detailed statement of a Claim which we prove in section 3.3. The validity of the claim
will be seen to ensure the validity  of equation (\ref{dcfinal}) thereby providing the desired proof 
of diffeomorphism covariance.
\subsection{Properties of symmetries of $c_0$.}

We continue to use the notation of section 2.
First  note that equation (\ref{2.40}) expresses the induced action of the symmetry $\psi$ on the tangent space
$T_v$ at the point $v$, $v$ being the nontrivial vertex of $c_0$. In detail, 
since the diffeomorphism $\psi$ is a symmetry of $c_0$, the nontrivial 
vertex $v$ is a fixed point of the action of $\psi$. Hence $\psi$ induces an invertible linear map 
$G({\psi}): T_v\rightarrow T_v$ so that  the vector $V^a \in T_v $ 
is mapped to the vector $G^a_{\;b}(\psi ) V^b \in T_v$. 
It is easy to check that
the components of $G^a_{\;b}(\psi )$ in the coordinate system $\{x \}$ 
are given precisely by equation (\ref{2.40}) so that 
\begin{equation}
G^{\mu }_{\;\nu}(\psi ) = G^a_{\;b}(\psi ) (dx^{\mu})_a (\frac{\partial \;}{\partial x^{\nu}})^b
=\frac{\partial x^{\mu}( \psi (p) )}{\partial x^{\nu}(p)}|_{p=v}
\label{gmunu}
\end{equation}
As noted earlier, since  $\psi$ is orientation preserving, it follows that 
\begin{equation}
\det G >0 .
\label{detgpos}
\end{equation}
From equation (\ref{gmunu}) it follows that 
under a change of coordinates from $\{x \}$ to $\{ {\bar  x}\}$  the components of $G$ transform as
\begin{equation}
G^{{\bar \mu} }_{\;{\bar \nu}}= G^{\mu }_{\;\nu} 
\frac{  \partial x^{\nu} }{  \partial {\bar x}^{\bar \nu}  }
\frac{  \partial {\bar x}^{\bar \mu}  }{  \partial x^{\mu} }.
\label{gxbarx}
\end{equation}
Set $\frac{  \partial x^{\nu} }{  \partial {\bar x}^{\bar \nu}  }=: h^{\nu}_{\;{\bar\nu}}$.
Denote the $GL(3,R)$ matrices $G^{\mu }_{\;\nu} ,G^{{\bar \mu} }_{\;{\bar \nu}},h^{\nu}_{\;{\bar\nu}}$
by  $G, {\bar G},h$. Then equation (\ref{gxbarx}) takes the form of the following relation between 
$GL(3,R)$ matrices:
\begin{equation}
{\bar G} = h^{-1} G h
\label{g=hgh}
\end{equation}

Next, note that since 
$\psi$ is a  symmetry of  $c_0$ we have that
\begin{equation}
|c^{\psi}_0 \rangle :=
{\hat U}(\psi )|c_0\rangle = |c_0\rangle .
\label{symmrefpsi}
\end{equation}
so that $\psi $ maps the set of edges at $v$ into itself. More in detail, 
let us fix the orientation of the graph underlying $c_0$ so that the edges at the nontrivial $M$ valent vertex $v$
are all outgoing. Let the edges at $v$ be $e_I, I=1,..,M$. 
Denote  the set of these edges
by ${\cal E}_v$ so that ${\cal E}_v= \{e_1,..,e_M\}$
Equation (\ref{symmrefpsi}) implies that 
\begin{equation}
\{\psi (e_I), I=1,..M\}= \{ e_{ \pi_{\psi}(I)}, I=1,.., M\}= {\cal E}_v
\label{permute}
\end{equation}
where we have used the notation of section 2.3 to indicate the permutation $\pi_{\psi}$ induced by $\psi$.
In addition,
let the edges be labelled in such a way that 
the preferred edge (see (iii) of section 2.1) is $e_M$ so that 
\begin{equation}
\psi (e_M) = e_M \;\;\;\pi_{\psi}(M) = M.
\label{psim=m}
\end{equation}

Next, we analyse the consequences of equations (\ref{permute}), (\ref{psim=m}) for the action of the map
$G^{a}_{\;b}(\psi )$ on the edge tangents at $v$.
Accordingly, 
choose $M$ vectors $V_1,.., V_M$ such that $V_I$ is tangent to the edge $e_I$ at $v$, $I=1,..,M$.
Equations (\ref{permute}), (\ref{psim=m}) imply that the tangent space map $G^{a}_{\;b}(\psi )$ induced by $\psi$
at $T_v$ is such that
\begin{eqnarray}
G^{a}_{\;b} (\psi ) V_M^b &= &c_M V^b_M, \;\;\; c_M>0 \label{Gvm}\\
G^{a}_{\;b} (\psi ) V_i^b &= &\lambda_i V^b_{ \pi_{\psi}(i)}, \;\;\lambda_i >0 , \;i=1,..,M-1\label{Gvi}
\end{eqnarray}

In the next section we abstract  
the properties described above to a slightly more general setting and formulate a claim whose proof is readily seen to 
validate the condition (\ref{dcfinal}) and, consequently,  ensure  diffeomorphism covariance.

\subsection{Statement of Claim.}

Let $p\in \Sigma$. 
Let $V_1, .., V_M\in T_p, \;M>3$
 be a set of vectors 
no triple of which is coplanar.  
\footnote{
Recall that nontriviality of a vertex implies that its valence is always greater than 3 (see (i) section 2.1).
}
Let us call such a set as being of Grot- Rovelli (GR) type. Let 
$G\equiv G^{a}_{\;b}$ be an orientation preserving, invertible linear map from $T_p$ to itself which 
preserves this set of vectors modulo possible positive rescalings and which preserves the $M$th vector
$V_M$ modulo rescaling. It follows that $G^{a}_{\;b}$ specifies a permutation $\pi$ of the set $\{1,..,M\}$
with $\pi (M)= M$ and a set of positive numbers $\{c_1,..,c_{M-1}, c\}$ through the following relations:
\begin{eqnarray}
G^{a}_{\;b}  V_M^b &= &c  V^b_M, \;\;\; c >0 \label{Gvmgen}\\
G^{a}_{\;b}  V_i^b &= &c_i V^b_{\pi(i)}, \;\;c_i >0 , \;i=1,..,M-1\label{Gvigen} .
\end{eqnarray}
Let the set of all such  maps $G$ be ${\cal G}$. 
Let $\{x\}$ be a coordinate system at $p$ and let the evaluation of $G \in {\cal G}$ in this coordinate system be
the matrix
$G^{\mu}_{\;\nu}$ where
\begin{equation}
G^{\mu}_{\;\nu} := G^{a}_{\;b} dx^{\mu}_a (\frac{\partial \;}{\partial x^{\nu}})^b.
\end{equation}
\\

\noindent{\bf Claim}: There exists a  
choice of the coordinate system $\{x\}$, not necessarily unique,  
for the {\em entire} set ${\cal G}$ such that 
$\forall G \in {\cal G}$, we have that 
\begin{equation}
G^{\mu}_{\;\nu} = C_{G} R^{\mu}_{\;\nu}(\theta_G ) 
\label{claim}
\end{equation}
where $C_G$ is some positive $G$ dependent constant, $\theta_G$ is some 
$G$ dependent angle  and $R^{\mu}_{\;\nu}(\theta_G )\in SO(3) $ is a rotation by 
$\theta_G$ about an axis in the direction of $V_M$.

\subsection{Proof of Claim}
The proof of the Claim of section 3.2 has three main steps:\\
\noindent{\em Step 1. Existence of $G$ dependent coordinates such that $G$ is proportional to a rotation.}
Let the  permutation $\pi$ associated with $G$ be of order $m$. It follows that $G^m$ maps each $V_I$
to itself modulo rescaling:
\begin{equation}
(G^m)^{a}_{\;b}V_I= d_I V_I, \;I =1,..,M, d_I>0,
\label{gm}
\end{equation}
where $(G^m)^{a}_{\;b} :=G^{a}_{\;a_1} G^{a_1}_{\;a_2}... G^{a_m}_{\;b}$. Since the set of vectors $V_I, I=1,..,M$
is GR, we may expand $V_M$ uniquely in terms of $V_1,V_2, V_3$:
\begin{equation}
V_M= \alpha_1 V_1 + \alpha_2 V_2 +\alpha_3 V_3, \alpha_i \neq 0 .
\end{equation}
Equation (\ref{gm}) then implies that
\begin{eqnarray}
G^m V_M& = &d_1 \alpha_1 V_1 + d_2\alpha_2 V_2 + d_3\alpha_3 V_3 \nonumber\\
       &=& d_M V_M= d_M(\alpha_1 V_1 + \alpha_2 V_2 +\alpha_3 V_3 ),
\end{eqnarray}
from which it follows that $d_1=d_2=d_3= d_M$. Applying the same argument to all triples of vectors not containing
$V_M$, we obtain $d_i= d_M, i=1,..,M-1$ so that
\begin{equation}
G^m = d_M {\bf 1} = c^m{\bf 1}
\label{gm=ci}
\end{equation}
where $d_M= c^m$ follows from (\ref{Gvmgen}). Next let us evaluate the map $G$ in some coordinate system $\{x\}$.
Then equation 
(\ref{gm=ci}) implies that 
the $GL(3,R)$ matrix $\frac{G^{\mu}_{\;\nu}}{c}$ generates a cyclic group of order $m$. But any finite order 
cyclic subgroup of $GL(3,R)_+$ (i.e. the group of $GL(3,R)$ matrices with positive determinant) is 
conjugate to a finite subgroup of $SO(3)$ 
\footnote{
See, for example, Theorem 2.2 (a), Chapter 9 of Reference \cite{gl3r}. While the theorem is proved for $GL(n,C)$
it is easy to see that the proof can be applied to $GL(n,R)_+$ and that the  result claimed here follows.} 
so that, in matrix notation,
\begin{equation}
G= c hRh^{-1}, R\in SO(3), h\in GL(3,R).
\end{equation}
Equation (\ref{g=hgh}) then implies that there exists coordinates  
for which $G= cR$. Of course these coordinates are dependent on the map $G$.
Further, equation (\ref{Gvmgen}) implies that in these coordinates $R$ is a rotation about the 
axis which points in the direction $V_M$.
\\

\noindent{\em Step 2. Invariance of angular order of transverse projections of $V_i, i=1,..M-1$}:
Fix a right handed coordinate chart about $p$. The basis of coordinate vectors at $p$ 
defines an isomorphism between
$T_p$ and $R^3$ with $p$ identified with the origin of $R^3$. 
Consider the  coordinate plane in $T_p\equiv R^3$ normal to $V_M$, project
$V_i, i=1,..M-1$ to this plane and arrange the projections so that they point outward from the 
origin. By virtue of their GR property, no two of these projections are collinear.
Clearly these projections acquire an ordering when one proceeds in an anticlockwise 
manner around the $V_M$ axis in this plane i.e. a given vector $V_k$ has a pair of nearest neighbours
where the notion of nearest is defined in terms of angular seperation.
We show that this ordering is independent of the choice of coordinates, the choice of transverse plane and
the action of any $GL(3,R)_+$ transformation. We proceed as follows.

We fix coordinates as above and use the natural Euclidean coordinate metric and volume form to define
angles, normals and  cross products between vectors.
Let the normal vector  to the plane containing $V_i$ and $V_M$ be $n_i$ so that in vector notation
${\vec n}_i= {\vec V}_M\times {\vec V}_i$. Let the set of these normals be ${\bf n}=\{ n_1,..,n_{M-1}\}$.
These normals all lie in the coordinate plane normal to
$V_M$ and acquire an ordering reflective of the one for $\{V_i, i=1,..,M-1\}$ in the previous paragraph.
Since $M-1> 2$, it is easy to see that 
each $n_i$ has a pair of nearest neighbours $n_{i_{>}}, n_{i_{<}} \in {\bf n}$ 
where $n_{i_{>}}$ is the first element of ${\bf n}$ encountered when proceeding anticlockwise from $n_i$
and 
where $n_{i_{<}}$ is the first element of ${\bf  n}$ encountered when proceeding clockwise from $n_i$.
Accordingly, these nearest neighbours are characterised by the following properties:\\
\noindent 1. Defining property of $n_{i_{>}}$:\\ $n_{i_{>}}$ is such that either\\
\noindent  Case A: 
No  $n \in {\bf  n}$ exists
for which we can find $\alpha, \beta >0$ such that 
\begin{equation}
{\vec n} = \alpha {\vec n}_i + \beta {\vec n}_{i_{>}},
\label{1a1}
\end{equation}
and there exists some $\lambda >0$ such that
\begin{equation}
{\vec n}_i \times {\vec n}_{i_{>}}= \lambda {\vec V}_M .
\label{1a2}
\end{equation}

or \\

\noindent Case B:
No $n \in {\bf  n}$ exists
for which we can find $\alpha, \beta $ such that either $\alpha <0, \beta >0$ or $\alpha < 0, \beta <0$ or
$\alpha >0, \beta <0$,  such that 
\begin{equation}
{\vec n} = \alpha {\vec n}_i + \beta {\vec n}_{i_{>}},
\label{1b1}
\end{equation}
and there exists some $\lambda <0$ such that
\begin{equation}
{\vec n}_i \times {\vec n}_{i_{>}}= \lambda {\vec V}_M .
\label{1b2}
\end{equation}

\noindent Note: Case A holds if the angle  between $n_{i_{>}}$ and $n_i$ is  less than $\pi$ and 
Case B holds if the angle  between $n_{i_{>}}$ and $n_i$ is  greater  than $\pi$.
The GR property ensures that this angle cannot be exactly $\pi$.

\noindent 2. Defining property of $n_{i_{<}}$:\\
$n_{i<}$ is such that either
\noindent Case A: 
No $n \in {\bf n}$ exists
for which we can find $\alpha, \beta >0$ such that 
\begin{equation}
{\vec n} = \alpha {\vec n}_i + \beta {\vec n}_{i_{<}},
\label{2a1}
\end{equation}
and there exists some $\lambda <0$ such that
\begin{equation}
{\vec n}_i \times {\vec n}_{i_{<}}= \lambda {\vec V}_M .
\label{2a2}
\end{equation}

or \\

\noindent Case B: 
No $n \in {\bf n}$ exists
for which we can find $\alpha, \beta $ such that either $\alpha <0, \beta >0$ or $\alpha < 0, \beta <0$ or
$\alpha >0, \beta <0$,  such that 
\begin{equation}
{\vec n} = \alpha {\vec n}_i + \beta {\vec n}_{i_{<}},
\label{2b1}
\end{equation}
and there exists some $\lambda >0$ such that
\begin{equation}
{\vec n}_i \times {\vec n}_{i_{<}}= \lambda {\vec V}_M .
\label{2b2}
\end{equation}

\noindent Note: Case A holds if the angle  between $n_{i_{<}}$ and $n_i$ is  less than $\pi$ and 
Case B holds if the angle  between $n_{i_{<}}$ and $n_i$ is  greater  than $\pi$.
The GR property ensures that this angle cannot be exactly $\pi$.

We now show that the defining properties (1) and (2) above can be expressed without reference to a coordinate
system. 
Let ${\eta}_{abc}$ be the alternating Levi- Civita tensor density of weight -1. Define the density weighted 2 form
$\eta_{ab}$ and the density weighted one forms $N_{ia}, i=1,..,M-1$ by
\begin{eqnarray}
\eta_{bc} &:= & \eta_{abc} V^a_M,  \\
N_{ia}= \eta_{ba}V^b_{i}
\end{eqnarray}
Let ${\cal N}:= \{N_{i}, i=1,..,M-1\}$ and let the nearest neighbours of $N_{i}$ be denoted by $N_{i_{<}},N_{i_{>}}$.
The defining properties (1) and (2) above can be rewritten as:\\
\noindent 1. Defining property of $N_{i_{>}a}$:\\ $N_{i_{>}a}$ is such that either\\
\noindent  Case A: 
No  $N \in {\cal N}$ exists
for which we can find $\alpha, \beta >0$ such that 
\begin{equation}
 N_a = \alpha N_{ia} + \beta  N_{i_{>}a},
\label{N1a1}
\end{equation}
and there exists some $\lambda >0$ such that
\begin{equation}
N_{i[a}N_{i_{>}b]}= \lambda \eta_{ab}.
\label{N1a2}
\end{equation}

or \\

\noindent Case B:
No $N \in {\cal N}$ exists
for which we can find $\alpha, \beta $ such that either $\alpha <0, \beta >0$ or $\alpha < 0, \beta <0$ or
$\alpha >0, \beta <0$,  such that 
\begin{equation}
 N_a = \alpha N_{ia} + \beta  N_{i_{>}a},
\label{N1b1}
\end{equation}
and there exists some $\lambda <0$ such that
\begin{equation}
N_{i[a}N_{i_{>}b]}= \lambda \eta_{ab}.
\label{N1b2}
\end{equation}

\noindent 2. Defining property of $N_{i_{<}a}$:\\ $N_{i_{<}a}$ is such that either\\
\noindent  Case A: 
No  $N \in {\cal N}$ exists
for which we can find $\alpha, \beta >0$ such that 
\begin{equation}
 N_a = \alpha N_{ia} + \beta  N_{i_{<}a},
\label{N2a1}
\end{equation}
and there exists some $\lambda <0$ such that
\begin{equation}
N_{i[a}N_{i_{<}b]}= \lambda \eta_{ab}.
\label{N2a2}
\end{equation}

or \\

\noindent Case B:
No $N \in {\cal N}$ exists
for which we can find $\alpha, \beta $ such that either $\alpha <0, \beta >0$ or $\alpha < 0, \beta <0$ or
$\alpha >0, \beta <0$,  such that 
\begin{equation}
 N_a = \alpha N_{ia} + \beta  N_{i_{<}a},
\label{N2b1}
\end{equation}
and there exists some $\lambda >0$ such that
\begin{equation}
N_{i[a}N_{i_{<}b]}= \lambda \eta_{ab}.
\label{N2b2}
\end{equation}

The coordinate invariant reformulation above  proves that the 
ordering of the vectors $V_i, i=1,..M-1$ is independent of the choice of coordinates.
It remains to show that the ordering is invariant under the action of any $GL(3,R)_+$
transformation. We proceed as follows. Let $G\in GL(3,R)_+$.
We use step 1 and transit to a (right handed) coordinate system in which $G$ is proportional to 
a rotation about $V_M$. Since rotations about $V_M$ cannot change the ordering of the transverse components
of $V_i$ and since this ordering is independent of coordinates, it follows that no $G\in GL(3,R)_+$
can change this ordering.
\\

\noindent{\em Step 3. Identification of $G_{min} \in {\cal G}$ with the smallest non- zero angle of rotation}: 
We relabel the vectors $V_i$ to reflect the ordering of Step 2, so that $V_{i+1} := V_{i_{>}}, i=1,..,M-2$.
Step 2 implies that no $V_i$ can be `dragged' past $V_{i+1}$ by any $G\in {\cal G}$. 
It then follows that the permutation $\pi$ associated with the element $G$ in equation (\ref{Gvigen})
must be a {\em translation} by an integer $a$, $-(M-2)\leq a \leq M-2$ so that:
\begin{equation}
\pi (i) = i+a, i=1,.., M-1
\label{translation}
\end{equation}
where in the case that  $i+a>M-1$ we use $i+a$ as a shorthand for  $i+a- (M-1)$.
From Step 2 it then follows that in appropriate coordinates, $G$ is proportional to a rotation about the axis
in the direction of $V_M$ by an angle $\theta = \frac{a}{M-1}2\pi$. To remind us that $a$ is associated with the 
map $G$, we write $a\equiv a_G$.

Next, we restrict attention to $a_G \neq 0$ and define $a_{min}$ by:
\begin{equation}
a_{min}= \inf_{g\in {\cal G}\;{\rm such \; that\; } a_G\neq 0} |a_G|= \inf_{g\in {\cal G}\; {\rm such \; that\; } a_G>0 } a_G
\label{defamin}
\end{equation}
where the second equality follows from the easily verifiable fact that if $G\in {\cal G}$ then $G^{-1} \in {\cal G}$
and $a_G= - a_{G^{-1}}$. Since $a_G$ can only take a finite number of values it follows that 
there exists some (not necessarily unique) $G_{min}$ such that $a_{min}= a_{G_{min}}$.
Using the arguments of Step 2, we choose our coordinates  $\{x \}$ 
to be adapted to $G_{min}$ so that $G_{min}$ is proportional to
a rotation in the chosen coordinates:
\begin{eqnarray}
G^{\mu}_{min \;\nu} &=& C_{G_{min}} R^{\mu}{\;\nu}(\theta_{min} ), C_{G_{min}} >0
\label{gmin}\\
\theta_{min} &= &\frac{a_{min}}{M-1}2\pi
\label{thetamin}
\end{eqnarray}
where $R^{\mu}{\;\nu}(\theta_{min} )$ is a rotation by the angle $\theta_{min}$ around the $V_M$ axis.
 
We now show that any other $G\in {\cal G}$ necessarily satisfies equation (\ref{claim}).
First note that if $G$ is such that  $a_G=0$ then a repetition of the argument of step 1 with $m=1$ leads to 
the conclusion that $G$ is proportional to the idenity {\em independent} of the choice of coordinates.
Hence we need only concentrate on $G$ such that $a_G\neq 0$.
Accordingly consider some $G\in {\cal G}$ with $a_G > 0$. 
\footnote{This entails no loss of generality; if  $a_G <0$, we apply the argument below to $G^{-1}$.}
We set
\begin{equation}
a_G= l a_{min} + n,   0\leq n < a_{min},\; l>0.
\end{equation}
Note that $l\neq 0$ else $a_G<a_{min}$ which is not possible.
If $n\neq 0$ then consider the map ${\hat G}:=(G_{min}^{-1})^l G$. Clearly ${\hat G} \in {\cal G}$
and $a_{\hat G}= n< a_{min}$. The only allowable possibility is then $n=0$.
But we have already shown that $a_{\hat G}= 0$ implies that ${\hat G}$ is a multiple of identity:
\begin{equation}
{\hat G}= (G_{min}^{-1})^l G = C_{\hat G} {\bf 1} ,\;\; C_{\hat G} >0
\end{equation}
which implies that 
\begin{equation}
G = C_{\hat G} (G_{min})^l .
\end{equation}
Equation (\ref{gmin}) then implies that in the $\{x\}$ coordinates we have that:
\begin{eqnarray}
G^{\mu}{\;\nu} &= & C_G R^{\mu}_{\;\nu} (\theta_G ),\\
C_G &=&(C_{G_{min}})^l C_{\hat G},\\
\theta_G &=& l \theta_{min},
\end{eqnarray}
which completes the proof of our claim.

\noindent{\em Remark}: Note that any rotation of the coordinate system $\{x\}$ results in another
acceptable coordinate system. We use this freedom
to choose the $\{x\}$ coordinates so that $V_M$ is along the 
$x^3$ axis.

\subsection{Effect of the choice of the set of preferred diffeomorphisms ${\cal D}_{[c_0]}$.}

First, note that the set ${\cal D}_{[c_0]}$ has {\em no} role to play in the  choice of coordinates for $c_0$.
This follows from the fact that 
the diffeomorphism covariance condition  (\ref{dcfinal}) as well the considerations of section 3.3 including
 the class of reference coordinate patches
chosen there only depend on the symmetries of $c_0$ and have nothing to do with the choice of ${\cal D}_{[c_0]}$. 
\footnote{More precisely the only dependence is on the graph structure in the vicinity of the nontrivial vertex
which determines the set ${\cal G}$ in section 3.2.}
As a result, as we show below, 
the continuum limit action of a single Hamiltonian constraint 
is independent 
of the choice of the set
${\cal D}_{[c_0]}$. Recall, from P1 (and, for example, the considerations leading upto equation (\ref{2.34}))
that this continuum limit action, when nontrivial, is given by 
\begin{equation}
\lim_{\delta\rightarrow 0} \Psi_{B_{VSA}}^f (\hat{C}_{\delta}(N)|c\rangle)
=\frac{\hbar}{2\mathrm{i}}\frac{3}%
{4\pi}N(\beta (v), \{x\}_{{\beta}})\nu_{v}^{-2/3}
 \sum_{I_{v},i} 
       q_{I_{v}}^{i}
{\hat e}^{a}_{I_{v} } \partial_a f(\beta(v))
\label{singlecont}
\end{equation}
where we have used the notation of section 2.2 and denoted the nontrivial vertex of $c_0$ by $v$. Thus 
$\beta \in {\cal D}_{[c_0]}$ is such that $c_0^{\beta}= c$ and the unit tangents ${\hat e}^{a}_{I_{v} }$ are
computed with respect to the coordinate system  $\{x\}_{c_0^{\beta}}$.  
A distinct choice ${\cal D}^{\prime}_{[c_0]}$ with $\beta^{\prime}\in{\cal D}^{\prime}_{[c_0]}$ such that
$c=c^{\beta^{\prime}}_0$ yields:
\begin{equation}
\lim_{\delta\rightarrow 0} \Psi_{B_{VSA}}^f (\hat{C}_{\delta}(N)|c\rangle)
=\frac{\hbar}{2\mathrm{i}}\frac{3}%
{4\pi}N(\beta (v), \{x\}_{{\beta^{\prime}}})\nu_{v}^{-2/3}
 \sum_{I_{v},i} 
       q_{\pi_{\bar{\beta}}(I_{v})}^{i}
{\hat e}^{\prime a}_{\pi_{\bar{\beta}}(I_{v}) } \partial_a f(\beta(v))
\label{singlecontprime}
\end{equation}
where ${\hat e}^{\prime a}_{\pi_{\bar{\beta}}(I_{v}) }$ is normalised with respect to the $\{x\}_{\beta^{\prime}}$
coordinates and we have defined ${\bar \beta}:= \beta^{\prime}\circ \beta^{-1}$ and used 
$\beta^{\prime}(v)= \beta (v)$. 
Define the Jacobian matrix $B$ by 
\begin{equation}
B^{\mu}_{\;\nu} = \frac{\partial x^{\mu}_{\beta}}{\partial x^{\nu}_{\beta^{\prime}}}|
_{\beta (v)}.
\label{defB}
\end{equation}
Clearly we have that 
\begin{equation}
N(v, \{ x \}_{  \beta^{\prime}  })= 
N(v, \{x\}_{\beta}) (\det B)^{-\frac{1}{3}}
\label{lapseB}
\end{equation}
Next, the following equations follow from an analysis similar to that in section 2.6:
\begin{equation}
{\hat e}_{ \pi_{\bar{\beta}}  (I_{v}) }^{\prime a}
=\lambda {\hat e}_{ \pi_{\bar{\beta}}(I_{v}) }^a, 
\label{deflambdaB}
\end{equation}
\begin{equation}
{\hat e}_{ \pi_{{\bar{\beta}}   } (I_{v}) }^{\prime\mu}=
{\hat e}_{ \pi_{ {\bar{\beta}}       }(I_{v}) }^{\prime {\nu}^{\prime}}
\frac{\partial x_{\beta}^{\mu}}{\partial x_{  \beta^{\prime}        }^{\nu^{\prime}}}
\label{normprimenormB}
\end{equation}
where ${\hat e}_{ \pi_{ {\bar{\beta}}   } (I_{v}) }^{\prime\mu},
{\hat e}_{ \pi_{ {\bar{\beta}}       }(I_{v}) }^{\prime {\nu}^{\prime}}$ are the components of 
${\hat e}_{ \pi_{ \bar{\beta} } (I_{v}) }^{\prime a}$ in the $\{x_{\beta}\}, \{x_{\beta^{\prime}}\}$ coordinates.
Since ${\hat e}_{ \pi_{  \bar{\beta}  }(I_{v}) }^{\prime a}= {\bar \beta}^* 
{\hat e}_{I_{v} }^{ a}$ it follows that 
\begin{equation}
{\hat e}_{ \pi_{  \bar{\beta}  }(I_{v}) }^{\mu^{\prime}}
={\hat e}_{I_{v} }^{ \mu}\delta^{\mu^{\prime}}_{\mu }.
\end{equation}
Using this in conjunction with equations (\ref{deflambdaB}) and (\ref{normprimenormB}) yields:
\begin{equation}
\lambda = \sqrt{\sum_{\mu}({\hat e}_{ \pi_{ {\bar{\beta}}    }(I_{v}) }^{\prime\mu})^2}
= \sqrt{
\sum_{\mu,\nu,\tau}({\hat e}_{I_v}^{\nu}B^{\mu}_{\;\;\nu}{\hat e}_{I_v}^{\tau}B^{\mu}_{\;\;\tau})}
\label{2.29B}
\end{equation}
We have that 
\begin{eqnarray}
B^{\mu}_{\;\nu}
&=&\frac{\partial x^{\mu}_{\beta}(p)}{\partial x^{\nu}_{\beta^{\prime}}(p)}|_{p=\beta (v)}
= \frac{\partial x^{\mu}({\beta}^{-1}p)}{\partial x^{\nu}((\beta^{\prime})^{-1}p)}|_{p=\beta (v)}\\
&=&\frac{\partial x^{\mu}({\beta}^{-1}\circ \beta^{\prime}p)}{\partial x^{\nu}(p)}|_{p=v},
\end{eqnarray}
so that we may set $\psi = {\beta}^{-1}\circ \beta^{\prime}$ in equation (\ref{2.40})
and conclude  from section 3.3 that $B$ is proportional to a rotation. It is then easy to
verify that this property of $B$, together with equations  (\ref{lapseB}),(\ref{2.29B})
ensure that the right hand sides of equations (\ref{singlecont}), (\ref{singlecontprime}) agree.

\section{Divergence of the commutator between a pair of Hamiltonian constraints.}

\subsection{Preliminary Remarks}
The commutator involves the action of two Hamiltonian constraints. Consequently, the charge net being acted
upon is subjected to successive deformations and the result is a linear combination of doubly deformed 
charge nets. The first deformation at parameter $\delta$ is defined in terms of the coordinate patch 
at the nontrivial vertex of the charge net, the deformed nontrivial vertex being at a coordinate 
distance of $O(\delta )$ from
the orginal one. 
The second deformation at parameter $\delta^{\prime}$ is then defined in terms of the coordinate patch at the 
deformed vertex produced by the first deformation. The continuum limit is obtained by first taking 
$\delta^{\prime}\rightarrow 0$ and then $\delta \rightarrow 0$. The $\delta^{\prime}\rightarrow 0$
limit yields derivatives of the vertex smooth
function with respect to
the unit edge tangents at the deformed vertex produced by the first deformation where 
the unit norm is defined with respect to the coordinate patch at this deformed vertex.
These derivatives are multiplied by the lapse which is evaluated at the  deformed vertex in the coordinate patch
associated with the deformed vertex. Subsequent to this, the $\delta \rightarrow 0$ is taken.
Hence the final result depends crucially on the behaviour of the coordinate patch around the deformed vertex
as $\delta \rightarrow 0$. In section 4.2 we analyse this limiting behaviour in the light of the 
diffeomorphism covariant choices of coordinate patches of section 2.2.

The position of the deformed vertex and the structure of the graph in its vicinity
depend on $\delta$. From P1, these structures at different values of $\delta$ are related by diffeomorphisms.
In P1, the coordinate patch around the deformed vertex at $\delta$ was assumed to be well behaved in the 
$\delta \rightarrow 0$ limit. Here the behaviour of these $\delta$- dependent  coordinate patches cannot be
prescribed freely. Instead, their behaviour is constrained by the requirements of diffeomorphism covariance.
Specifically, the coordinate patch around the deformed vertex at any $\delta$ is related by a $\delta$ dependent 
diffeomorphism to that at fixed parameter value $\delta_0$. Since this diffeomorphism has singular behaviour
in the $\delta \rightarrow 0$ limit, the coordinate patch goes bad in the $\delta \rightarrow 0$ limit
and the commutator diverges.

In more detail,
the continuum limit of the finite triangulation commutator is:
\begin{equation}
\lim_{\delta\rightarrow 0}
\lim_{\delta^{\prime}\rightarrow 0}
\Psi_{B_{VSA}}^f (({\hat C}(M)_{\delta^{\prime}}{\hat C}(N)_{\delta}- N \leftrightarrow M)|c\rangle)
\label{4.1}
\end{equation}
Taking the $\delta^{\prime}\rightarrow 0$ limit yields, the $\delta$- dependent object:
\begin{equation}
\lim_{\delta^{\prime}\rightarrow 0}\Psi_{B_{VSA}}^f (({\hat C}(M)_{\delta^{\prime}}{\hat C}(N)_{\delta}- N \leftrightarrow M)|c\rangle).
\label{4.2}
\end{equation}
The above equation computes,  at $\delta\neq 0$, the continuum limit action of 
${\hat C}(M)_{\delta^{\prime}}$ on the state ${\hat C}(N)_{\delta}|c\rangle$ minus the same action with 
$N$ and $M$ interchanged. From P1 and the analysis of section 2 (see for example equations
(\ref{2.33}) and (\ref{2.34})), the 2 key coordinate dependent objects which appear in the continuum 
limit action of ${\hat C}(M)_{\delta^{\prime}}$
are the set of unit norm 
tangent vectors at the deformed vertices (one for each deformed chargenet produced by the action of 
${\hat C}(N)_{\delta}$ on $|c\rangle$) and the densitized lapse $M$. 
The unit norms of the former and the evaluation of the latter are with respect to the 
coordinate metric and coordinate volume form respectively, the coordinates being those associated with the deformed
vertices created by ${\hat C}(N)_{\delta}$. 

We show, in section 4.2, that while the unit tangents have well defined $\delta\rightarrow 0$
limits, the volume form at each deformed vertex 
degenerates in this limit. In section 4.3 we show that this degeneracy is responsible for a divergence in the continuum limit 
commutator. Section 4.4 is devoted to a technicality concerning the choice of coordinates around the deformed 
vertices created by the action of ${\hat C}(N)_{\delta}$ on $|c\rangle$.

\subsection{Behaviour of coordinate patches as $\delta\rightarrow 0$}

\subsubsection{Notation and conventions.}

In what follows  we use the following notation. 
The coordinate patch associated with the nontrivial
vertex $v_0$ of the reference charge net $c_0$  is $\{x_0\}_{v_0}$. 
The coordinate patch associated with the nontrivial
vertex $v$ of the charge net $c$  is $\{x\}_v$. The reference diffeomorphism
$\beta \in {\cal D}_{[c_0]}$  maps $c_0$ to $c$
so that $\{x\}_v= \beta^*\{x_0\}_{v_0}$. 
The deformations, generated by the Hamiltonian constraint at parameter $\delta_0$, 
of any reference chargenet $c_0$ are defined as in P1 and result in the chargenets
$c_0(i,v_{I_{v_0},\delta}^{\prime})$. As in P1, the deformed chargenet $c_0(i,v_{I_{v_0},\delta}^{\prime})$
at parameter $\delta$ is the image by a diffeomorphism (see Appendix C.4, P1) of the deformed chargenet
$c_0(i,v_{I_{v_0},\delta_0}^{\prime})$ at parameter $\delta_0$.
We shall denote this  diffeomorphism by $\phi_{v_{I_{v_0},\delta}^{\prime},v_{I_{v_0},\delta_0}^{\prime}}$.
If the context is clear we shall use an abbreviated notation (similar to that in P1) and 
write  $\phi_{v_{I_{v},\delta}^{\prime},v_{I_{v},\delta_0}^{\prime}}\equiv\phi_{\delta}$
so that, we may write:
\begin{equation}
{\hat U}(\phi_{\delta}) |c_0(i,v_{I_{v_0},\delta_0}^{\prime})\rangle
=| c_0(i,v_{I_{v_0},\delta}^{\prime})\rangle
\label{abbrev}
\end{equation}

The deformations $c(i,v_{I_{v},\delta}^{\prime})$ of any chargenet $c$ diffeomorphic to $c_0$
are defined, as in section 2.4, to be the image of the corresponding deformations $c_0(i,v_{I_{v_0},\delta}^{\prime})$ 
of $c_0$ by the reference 
diffeomorphism $\beta \in {\cal D}_{[c_0]}$ which maps $c_0$ to $c$.
It follows that the deformation of $c$ at parameter $\delta$ is the diffeomorphic image of its deformation
at parameter $\delta_0$. 
In abbreviated notation similar to that in equation (\ref{abbrev})
we denote this diffeomorphism as $\phi_{\beta, \delta}$. It follows that
\begin{equation}
\phi_{\beta, \delta} = \beta \circ \phi_{\delta} \circ \beta^{-1},
\label{defphibeta}
\end{equation}
so that 
\begin{equation}
{\hat U}(\phi_{\beta,\delta}) |c(i,v_{I_{v},\delta_0}^{\prime})\rangle
=| c(i,v_{I_{v},\delta}^{\prime})\rangle
\label{betaabbrev}
\end{equation}
The patch associated with the vertex $v_{I_{v},\delta}^{\prime}$ of the deformed
charge network $c(i,v_{I_{v},\delta}^{\prime})$ is $\{x^{\prime }\}_{v_{I_{v},\delta}^{\prime}}$.
\footnote{
These coordinates are obtained as the image of the reference coordinates associated with
 the reference chargenets for the diffeomorphism class $[c(i,v_{I_{v},\delta}^{\prime})]$. The 
`sibling' restrictions imply that the reference coordinates are independent of $i$. However their images
may be $i$- dependent through the $i$- dependence of the reference diffeomorphism sets 
${\cal D}_{[c(i,v_{I_{v},\delta}^{\prime})}]$. Here we assume the images are $i$- independent; this assumption
is justified in section 4.4
\label{footnoteidep}}
If the context is clear we shall drop the subscripts which denote the vertex 
and write  $\{x\}_v \equiv \{x \}$,  and $\{x^{\prime }\}_{v_{I_{v},\delta}^{\prime}}\equiv \{x^{\prime}_{\delta}\}$.
Given a deformation along the edge $I_v$ the following  3 sets of coordinates will play important roles 
in the commutator calculation:
$\{x\}$ associated with $c$, $\{x^{\prime}_{\delta}\}$ associated with $c(i,v_{I_{v},\delta}^{\prime})$
and $\{x^{\prime}_{\delta_0}\}$ associated with $c(i,v_{I_{v},\delta_0}^{\prime})$.

Note also that
from P1, $\delta_0$ is chosen small enough that $\{x\}$ also covers a neighbourhood of $v_{I_{v},\delta_0}^{\prime}$.
The discussion in section 3.4 implies that  we may define $\{x^{\prime}_{\delta}\}$ to be obtained as the pushforward
of $\{x^{\prime}_{\delta_0}\}$ 
by the diffeomorphism $\phi_{\beta, \delta}$ (see equations (\ref{defphibeta}), (\ref{betaabbrev})).

As in P1, the $J_v$th edge emanating from the nontrivial vertex $v$ of $c$ is denoted by $e_{J_v}$
and its deformed counterpart emanating from the  nontrivial vertex $v_{I_{v},\delta}^{\prime}$ of 
$c(i,v_{I_{v},\delta}^{\prime})$ is denoted by ${\tilde e}_{J_v}$.
As in P1 (and section 2.6 of this paper),  the unit edge tangent to $e_{J_v}$ at $v$ 
is denoted by ${\hat e}_{J_v}^a$, the (unprimed) hat denoting 
normalization with respect to the coordinate metric defined by the  $\{x\}$ coordinates. Since $\{x\}$ 
covers a neighbourhood of $v_{I_{v},\delta}^{\prime}$ for all $\delta \leq \delta_0$, we may also evaluate
the edge tangent to ${\tilde e}_{J_v}$ at $v_{I_{v},\delta}^{\prime}$ normalized  in the 
$\{x\}$ coordinate metric. We denote this unit edge tangent by 
${\hat {\tilde e}}^{ a}_{J_v}|_{v_{I_{v},\delta}^{\prime}}$, 
where, once again, the (unprimed) hat denotes normalization in $\{x\}$ coordinates.

The deformations of the reference charge net (defined as in P1) are such that the $I_{v_0}$th unit tangent 
${\hat {\tilde e}}^{ a}_{J_{v_0}=I_{v_0}}|_{v_{I_{v_0},\delta}^{\prime}}$ of the deformed charge net 
$c_0(i,v_{I_{v_0},\delta}^{\prime})$, at its nontrivial  vertex  $v_{I_{v_0},\delta}^{\prime}$ 
has the same components in the $\{x_0\}$ coordinates
as the $I_{v_0}$th unit tangent ${\hat e}_{I_{v_0}}^a|_{v_0}$ of the undeformed charge net $c_0$ at {\em its} 
nontrivial vertex $v_0$. Cleary the same results hold in the $\{x\}$ coordinates for the 
corresponding deformations of  $c$ which is the image of $c_0$ by the reference diffeomorphism $\beta$. 
It follows  that:
\begin{equation}
{\hat {\tilde e}}_{I_v}^{\mu}|_{v_{I_{v},\delta }^{\prime}}=
{\hat {e}}_{I_v}^{\mu}|_{v}
\label{ei=etildei}
\end{equation}

Since a neighbourhood of the vertex $v_{I_{v},\delta }^{\prime}$ is also covered by the 
$\{x^{\prime}_{\delta}\}$ coordinates, we 
may also normalize the edge tangent along ${\tilde e}_{J_v}$ at $v_{I_{v},\delta}^{\prime}$ in the 
coordinate metric associated with the $\{x^{\prime}_{\delta}\}$ coordinates. We denote the edge tangent so 
normalized by ${\hat {\tilde e}}^{\prime a}_{J_v}|_{v_{I_{v},\delta}^{\prime}}$, 
the primed hat,  ${\hat \;}^{\prime}$
denoting normalization with respect to the 
$\{x^{\prime}_{\delta}\}$ metric. Since ${\hat {\tilde e}}^{\prime a}_{J_v}|_{v_{I_{v},\delta}^{\prime}}$
and ${\hat {\tilde e}}^{ a}_{J_v}|_{v_{I_{v},\delta}^{\prime}}$ are along the same edge, we have that:
\begin{equation}
{\hat {\tilde e}}^{\prime a}_{J_v}|_{v_{I_{v},\delta}^{\prime}}= \alpha_{J_v, \delta }
{\hat {\tilde e}}^{ a}_{J_v}|_{v_{I_{v},\delta}^{\prime}} , \;\;\alpha_{J_v, \delta }>0 .
\label{defalpha}
\end{equation}
In section 4.2.2 we will be interested in computing  $\lim_{\delta\rightarrow 0}\alpha_{J_v, \delta }$.

It is convenient, for the purposes of that computation, to rotate the coordinate system
$\{x_0\}$ at $v_0$ so that $x_0^{3}$ points along the $I_{v_0}$th edge  in $c_0$. We shall slightly 
abuse notation  and continue to use $\{x_0\}$ to denote  the rotated system and continue to use $\{x\}$
for the image of this rotated system by the diffeomorphism $\beta$. It is easy to check that $\{x\}$
is related by the same rotation matrix to the $\beta$ image of the unrotated reference coordinates which
rotates the unrotated reference coordinates to their rotated image. As a consequence, it follows that
in the rotated system, 
From the remsrk at the end of section 3.3 it follows that $x^{\prime 3}_{\delta_0}$ 
points along the $I_v$th edge  in  $c(i,v_{I_{v},\delta_0}^{\prime})$.  
Since $\phi_{\delta}^*\{x^{\prime}_{\delta_0} \}= \{x^{\prime}_{\delta }\}$ and since  $\phi_{\delta}$ maps 
the $I_v$th edge of $c(i,v_{I_{v},\delta_0}^{\prime})$  at   
$v_{I_{v},\delta_0}^{\prime}$  
to
the $I_v$th edge  of $c(i,v_{I_{v},\delta}^{\prime})$ at  $v_{I_{v},\delta}^{\prime}$,
 it follows that $x^{{\prime}3}_{\delta}$ lies along the $I_v$th edge tangent at
$v_{I_{v},\delta}^{\prime}$ in $c(i,v_{I_{v},\delta}^{\prime})$.
It follows that in these coordinates we have that:
\begin{eqnarray}
{\hat e}_{I_v}^a|_v &= &(\frac{\partial \;}{\partial x^3})^a|_v \label{eicx3}\\
{\hat {\tilde e}}^{\prime a}_{I_v}|_{v_{I_{v},\delta}^{\prime}}
&= &(\frac{\partial \;}{\partial x_{\delta}^{\prime 3}})^a|_{v_{I_{v},\delta}^{\prime}} \label{eicx3prime}.
\end{eqnarray}
It follows from (\ref{eicx3}) and (\ref{ei=etildei}) that 
\begin{equation}
{\hat {\tilde e}}^{ a}_{I_v}|_{v_{I_{v},\delta}^{\prime}}= 
(\frac{\partial \;}{\partial x^3})^a|_{v_{I_{v},\delta}^{\prime}} .
\label{tildeex3}
\end{equation}

\subsubsection{Edge tangents}
Note that since rotations are norm preserving, we may directly compute $\alpha_{J_v,\delta}$ in 
equation (\ref{defalpha}) by using the rotated coordinates as follows.

Let $V(p)$ be a unit vector at $p=v_{I_{v},\delta_0}^{\prime}$ in the $\{x^{\prime}_{\delta_0} \}$ coordinates. 
It is easy to see that $(\phi^*_{\beta, \delta}V)(p)=: V_{\delta}(p)$ at $p=v_{I_{v},\delta}^{\prime}$
is a unit vector in the $\{x^{\prime}_{\delta } \}$ coordinates.
We are interested in relating the components $V^{\mu}_{\delta}(p)$ of $V_{\delta}(p)$  in the $\{x\}$ coordinates to
its components $V^{\mu^{\prime}}_{\delta}$ in the $\{x^{\prime}_{\delta} \}$ coordinates when 
$p= v_{I_{v},\delta}^{\prime}$. Accordingly, at the point $p= v_{I_{v},\delta}^{\prime}$, we have that
\begin{eqnarray}
V^{\mu}_{\delta} &= & V_{\delta}^{\nu^{\prime}} 
\frac{\partial x^{\mu}}{\partial (\phi_{\beta, \delta}^*x^{\prime}_{\delta_0})^{\nu^{\prime}}}
\label{4.6a}
\\
&=&V_{\delta}^{\nu^{\prime}} \frac{\partial x^{\mu}}{\partial (\phi_{\beta, \delta}^*x)^{\lambda}}
\frac{\partial (\phi_{\beta, \delta}^*x)^{\lambda}}{\partial (\phi_{\beta, \delta}^*x^{\prime}_{\delta_0})^{\nu^{\prime}}},
\label{4.6}
\end{eqnarray}
where the coordinates $\phi_{\beta, \delta}^*\{x\}$ (in the vicinity of $v_{I_{v},\delta}^{\prime}$) 
in the last line refer to the pushforward of the coordinates
$\{x \}$ (in the vicinity of $v_{I_{v},\delta_0}^{\prime}$) by the diffeomorphism $\phi_{\beta, \delta}$.
Next, note that the components $V_{\delta}^{\nu^{\prime}}(p)$ of $V_{\delta}(p)$ in the $\{x^{\prime}_{\delta} \}$
 coordinates
are equal to the components $V^{\nu^{\prime}} (\phi^{-1}_{\beta, \delta }(p)) $ of $V(\phi^{-1}_{\beta,\delta }(p))$ in the 
$\{x^{\prime}_{\delta_0} \}$ system i.e.
\begin{equation}
V_{\delta}^{\nu^{\prime}}(p)= V^{\nu^{\prime}} (\phi^{-1}_{\beta, \delta }(p))
\label{4.7}
\end{equation}
Also note that 
\begin{equation}
\frac{\partial (\phi_{\beta, \delta}^*x)^{\lambda}}{\partial (\phi_{\beta, \delta}^*x^{\prime}_{\delta_0})^{ \nu^{\prime} } }|_p
:=
\frac{\partial x^{\lambda}}{\partial x_{\delta_0}^{\prime\nu^{\prime}}}|_{\phi^{-1}_{\beta, \delta}(p)}.
\label{4.8}
\end{equation}
Using this in equation (\ref{4.6}) and restoring the arguments indicative of the point at which the 
object in question is being evaluated, it follows that:
\begin{equation}
V^{\mu}_{\delta} (v_{I_{v},\delta}^{\prime} ) = 
V^{\nu^{\prime}}(v_{I_{v},\delta_0}^{\prime})
\frac{\partial x^{\mu}}{\partial (\phi_{\beta,\delta}^*x)^{\lambda}}|_{v_{I_{v},\delta}^{\prime}}
\frac{\partial x^{\lambda}}{\partial x_{\delta_0}^{\prime\nu^{\prime}}}|_{v_{I_{v},\delta_0}^{\prime}}.
\label{4.9}
\end{equation}
Note that the last set of partial derivatives are independent of $\delta$.
Next, 
from the definition of the diffeomorphism $\phi_{\delta}$ in 5. of  Appendix C4, 
P1, from equation (\ref{defphibeta}) and from the fact that $\{x\}= \beta^*\{x_0\}$, we have that:
\begin{eqnarray}
\frac{\partial x^{3}}{\partial (\phi_{\beta, \delta}^*x)^{\lambda}}|_{v_{I_{v},\delta}^{\prime}} 
= \frac{\partial x_0^{3}}{\partial (\phi_{ \delta}^*x_0)^{\lambda}}|_{v_{I_{v_0},\delta}^{\prime}} 
&=& \delta^3_{\lambda}
\label{4.10}
\\
\frac{\partial x^{\mu\neq 3}}{\partial (\phi_{\delta}^*x)^{\lambda}}|_{v_{I_{v},\delta}^{\prime}}
=\frac{\partial x_0^{\mu\neq 3}}{\partial (\phi_{\delta}^*x_0)^{\lambda}}|_{v_{I_{v_0},\delta}^{\prime}}
&=& \delta^{q-1}
\delta^{\mu}_{\lambda}
\label{4.11}
\end{eqnarray}
Using (\ref{4.10}), (\ref{4.11}) in (\ref{4.9}) we obtain:
\begin{eqnarray}
V_{\delta}^{\mu =3} (v_{I_{v},\delta}^{\prime} )& =&
V^{\nu^{\prime}}(v_{I_{v},\delta_0}^{\prime}) 
\frac{\partial x^{3}}{\partial x_{\delta_0}^{\prime\nu^{\prime}}}|_{v_{I_{v},\delta_0}^{\prime}}
\label{4.12},
\\
V_{\delta}^{\mu \neq 3} (v_{I_{v},\delta}^{\prime} ) &=& O(\delta^{q-1}) .
\label{4.13}
\end{eqnarray}

Setting
$V^a|_{v_{I_{v},\delta_0}^{\prime}} =:{\hat {\tilde e}}_{J_v}^{\prime a}|_{v_{I_{v},\delta_0}^{\prime}}$,
it follows from equations (\ref{4.12}), (\ref{4.13}) 
that the edge tangents ${\hat {\tilde e}}_{J_v}^{\prime a}|_{v_{I_{v},\delta}^{\prime}}$
all admit well defined limits as $\delta\rightarrow 0$.
For example, for $J_v= I_v$, it follows from (\ref{eicx3}), (\ref{eicx3prime})  that 
\begin{equation}
\lim_{\delta \rightarrow 0}
{\hat {\tilde e}}_{I_v}^{\prime a}|_{v_{I_{v},\delta}^{\prime}}
= \lambda_{I_v}{\hat e}_{I_v}^{a}|_v
\label{eprimeivcont}
\end{equation}
where we have defined $\lambda_{I_v}$ to be
\begin{equation}
\lambda_{I_v}:= \frac{\partial x^{3}}{\partial x_{\delta_0}^{\prime 3}}|_{v_{I_{v},\delta_0}^{\prime}}
\label{lambdaiv}
\end{equation}

\subsubsection{The determinant of the Jacobian}
Next, we show that the volume form in the $\{x^{\prime}_{\delta}\}$ coordinates degenerates in the continuum 
limit. Accordingly, we compute the determinant of the Jacobian between the $\{x\}$ and 
$\{x^{\prime}_{\delta}\}$ coordinates at $v_{I_{v},\delta}^{\prime}$.
\footnote{As for the norm, the determinant is also insensitive to rotations and we can directly use the 
`rotated' coordinates of section 4.2.1 to compute the determinant.}
 We have that 
\begin{eqnarray}
\frac{\partial x^{\mu}}{\partial x_{\delta}^{ {\prime}\nu^{\prime} } }|_{v_{I_{v},\delta}^{\prime}}
&=&
\frac{\partial x^{\mu}}{\partial (\phi_{\beta, \delta}^*x)^{ \lambda } }|_{v_{I_{v},\delta}^{\prime}}\;\;\;\;
\frac{\partial (\phi_{\beta, \delta}^*x)^{ \lambda } }
     {\partial (\phi_{\beta, \delta}^*x^{\prime}_{\delta_0})^{\nu^{\prime}}}|_{v_{I_{v},\delta}^{\prime}}
\nonumber
\\
&=&
\frac{\partial x^{\mu}(p)}{\partial x^{ \lambda } (\phi^{-1}_{\beta, \delta}(p))   }|_{p=v_{I_{v},\delta}^{\prime}}
\;\;\;\;
\frac{\partial x^{ \lambda } }
     {\partial x_{\delta_0}^{{\prime}\nu^{\prime}}}|_{v_{I_{v},\delta_0}^{\prime}}
\nonumber
\\
&=&
\frac{\partial x^{\mu}(\phi_{\beta, \delta}(p))}{\partial x^{ \lambda } (p)   }|_{p=v_{I_{v},\delta_0}^{\prime}}
\;\;\;\;
\frac{\partial x^{ \lambda } }
     {\partial x_{\delta_0}^{{\prime}\nu^{\prime}}}|_{v_{I_{v},\delta_0}^{\prime}}\\
\Rightarrow
\det 
\frac{\partial x^{\mu}}{\partial x_{\delta}^{ {\prime}\nu^{\prime} } }|_{v_{I_{v},\delta}^{\prime}}
&=& \det\frac{\partial x^{\mu}_0(\phi_{\delta}(p))}{\partial x_0^{ \lambda } (p)   }|_{p=v_{I_{v},\delta_0}^{\prime}}
\;\;\;\;
\det\frac{\partial x^{ \lambda } }
     {\partial x_{\delta_0}^{{\prime}\nu^{\prime}}}|_{v_{I_{v},\delta_0}^{\prime}}\nonumber\\
&=& (\delta^{q-1})^2 \det (H_{I_v})
\label{detjacobian}
\end{eqnarray}
In equation (\ref{detjacobian})
we have used (\ref{defphibeta}) together with 
the definition of $\phi_{\delta}$ in P1 (see also (\ref{4.10}), (\ref{4.11})) above)
and defined the matrix $H_{I_v}$ as
\begin{equation}
H_{I_v\nu^{\prime}}^{\lambda}= \frac{\partial x^{ \lambda } }
     {\partial x_{\delta_0}^{{\prime}\nu^{\prime}}}|_{v_{I_{v},\delta_0}^{\prime}}.
\label{defH}
\end{equation}
Clearly, $H_{I_v}$ is independent of $\delta$ so that the Jacobian vanishes as $\delta^{2q-2}$ (recall from P1 that 
$q\geq 2$)
in the continuum limit. This shows that the coordinate volume form in the $\{x^{\prime}_{\delta}\}$ coordinates
is ill behaved in the continuum limit.

\subsection{Divergence of the commutator.}

It is easy to see that the computation of the 
$\delta^{\prime}\rightarrow 0$ limit of the finite triangulation commutator
goes through exactly as in section 4, P1 with the proviso that we choose
the $\{x^{\prime}\}$ coordinates
of section 4, P1 
to be the $\{x^{\prime}_{\delta}\}$ coordinates of this work. 
Accordingly, from section 4, P1 we have that:
\begin{align}
&  \lim_{\delta^{\prime}\rightarrow0}(\Psi_{B_{\mathrm{VSA}}}|(\hat{C}%
_{\delta^{\prime}}[M]\hat{C}_{\delta}[N]-\left(  N\leftrightarrow M\right)
)|c\rangle\nonumber\\
&  =\left(  \frac{\hbar}{2\mathrm{i}}\frac{3}{4\pi}\right)  ^{2}\sum_{v}%
\nu_{v}^{-2/3}\nonumber\\
&  \times\sum_{I_{v}}\{N(v,\{x\})\hat{e}_{I_{v}}^{a}\partial_{a}M(v,\{x\})-\left(
N\leftrightarrow M\right)  +O(\delta)\}\left[  \det\left(  \frac{\partial
x}{\partial x^{\prime}_{\delta}}\right)  _{v_{I_{v},\delta}^{\prime}}\right]
^{-1/3}\nu_{v_{I_{v}}}^{-2/3}\{\cdots\}_{I_{v},\delta} \label{cmcnbrkt2}%
\end{align}
where 
\begin{equation}
\{\cdots\}_{I_{v},\delta}:=\sum_{i}q_{I_{v}}^{i}\nu_{v_{I_{v}}}^{-2/3}%
\sum_{J_{v},i^{\prime}}\left.  ^{(i)}\!q_{J_{v}}^{i^{\prime}}\right.
({\hat{\tilde{e}}}{}_{J_{v}}^{\prime})^{a}\partial_{a}f(v_{I_{v},\delta
}^{\prime}) \label{defcurly}%
\end{equation}
From (\ref{4.12})(\ref{4.13}) and (\ref{eprimeivcont}) it follows that for generic $f, |c\rangle$,
we have that $\lim_{\delta \rightarrow 0}\{\cdots\}_{I_{v},\delta}:= \{\cdots\}_{I_{v},\delta=0}$ is well defined and non- vanishing. From (\ref{4.12}), (\ref{4.13}), the fact that  $v_{I_{v},\delta}^{\prime}$ and $v$ are seperated
by an $\{x\}$ coordinate distance of $O(\delta )$ and the fact that $q\geq 2$  we have that 
\begin{equation}
\{\cdots\}_{I_{v},\delta} =\{\cdots\}_{I_{v},\delta= 0} + O(\delta ) .
\label{curlydelta}
\end{equation}

Using (\ref{curlydelta}) and (\ref{detjacobian}) in (\ref{cmcnbrkt2}), we obtain:
\begin{eqnarray}
&&\lim_{\delta^{\prime}\rightarrow0}(\Psi_{B_{\mathrm{VSA}}}|(\hat{C}%
_{\delta^{\prime}}[M]\hat{C}_{\delta}[N]-\left(  N\leftrightarrow M\right)
)|c\rangle\nonumber\\
&&  =
\Big[ \;\;
\left(  \frac{\hbar}{2\mathrm{i}}\frac{3}{4\pi}\right)  ^{2}\sum_{v}%
\nu_{v}^{-2/3}
\delta^{-\frac{2}{3}(q-1)}\nonumber\\
&&\sum_{I_{v}} \det (H_{I_v})
\{N(v,\{x\})\hat{e}_{I_{v}}^{a}\partial_{a}M(v,\{x\})-\left(
N\leftrightarrow M\right)\}
\{\cdots\}_{I_{v},\delta= 0}\;\;
\Big]
\nonumber\\
&&\;\;\;\;\;\;\;\;\;\;\;\;\;\;\;\;\;\;+ \;\;\;\;\;O(\delta^{1-\frac{2}{3}(q-1) })
\label{commdiv}
\end{eqnarray}
It follows that 
the commutator diverges as $\delta^{-\frac{2}{3}(q-1)}$ in the continuum limit $\delta \rightarrow 0$
and that this divergence is due to the singular behaviour of the coordinate volume form in the
$\{x^{\prime}_{\delta}\}$ coordinates.

\subsection{The dependence of the primed coordinates on $i$}
Note that, as explained in Footnote \ref{footnoteidep}, the coordinates $\{x^{\prime}\}_{\delta}$
at $v_{I_{v},\delta}^{\prime}$ in $c(i,v_{I_{v},\delta}^{\prime})$ inherit an $i$- dependence from the choice of
reference diffeomorphism sets ${\cal D}_{[c(i,v_{I_{v},\delta}^{\prime})     ]}$. 
As we show below, we can nevertheless, assume that these coordinates are $i$ independent for the purposes
of the computation of the commutator continuum limit of section 4.3. 
The commutator calculation involves a $\delta^{\prime}\rightarrow 0$ limit followed by a $\delta\rightarrow 0$
limit. We remark on the $i$- independence assumption of $\{x^{\prime}_{\delta}\}$ for each of these limits:\\

\noindent (a) The $\delta^{\prime}\rightarrow 0$ limit: This limit computes the continuum limit of the
{\em single} action of the finite triangulation constraint at triangulation $\delta^{\prime}$ on the charge networks 
$c(i,v_{I_{v},\delta}^{\prime})$ produced by the finite triangulation Hamiltonian constraint at triangulation 
parameter $\delta$. From section 3 it follows that this continuum limit is independent of the choice
of reference diffeomorphism sets ${\cal D}_{[c(i,v_{I_{v},\delta}^{\prime})     ]}$. From section 2,
the sibling restriction implies that the reference charge networks for 
$[c(i,v_{I_{v},\delta}^{\prime})     ]$ have the same reference charge coordinates. 
It is also straightforward to check that the definition of siblings implies  that given any diffeomorphism 
$d$ which maps ${\bar c}_i$ to $c(i,v_{I_{v},\delta}^{\prime})$ for some value of $i$, the same
diffeomorphism $d$ also maps  ${\bar c}_j$ to $c(j,v_{I_{v},\delta}^{\prime})$ for any $j\neq i$.
It then follows that for the 
$\delta^{\prime}\rightarrow 0$ limit, we are justified in assuming that $\{x^{\prime}_{\delta}\}$ can be chosen
to be $i$- independent.
\\

\noindent (b)The $\delta \rightarrow 0$ limit: This limit only involves the properties of the 
diffeomorphisms $\phi_{\delta}$ of P1.  All these properties (see Appendix C4 of P1) such as the placement of 
various kink vertices and the nontrivial vertex, as well as the scrunching of edge tangents are
independent of the choice of $\{x^{\prime}_{\delta}\}$. Hence the 
$\delta \rightarrow 0$ limit is also independent of the choice of ${\cal D}_{[c(i,v_{I_{v},\delta}^{\prime})     ]}$.
\\

The considerations (a) and (b) above imply that we may assume that $\{x^{\prime}_{\delta}\}$ can be chosen
to be $i$- independent for the purposes of section 4.3.

\section{Modified deformations.}
In this section, the structure of the deformations generated by the finite triangulation Hamiltonian constraint
and  the finite triangulation electric diffeomorphism operator (see section 5,P1) 
are modified relative to P1.
These modifications result in 
the well
defined anomaly free commutator of section 7. As in section 2,
these deformations are defined in a diffeomorphism covariant manner through the following steps:\\
(i) A set of fixed reference structures is chosen.\\
(ii) The deformations  at parameter value $\delta_0$ of the reference charge networks chosen in (i) 
are constructed. \\
(iii)The diffeomorphisms $\phi_{\delta}$ are constructed and their action on
the deformations of a reference charge net at $\delta_0$ result in deformations of the reference charge net
at $\delta <\delta_0$.\\
(iv)The deformations of the  charge networks diffeomorphic to a specific reference charge network
are obtained as diffeomorphic images of the deformations of the reference charge network using the reference 
diffeomorphisms of (i).

We proceed more or less in the order (i) - (iii), step (iv) being a trivial consequence of applying the 
reference diffeomorphisms of (i) to the deformations of the reference charge networks defined through (ii) and (iii). 
In section 5.1 we
review the fixed reference structures of step (i) and introduce a key additional ``conicality''
 restriction on their choice.
In sections 5.2 and 5.3,
we implement steps (ii) and (iii) for the deformations generated by the Hamiltonian constraint.
The steps (ii)- (iii) for the deformations produced by the electric diffeomorphism constraint 
follow trivially from sections 5.2 and 5.3 and we describe them in section 5.4. 

The modifications do not affect the continuum limit action of the Hamiltonian constraint in any deep way.
Their effect is primarily on the continuum limit of the commutator. 
The key modifications pertain to step (ii) in section 5.2  and step (iii) in section 5.3.
 In section 5.2 the deformations of the reference charge networks
at $\delta =\delta_0$ are modified so as to acquire a certain ``conical'' structure and in section 5.3
the scrunching of edges is done delicately  enough that this conical structure
manifests intact at any $\delta$. As we shall see in section 7, this conical structure plays a key role
in our demonstration of the anomaly free property of the commutator. The considerations of section 7 
require a knowledge of the behaviour of the edge tangents at the nontrivial vertices  of the deformed charge nets
as  $\delta\rightarrow 0$.
We compute this behaviour in section 5.5.

\subsection{Fixed Reference Structures}
These structures are identical to those of section 2.2. We repeat the contents of that section in slightly 
modified notation appropriate to our considerations here. In each diffeomorphism class of nontrivial chargenets,
fix a reference charge net $c_0$. Let its nontrivial vertex be $v_0$. Fix the reference coordinate
chart $\{x_0\}$ around $v_0$ in accordance with the considerations of section 3. 
The choice of reference charge networks and reference coordinate patches are subject to the 
``sibling'' restrictions of section 2.2.
\footnote{It is easy to check that the choice of reference coordinates in section 3 is consistent with the 
sibling restrictions of section 2.2.}

Let $[c_0]$ be the diffeomorphism
class of $c_0$. 
Let ${\cal D}_{[c_0]}$ be a set of reference diffeomorphisms such that for every distinct 
$c\in [c_0]$ there is a unique $\beta \in {\cal D}_{[c_0]}$ with $c =c_0^{\beta}$ being the image of $c_0$ by
$\beta$. 
Deformations of $c_0^{\beta}$ at parameter $\delta$ are defined to be the images of the corresponding deformations
of $c_0$ by $\beta$.
Deformations generated by the Hamiltonian constraint at parameter $\delta_0$ are defined in the next section.

In the case where the edge tangent set has symmetries and $a_{min}\geq 1$, the reference coordinates are 
chosen as in section 3.3. If the edge tangent set has no symmetries, diffeomorphism covariance places no 
restrictions on the choice of reference coordinates. However, in order to obtain an anomaly free commutator
in section 7, it turns out that we need to restrict the choice of reference coordinates in the case of 
no symmetries as follows. Let the preferred edge in the reference charge net $c_0$ at its nontrivial vertex 
$v_0$ be $e_{I_{v_0}}$. Similar 
to the remark at the  end of section 3.3, we 
choose reference  coordinates $\{x_0\}$ such that at $v_0$ the $x^3_0$ axis is along the tangent
to the preferred edge.

Next, consider those edge tangent
sets without symmetries for which there exists a choice of coordinates such that 
the angle between each edge tangent 
${\dot e}_{J_{v_0}\neq I_{v_0}}|_{v_0}$ and the $x^3_0$ axis is the same (i.e. independent of $J_{v_0}$).
We shall refer to such edge tangent sets as {\em conical}. We require that if an edge tangent set 
is conical, the associated reference coordinate system should be chosen so as to make this 
conicality manifest i.e. the reference coordinate system is such that 
the angle $\alpha$  between each edge tangent 
${\dot e}_{J_{v_0}\neq I_{v_0}}|_{v_0}$ and the $x^3_0$ axis is independent of $J_{v_0}$.
Further, if there is a coordinate system in which this conicality is downward (i.e. the angle 
$\alpha$ is greater than $\frac{\pi}{2}$), then the associated reference system is to be chosen so
as to make this downward conicality manifest.

Since conicality is an important concept we make the following definition:\\

\noindent{\em Definition. Conical Chargenet:} A nontrivial chargenet will be said to be conical 
iff there exists some coordinate system around its nontrivial vertex in which, at this vertex, 
the edge tangents to all the edges other than the preferred one subtend the same angle
with the edge tangent along the preferred edge.
\\

It is easy to see that the property of conicality is diffeomorphism invariant since, if conicality
is manifest for a charge net in some coordinate system, conicality manifests for a diffeomorphic 
charge net in the diffeomorphic image of this coordinate system.

\subsection{Deformations produced by ${\hat H}_{\delta_0}(N)$ on $c_0$.}
 
The action of ${\hat H}_{\delta_0}(N)$ on $c_0$ produces the deformed charge nets 
$c_0(i, v_{I_{v_0},\delta_0}^{\prime})$. An important property of the deformation is that 
$v_{I_{v_0},\delta_0}^{\prime}$ is required to be GR in $c_0(i, v_{I_{v_0},\delta_0}^{\prime})$, if 
$v_0$ is GR in $c_0$. 
In P1 we {\em assumed} that the deformed vertex 
$v_{I_{v_0},\delta_0}^{\prime}$ in the deformed charge net $c_0(i, v_{I_{v_0},\delta_0}^{\prime})$ is GR if 
the vertex $v_0$ is GR in $c_0$, and left a validation of 
this assumption for future work. In more detail, recall that at the end of Step 3, Appendix C.2, P1
we were unable to ascertain if the deformed vertex $v_{I_{v_0},\delta_0}^{\prime}$  is GR. 
As a result, in Step 4, C2, P1 we assumed
the existence of a suitable  prescription  which converted the possibly non GR vertex of Step 3, C2, P1
into an GR one. Our first task in this section is to construct such a prescription. We do so by a further deformation
of the charge net provided by Step 3, C2, P1 in a small vicinity of 
$v_{I_{v_0},\delta_0}^{\prime}$, which results in the 
downward conicality (with respect to the $\{x_0\}$ coordinates) of the so deformed
edge tangent set at 
$v_{I_{v_0},\delta_0}^{\prime}$.
We describe this deformation in section 5.2.1 and Appendix A.1, and show in Appendix A.1 that the deformed
edge tangent set is GR.

For the purposes of section 7, it turns out that the desired deformation of $c_0$ is required to be conical 
in the coordinates $\{x^{\prime}_{\delta_0}\}$ associated with $v_{I_{v_0},\delta_0}^{\prime}$. These 
coordinates are, in general, different from the $\{x_0\}$ coordinates appropiate to $v_0$ in $c_0$.
Hence we need to further deform the chargenets which are conical in $\{x_0\}$ to achieve
conicality with respect to $\{x^{\prime}_{\delta_0}\}$. In section 5.2.2 we apply an iterative procedure to
the charge nets provided by section 5.2.1 and Appendix A.1 so as to obtain the desired conicality with 
respect to $\{x^{\prime}_{\delta_0}\}$. Technicalities pertaining to section 5.2.2 are detailed in Appendix A.2.

The considerations of sections 5.1.1 and especially of 5.2.2 are technically quite involved. The reader may find
it easier to skip directly to section 5.3 on a first reading and assume the main result of section 5.2 which is
that the deformations of the reference chargenet at $\delta =\delta_0$ are such that the 
edge tangents at the deformed vertex are arranged in a downward coordinate cone with axis opposite to the preferred edge
tangent, the coordinates being the ones appropriate to the deformed vertex. Thus, in the 
$\{x^{\prime}_{\delta_0}\}$ coordinates, all the unit edge tangents 
(other than the preferred one) make the {\em same} angle (of less than 90 degrees) with respect to the 
cone axis. This conicality is crucial for our demonstration of anomaly free- ness of the commutator in section 7.

\subsubsection{Conicality in $\{x_0\}$.}
Set $c=c_0$
and $\{x\}= \{x_0\}$ in 
Appendix C2,P1 and  
Step 3, Appendix C4,P1.
We slightly abuse notation and refer to the chargenets provided by Step 3, C2, P1 as
$c_0(i, v_{I_{v_0},\delta_0}^{\prime})$ even though $v_{I_{v_0},\delta_0}^{\prime}$ may not be 
GR.

We deform $c_0(i, v_{I_{v_0},\delta_0}^{\prime})$ so that the deformation in a small vicinity of
of $v_{I_{v_0},\delta_0}^{\prime}$ puts the deformed edge tangents (other than the $I_{v_0}$th one) 
emanating from $v_{I_{v_0},\delta_0}^{\prime}$
along a downward cone in the $\{x_0\}$ coordinates. This is done as follows.
The edges (other than the $I_{v_0}$th one) 
emanating from $v_{I_{v_0},\delta_0}^{\prime}$ in $c_0(i, v_{I_{v_0},\delta_0}^{\prime})$
point downwards and hence each such edge is at an angle less than $\frac{\pi}{2}$ 
with respect to the negative 
$x_0^{3}$ axis. Let the least of these angles be $\Theta_{x_0}$. We 
rotate each edge tangent within the plane containing the edge tangent and the $x_0^{3}$ axis down to the
angle $\Theta_{x_0} $. This is achieved by the  procedure of Appendix A.1. Note that this procedure
leaves the $I_{v_0}$th edge untouched.  
We denote the resulting chargenet by $c^{(0)}_0(i, v_{I_{v_0},\delta_0}^{\prime})$. As shown in 
Appendix A.1, by virtue of the Lemma of Appendix D, $v_{I_{v_0},\delta_0}^{\prime}$ is GR in 
$c^{(0)}_0(i, v_{I_{v_0},\delta_0}^{\prime})$.

To summarise: The chargenet $c^{(0)}_0(i, v_{I_{v_0},\delta_0}^{\prime})$ is downward conical 
in the coordinates $\{x_0\}$ associated with $c_0$, and $v_{I_{v_0},\delta_0}^{\prime}$
is GR in $c^{(0)}_0(i, v_{I_{v_0},\delta_0}^{\prime})$.

\subsubsection{Conicality in $\{x^{\prime}_{\delta_0}\}$}

In this section
we further deform $c^{(0)}_0(i, v_{I_{v_0},\delta_0}^{\prime})$ of section 5.2.1 to obtain the desired 
$c_0(i, v_{I_{v_0},\delta_0}^{\prime})$ of this work.
These further deformations of $c^{(0)}_0(i, v_{I_{v_0},\delta_0}^{\prime})$ of P1 are generated through 
an iterative procedure 
which terminates in a finite
number of steps. Each step alters the deformation  produced by the previous step in an arbitrarily
small neighbourhood of 
the nontrivial vertex
$v_{I_{v_0},\delta_0}^{\prime}$, without changing the position of this nontrivial vertex and without deforming
the $I_{v_0}$th edge. 

Our notation is as follows. The chargenet obtained 
at the end of the $k$th  step is denoted by 
$c^{(k)}_0(i, v_{I_{v_0},\delta_0}^{\prime})$. The edges emanating from its nontrivial vertex, 
$v_{I_{v_0},\delta_0}^{\prime}$ are denoted by ${\tilde e}^{(k)}_{J_{v_0}}$ for $J_{v_0}\neq I_{v_0}$.
Since the $I_{v_0}$th edge is untouched at each step, we continue to denote this edge by 
${\tilde e}_{I_{v_0}}$. The coordinate patch associated with 
$v_{I_{v_0},\delta_0}^{\prime}$ in $c^{(k)}_0(i, v_{I_{v_0},\delta_0}^{\prime})$
is denoted by $\{x^{\prime (k)}_{\delta_0}\}$. 
These coordinates are chosen as described in section 5.1. 
In particular, the remarks at the end of section 3 imply that 
$(\frac{\partial \;}{\partial x^{\prime (k)3}_{\delta_0}})^a$ points along the tangent to ${\tilde e}_{I_{v_0}}$
at $v_{I_{v_0},\delta_0}^{\prime}$.
The total number of steps required to accomplish
the desired deformation is designated as $m$.
Our aim is to obtain a deformed chargenet 
$c^{(m)}_0(i, v_{I_{v_0},\delta_0}^{\prime})$ at the end of $m$ steps
 wherein all the edges 
${\tilde e}^{(m)}_{J_{v_0}\neq I_{v_0}}$ are arranged such that their tangents at $v_{I_{v_0},\delta_0}^{\prime}$
are  exactly along a coordinate cone in the coordinates 
$\{x^{\prime (m)}_{\delta_0}\}$
with axis 
opposite to the tangent to the $I_{v_0}$th edge. 
In what follows, we shall refer to the direction opposite to the $I_{v_0}$th edge tangent as the ``downward''
direction. 

We start at the $k=0$th step with the chargenet  $c^{(0)}_0(i, v_{I_{v_0},\delta_0}^{\prime})$ of section 5.2.1.
We denote its associated coordinate patch at 
$v_{I_{v_0},\delta_0}^{\prime}$ 
by $\{x^{\prime (0) }_{\delta_0}\}$ with 
$(\frac{\partial \;}{\partial x^{\prime (0)3}_{\delta_0}})^a$  pointing  along ${\tilde e}_{I_{v_0}}$, and 
its edges emanating from $v_{I_{v_0},\delta_0}^{\prime}$ by ${\tilde e}^{(0)}_{J_{v_0}}$. 
\footnote{
We apologise for the detailed notation and the use of the number $0$ to denote, on the one hand, the 
structures associated with the zeroth step of the iterative procedure, and on the other, the structures associated
with the undeformed reference charge network. In this regard, we emphasize that whereas the {\em subscript}
`$0$' is associated with {\em reference} structures, the {\em superscript} `$(0)$' labels structures 
at associated with the $0$th step of the iterative procedure.
}
Suppose the edge tangent configuration at the nontrivial vertex of 
$c^{(0)}_0(i, v_{I_{v_0},\delta_0}^{\prime})$ has {\em no} symmetries. 
Then we terminate the procedure and define the deformation at $\delta =\delta_0$
to be $c^{(0)}_0(i, v_{I_{v_0},\delta_0}^{\prime})$. 
Note that the $c^{(0)}_0(i, v_{I_{v_0},\delta_0}^{\prime})$ is conical (see the discussion at the end
of section 5.1) because, by construction, its (downward) conicality is manifest in the $\{x_0\}$ coordinates.
By virtue of 
the restriction on the choice of reference coordinates at the end of section 5.1
it follows that $c^{(0)}_0(i, v_{I_{v_0},\delta_0}^{\prime})$ is `downwardly conical' also in the 
$\{x^{\prime (0) }_{\delta_0}\}$ coordinates associated with 
$c^{(0)}_0(i, v_{I_{v_0},\delta_0}^{\prime})$. 
\footnote{Specifically,  $c^{(0)}_0(i, v_{I_{v_0},\delta_0}^{\prime})$ is the diffeomorphic image of
its reference charge network by some reference diffeomorphism. Applying the inverse of this diffeomorphism
to $\{x_0\}$ yields a coordinate system in which the reference charge network  exhibits
downward conicality implying that downward conicality manifests in its reference coordinates and hence
implying downward conicality of its diffeomorphic image, $c^{(0)}_0(i, v_{I_{v_0},\delta_0}^{\prime})$,
in the diffeomorphic image of these reference coordinates.}

If the edge tangent configuration in $c^{(0)}_0(i, v_{I_{v_0},\delta_0}^{\prime})$ at $v_{I_{v_0},\delta_0}^{\prime}$ 
has symmetries, 
then it is characterised
by some $a_{min}:= a_{min}^{(0)} \geq 1$ and, from section 3, it must be the case that 
a rotation by $\theta^{(0)}_{min}$ (see equation (\ref{thetamin})) about the 3rd axis
maps the edge tangents onto each other. It is then straightforward to see that 
in the $\{x^{\prime (0) }_{\delta_0}\}$ coordinates there is at 
least one edge which points ``downwards'' i.e. there is some $J_{v_0}$ for which
the unit edge tangent in these coordinates, 
${\hat \tilde e}_{J_{v_0}}^{\prime (0) a}$, has a negative projection along the $x^{\prime (0)3}_{\delta_0}$ axis. 
To
see this recall that the deformation of section 5.2.1  is such that in the $\{x_0\}$ coordinates, the edge tangents 
$\{{\hat {\tilde e}}_{J_{v_0}}^{(0)a}, J_{v_0}\neq I_{v_0}\}$ all point ``downwards'' 
at $v_{I_{v_0},\delta_0}^{\prime}$,
the hat as usual denoting unit norm with respect to the $\{x_0\}$ coordinates.
As shown in Appendix C, this ``downwardness'' in $\{x_0\}$ together with the 
the GR property of $v_{I_{v_0},\delta_0}^{\prime}$ and 
the fact that $a^{(0)}_{min}\geq 1$ implies that 
there exists a 
triple of edges ${\tilde e}^{(0)}_{J^i_{v_0}}, i=1,2,3, J^i_{v_0}\neq I_{v_0}$ such that:
\begin{equation}
\sum_i\alpha^{}_i{\hat {\tilde e}}_{J^i_{v_0}}^{^{\prime}(0) a} =  -\gamma{\hat {\tilde e}}_{I_{v_0}}^{^{\prime}(0)a},
\label{negativeprojprime}
\end{equation} 
for some $\alpha_i >0, \gamma>0$,
from which it follows that at least one of this triplet of edges points downwards in the 
$\{x^{\prime (0)}_{\delta_0}\} $ coordinates.

Next, consider all the edges which have negative projections along the 3rd axis in the $\{x^{\prime (0)}_{\delta_0}\}$ 
coordinates. Each of these edges is at angle less than $\frac{\pi}{2}$ with respect to the negative 
$x^{\prime (0) 3}_{\delta_0}$ axis. Let the least value of
this angle be $\Theta$. We `rotate'  each edge (except the $I_{v_0}$th one) down to this angle so that 
all the edge tangents
point downwards at this angle. 
The detailed construction of this `downward rotation'
is described in  Appendix A.2 
As a result, 
the edges are now aligned in such a way that, apart from the $I_{v_0}$th edge, they 
{\em all} point exactly along a downward cone of angle $\Theta$.
Denote the resulting chargenet by $c^{(1)}_0(i, v_{I_{v_0},\delta_0}^{\prime})$.

The procedure described in Appendix A does not necessarily correspond to the action of a {\em single}
diffeomorphism on $c^{(0)}_0(i, v_{I_{v_0},\delta_0}^{\prime})$. Hence the new configuration of edge tangents
may be characterised by an $a_{min}:= a_{min}^{(1)}$ (see equation (\ref{defamin})) different from $a_{min}^{(0)}$
We now argue that $1\leq a_{min}^{(1)}\leq a_{min}^{(0)}$.
%
Let $v_{I_{v_0},\delta_0}^{\prime}$ 
be $M$ valent.  It follows from the initial part of Appendix C
that in $c^{(0)}_0(i, v_{I_{v_0},\delta_0}^{\prime})$ there
are ${a^{(0)}_{min}}$ disjoint sets of edge tangents, $s_{\alpha}, \alpha =1,.., {a^{(0)}_{min}}$,
such that the elements of each $s_{\alpha}$ are mapped to each other by the rotation $R(\theta^{(0)}_{min})$ through
 the angle 
$\theta^{(0)}_{min}$ about the $x^{\prime (0) 3}_{\delta_0}$ axis. 
Clearly the elements of each $s_{\alpha}$ lie on a cone 
of some angle $\Theta_{\alpha}$ about the 3rd axis. 
From our considerations above, it follows that the new configuration of edge tangents is obtained by 
rotating each edge tangent in the  planar subspace of the tangent space at  $v_{I_{v_0},\delta_0}^{\prime}$
which contains itself and the 
$x^{\prime (0) 3}_{\delta_0}$ direction, from $\Theta_{\alpha}$
down to $\Theta$. It follows that the rotation  $R(\theta^{(0)}_{min})$ continues to be a symmetry of 
the edge tangents obtained by rotating those in $s_{\alpha}$ from $\Theta_{\alpha}$ to $\Theta$.
Moreover these transformations which change $\Theta_{\alpha}$ to $\Theta$ do not alter the ordering of the 
edges $1,..,M-1$ (see Step 2 of section 3.2). It then follows that the  $a_{min}=a^{(1)}_{min} $ for 
$c^{(1)}_0(i, v_{I_{v_0},\delta_0}^{\prime})$, 
is such that $1 \leq a^{(1)}_{min} \leq a^{(0)}_{min}$.
If $a^{(1)}_{min}= a^{(0)}_{min}$, it follows that 
$\{x^{\prime (0)}_{\delta_0}\}$ is an acceptable set of 
coordinates (by which we mean that symmetries of the edge tangent set are proportional to rotations in these
coordinates) and we terminate the procedure. Note that the deformation is manifestly (downwardly) conical 
with respect to $\{x^{\prime (0)}_{\delta_0}\}$. 
Note that this coordinate system is not necessarily the one associated with 
$c^{(1)}_0(i, v_{I_{v_0},\delta_0}^{\prime})$ through the choice of reference coordinate systems and reference
diffeomorphism sets. However,
from Appendix B, it follows that downward conicality is also 
manifest in the coordinate sytem (let us call it $\{x^{\prime (1)}_{\delta_0}\}$)
associated with $c^{(1)}_0(i, v_{I_{v_0},\delta_0}^{\prime})$.

If $a^{(1)}_{min}< a^{(0)}_{min}$ we proceed as follows.
If it so happens that $\{x^{\prime (0)}_{\delta_0}\}$ is an acceptable set of 
coordinates  for $c^{(1)}_0(i, v_{I_{v_0},\delta_0}^{\prime})$ 
(by which we mean that symmetries of the edge tangent set are proportional to rotations in these
coordinates), we terminate the procedure. Similar to the paragraph immediately before this one, 
Appendix B ensures that conicality is manifest with respect to the coordinates
$\{x^{\prime (1)}_{\delta_0}\}$)
associated with $c^{(1)}_0(i, v_{I_{v_0},\delta_0}^{\prime})$.
If this is not the case, we iterate the procedure starting from the new deformed charge net 
$c^{(1)}_0(i, v_{I_{v_0},\delta_0}^{\prime})$ with coordinate
patch $\{x^{\prime(1)}_{\delta_0}\}$ and with the new edge tangent configuration 
characterised by $a_{min}= a^{(1)}_{min}$. Since the edge tangents for $J_{v_0}\neq I_{v_0}$ at 
${v_{I_{v_0},\delta_0}^{\prime}}$ in $c^{(1)}_0(i, v_{I_{v_0},\delta_0}^{\prime})$ all point downwards
in the coordinate system $\{x^{\prime}_{\delta_0}\}$, an  arguementation similar to that resulting in equation
(\ref{negativeprojprime}) 
allows us to conclude that there is at least one
edge at ${v_{I_{v_0},\delta_0}^{\prime}}$ in $c^{(1)}_0(i, v_{I_{v_0},\delta_0}^{\prime})$
 which points downwards in the $\{x^{\prime(1)}_{\delta_0}\}$ coordinates. As a result there is, once again, a least
cone angle $\Theta^{(1)}$ for the set of all downward pointing edges in the $\{x^{\prime(1)}_{\delta_0}\}$ coordinates.
We iterate the procedure, using Appendix A to  rotate all 
the edge tangents other than the $I_{v_0}$th one down to $\Theta^{(1)}$
thus obtaining the new chargenet $c^{(2)}_0(i, v_{I_{v_0},\delta_0}^{\prime})$ characterised by 
$a_{min} = a^{(2)}_{min}$. If $a^{(2)}_{min}= a^{(1)}_{min}$ or if $a^{(2)}_{min} <a^{(1)}_{min}$
and the $\{x^{\prime(1)}_{\delta_0}\}$ coordinates happen to be acceptable coordinates, we terminate the procedure
else we iterate.

Since $a_{min}$ is bounded below by 1 and since $a^{(i+1)}_{min} \leq a^{(i)}_{min}$, this procedure necessarily
terminates in a finite number of steps $m$ with all edge tangents exactly along a downward coordinate cone
of some angle $\Theta$ where the coordinates $\{x^{\prime (m)}_{\delta_0}\}$ are appropriate to this
very same `conical' configuration of edge tangents.
We define this end point to be the deformation of the reference charge net at $\delta_0$ i.e. we set
$c_0(i, v_{I_{v_0},\delta_0}^{\prime})\equiv c^{(m)}_0(i, v_{I_{v_0},\delta_0}^{\prime})$.

We point out three important features of our construction above. First, as shown in Appendix A.2, the 
rotation of the edge tangents into conicality preserves the GR nature of $v_{I_{v_0},\delta_0}^{\prime}$
at each step of the procedure which allows an application of Appendix A  in the next step. Second, 
our construction above is {\em independent} of the choice of reference diffeomorphisms.
More in detail, note that while $c_0$ is a reference charge net, it is not necessary that 
the deformed charge net
$c^{(0)}_0(i, v_{I_{v_0},\delta_0}^{\prime})\equiv c^{}_0(i, v_{I_{v_0},\delta_0}^{\prime})$ of 
P1 or the deformed chargenets $c^{(k>0)}_0(i, v_{I_{v_0},\delta_0}^{\prime})$
above
are reference charge nets. Hence the coordinates $\{x^{\prime (k\geq 0)}_{\delta_0}\}$ 
are, in general, images of appropriate reference coordinates by reference diffeomorphisms.
Let us consider 2 distinct choices of these reference diffeomorphisms. From section 3, their 
action at the vertex $v_{I_{v_0},\delta_0}^{\prime}$ differs by a constant times a rotation.
A constant times a rotation cannot change cone angles. It is then straightforward to see 
that each step in our construction is independent of the choice of reference diffeomorphisms. 

Third, the charge nets
$\{c^{(m)}_0(i, v_{I_{v_0},\delta_0}^{\prime})\equiv c^{}_0(i, v_{I_{v_0},\delta_0}^{\prime}), \;i=1,2,3\}$
obtained at the end of the procedure are also flipped $I_{v_0}$ siblings of each other.
This follows from the following easily verifiable facts: \\
\noindent (a) The deformed chargenets $\{c^{(0)}_0(i, v_{I_{v_0},\delta_0}^{\prime}), \;i=1,2,3\}$ of P1 are flipped 
$I_{v_0}$ siblings 
\\
\noindent
(b) As in (a), section 4.4, the restriction on the reference classes for siblings 
and the properties of siblings, imply that given a triplet of flipped siblings and their
reference chargenets,  any reference diffeomorphism for the $i$th sibling is also 
an acceptable reference diffeomorphism for the $j$th sibling. 
\\
\noindent
(c)  As observed above, cone angles are independent
of the choice of reference diffeomorphisms. 
\\
\noindent
(d)Each step  in our iterative procedure 
only depends on the graph structure in a very small vicinity
of its vertex $v_{I_{v_0},\delta_0}^{\prime}$ and not on its coloring. 

\subsection{Construction of $\phi_{\delta}$}

As mentioned above,
we modify only that part of $\phi_{\delta}$ in P1 which scrunches together the edges
at $v_{I_{v},\delta}^{\prime}$ i.e only the second equation of 5.,Appendix C,P1 is modified below.
As we shall see, the modification involves structures associated with the choice of reference diffeomorphism
sets ${\cal D}_{[c_0(i, v_{I_{v_0},\delta_0}^{\prime})]}$. As a result, in contrast to section 4 (see especially 
Footnote \ref{footnoteidep}, section 4.2.1 as well as property (b),section 4.4) 
we shall be careful about the $i$- dependence
of the choice of coordinates associated with the nontrivial vertices of elements of 
$[c_0(i, v_{I_{v_0},\delta_0}^{\prime})]$.

More in detail, 
the action of the finite triangulation  Hamiltonian constraint at parameter $\delta =\delta_0$  on the
reference  chargenet $c_0$
yields the $(i, I_{v_0})$- flipped siblings  $c_0(i, v_{I_{v_0},\delta_0}^{\prime})$ 
(see sections 2.2 and 5.2).
It is not necessary that each $c_0(i, v_{I_{v_0},\delta_0}^{\prime})$ be a reference charge net.
Let ${\bar c}_{(i)0}$ be the reference chargenet diffeomorphic to 
 $c_0(i, v_{I_{v_0},\delta_0}^{\prime})$. 
From section 2.2 the position of the nontrivial vertex and the reference coordinate chart is the same 
for ${\bar c}_{(i)0}, i=1,2,3$. Let the position of the nontrivial vertex be
${\bar v}$ and the reference coordinate patch be
$\{\bar x\}$. Let ${\bar \gamma}_{(i)}\in {\cal D}_{ [  {\bar c}_{(i)0}  ]   }$ map
${\bar c}_{(i)0}$ to 
$c_0(i, v_{I_{v_0},\delta_0}^{\prime})$. Diffeomorphism covariance requires that
the coordinate patch $\{x^{\prime}_{(i)\delta_0}\}$ at  $v_{I_{v_0},\delta_0}^{\prime}$ be chosen as:
\begin{equation}
\{x^{\prime}_{(i)\delta_0}\} := {\bar \gamma}_{(i)}^*\{{\bar x}\}.
\label{xprimexbar}
\end{equation}
Having made this choice, 
we define $\phi_{\delta}$ as follows.

Let the deformation be such that 
the position of  $v_{I_{v},\delta}^{\prime}, \delta\leq \delta_0$
is as in P1. Rotate the coordinates $\{x_0\}$ as in section 4.2.1. Since the modification of section 5.2 leaves the 
$I_{v_0}$th deformed edge untouched relative to P1, we have that as in P1,
equations (\ref{ei=etildei}), (\ref{eicx3})- (\ref{tildeex3}) hold at parameter value $\delta= \delta_0$.
As in P1 we first translate $v_{I_{v_0},\delta_0}^{\prime}$ to $v_{I_{v_0},\delta}^{\prime}$ rigidly
along the straight line joining them in the $\{x_0\}$ coordinates. Let this translation be $T$
so that 
\begin{equation}
x^{\mu}_0 (T( p))=  x^{\mu}_0 (p) - a^{\mu}_{\delta}
\label{defT}
\end{equation}
where ${\vec a}_{\delta}$ is the constant coordinate vector:
\begin{equation}
a^{\mu}_{\delta}:= x^{\mu}_0 (v_{I_{v_0},\delta_0}^{\prime}) - x^{\mu}_0 (v_{I_{v_0},\delta }^{\prime}) .
\end{equation}

Next similar to the Appendix of P1, we scrunch the vectors at $v_{I_{v_0},\delta }^{\prime}$ together
by a diffeomorphism which is identity outside a small enough neighbourhood of $v_{I_{v_0},\delta }^{\prime}$
and which, within a small enough neighbourhood inside this neighbourhood, reduces to a `scrunching'
linear transformation. This transformation will be defined differently for each $i$, so that in contrast to a
single $\phi_{\delta}$ we have three different sets of diffeomorphisms,$\phi_{(i),\delta},\; i=1,2,3$, one for each
$c_0(i, v_{I_{v_0},\delta_0}^{\prime})$. Accordingly,
let us denote the `scrunching' diffeomorphism by 
$S_{(i)}$ and the associated linear transformation by $S_{(i)L}$. 
Then, for any point ${p}$ 
in a small enough  neighbourhood of  $v_{I_{v_0},\delta }^{\prime}$ we have that:
\begin{equation}
x^{\mu}_0 (S_{(i )}(p))=  S_{(i)L\;\nu}^{\mu}(x^{\nu}_0 (p) - x^{\nu}_0(v_{I_{v_0},\delta }^{\prime}))
+x^{\mu}_0(v_{I_{v_0},\delta }^{\prime})
\label{defS}
\end{equation}
where we have chosen $S_{(i)L}$ to be given by:
\begin{equation}
S_{iL\;\nu}^{\mu} = G^{\mu}_{\;\tau^{\prime}} (H_{(i)I_{v=v_0}}^{-1})^{\tau^{\prime}}_{\nu}
\label{defSL}
\end{equation}
where the matrix $H_{(i)I_{v_0}}$ has been defined, similar to equation (\ref{defH}), to be 
\begin{equation}
H_{(i)I_{v_0}\nu^{\prime}}^{\lambda}= \frac{\partial x_0^{ \lambda } }
     {\partial x_{(i)\delta_0}^{{\prime}\nu^{\prime}}}|_{v_{I_{v_0},\delta_0}^{\prime}},
\label{defHi}
\end{equation}
and
the matrix $G^{\mu}_{\;\tau^{\prime}}$ is the same as the 
matrix $G$ of P1:
\begin{eqnarray}
G^{\mu}_{\;\tau^{\prime}} &=& 0 \;{\rm iff}\; \mu \neq \tau^{\prime}\nonumber\\
&=& 1 \;{\rm iff}\;\mu=\tau^{\prime}= 3\nonumber \\
&=& \delta^{q-1}\;{\rm iff}\;\mu=\tau^{\prime}= 1,2
\label{defGdelta}
\end{eqnarray}
We define $\phi_{(i)\delta}$ to be $S_{(i)}\circ T$ so that 
\begin{equation}
x^{\mu}_0 (\phi_{(i)\delta}(p))=  S_{(i)L\;\nu}^{\mu}(x^{\nu}_0 (p) -x^{\nu}_0 (v_{I_{v_0},\delta_0}^{\prime})  )
+x^{\mu}_0(v_{I_{v_0},\delta }^{\prime})
\label{phi0st}
\end{equation}

\subsection{Deformations produced by ${\hat D}_{\delta}({\vec N}_i)$}
The action of  ${\hat D}_{\delta_0}({\vec N}_i)$ on the reference chargenet $c_0$ produces the deformed
chargenets $c_0(v_{I_{v_0},\delta_0}^{\prime})$. 
We define the `conically' deformed chargenet 
$c_0(v_{I_{v_0},\delta_0}^{\prime})$ to be the (unique) unflipped sibling of 
$c_0(i, v_{I_{v_0},\delta_0}^{\prime})$ where the latter are the flipped $I_{v_0}$ siblings 
defined at the end of section 5.2.\\

Next we turn to the definition of the deformations of the reference chargenet $c_0$ at any $\delta$ as
diffeomorphic images of the deformations at $\delta =\delta_0$.
Note that 
it is not necessary that $c_0(v_{I_{v_0},\delta_0}^{\prime})$ be a reference charge net.
Let ${\bar c}_{   ({}_{\hat{\;}})    0}$ be the reference chargenet diffeomorphic to 
 $c_0( v_{I_{v_0},\delta_0}^{\prime})$. 
Since $c_0( v_{I_{v_0},\delta_0}^{\prime})$ is the unflipped sibling of $c_0( i, v_{I_{v_0},\delta_0}^{\prime})$
it follows from
section 2.2 that the position of the nontrivial vertex of ${\bar c}_{({}_{\hat{\;}})     0}$ as well as 
the reference coordinate chart are the same as 
for ${\bar c}_{i0}, i=1,2,3$ (see section 5.3). Hence the position of the nontrivial vertex, 
${\bar v}$, and the reference coordinate patch 
$\{\bar x\}$ are the same as in section 5.3. Let ${\bar \gamma}_{({}_{\hat{\;}})}\in 
{\cal D}_{[{\bar c}_{({}_{\hat{\;}})  0}]}$ map
${\bar c}_{({}_{\hat{\;}})0}$ to 
$c_0(v_{I_{v_0},\delta_0}^{\prime})$. Diffeomorphism covariance requires that
the coordinate patch $\{x^{\prime}_{({}_{\hat{\;}})\delta_0}\}$ at  $v_{I_{v_0},\delta_0}^{\prime}$ be chosen as:
\begin{equation}
\{x^{\prime}_{({}_{\hat{\;}})\delta_0}\} := {\bar \gamma}_{({}_{\hat{\;}})}^*\{{\bar x}\}.
\label{xhatxbar}
\end{equation}

We proceed exactly as in section 5.2 except that we replace $S_{(i)}, S_{(i)L}$ there 
$S_{({}_{\hat{\;}})}, S_{({}_{\hat{\;}})L}$ where 
 where we have chosen $S_{({}_{\hat{\;}})L}$ to be given by:
\begin{equation}
S_{({}_{\hat{\;}})L\;\nu}^{\mu} = G^{\mu}_{\;\tau^{\prime}} (H_{({}_{\hat{\;}})I_{v=v_0}}^{-1})^{\tau^{\prime}}_{\nu}
\label{defShatL}
\end{equation}
where the matrix $H_{({}_{\hat{\;}})I_{v}}$ is defined 
to be 
\begin{equation}
H_{({}_{\hat{\;}})I_v\nu^{\prime}}^{\lambda}= \det\frac{\partial x_0^{ \lambda } }
     {\partial x_{({}_{\hat{\;}})\delta_0}^{{\prime}\nu^{\prime}}}|_{v_{I_{v},\delta_0}^{\prime}},
\label{defHhat}
\end{equation}
so that 
for any point ${p}$ 
in a small enough  neighbourhood of  $v_{I_{v_0},\delta }^{\prime}$ we have that:
\begin{equation}
x^{\mu}_0 (S_{({}_{\hat{\;}})}(p))=  S_{({}_{\hat{\;}})L\;\nu}^{\mu}(x^{\nu}_0 (p)- 
x^{\nu}_0(v_{I_{v_0},\delta }^{\prime})).
\label{defShat}
\end{equation}
Analogous to section 5.2 we define $\phi_{({}_{\hat{\;}})\delta}$ to be $S_{({}_{\hat{\;}})}\circ T$ so that 
\begin{equation}
x^{\mu}_0 (\phi_{({}_{\hat{\;}})\delta}(p))=  S_{({}_{\hat{\;}})L\;\nu}^{\mu}
(x^{\nu}_0 (p) -x^{\nu}_0(v_{I_{v_0},\delta_0 }^{\prime})  )
+x^{\mu}_0(v_{I_{v_0},\delta }^{\prime})
\label{phi0hatst}
\end{equation}

\subsection{Behaviour of coordinate structures as $\delta\rightarrow 0$.}
As seen in section 4.3 (and as we shall see again in section 7) the 
coordinate structures which play a key role in the evaluation of the commutator are the edge tangents
and the Jacobians at the deformed vertex $v_{I_{v},\delta}^{\prime}$ of the charge nets
$c(i,v_{I_{v},\delta}^{\prime}), c(v_{I_{v},\delta}^{\prime})$, the Jacobians being  between the 
coordinates associated with $c$ and those associated 
with $c(i,v_{I_{v},\delta}^{\prime}), c(v_{I_{v},\delta}^{\prime})$.
In section 5.5.1, We evaluate these Jacobians and analyse the 
behaviour of the edge tangents as $\delta\rightarrow 0$ for 
the case when $c$ is a reference charge net.
In section 5.5.2 we analyse the case when $c$ is not a reference chargenet.

\subsubsection{$c=c_0$}
For notational convenience we suppress the subscripts $ ({}_{\hat{\;}})$ of section 5.4 and $ (i)$ of sections 5.2 and
5.3. The considerations below are valid for the cases corresponding to each of these subscripts in turn.
For example for the case of the deformations generated by the Hamiltonian constraint,
the considerations below go through for the structures $\phi_{(i)\delta}, \{ x^{\prime}_{(i)\delta_0} \}$ for each fixed
$i=1,2,3$ and, accordingly, the subscript $(i)$ can be added to the appropriate quantities below.
Note also that the only difference in the deformation structure in the vicinity of the displaced
vertex $v_{I_{v},\delta}^{\prime}$ for the deformed charge nets $c(i,v_{I_{v},\delta}^{\prime}),
c(v_{I_{v},\delta}^{\prime})$ is in the choice of reference diffeomorphisms $\gamma_{(i)},\gamma_{({}_{\hat{\;}})}$
of equations  (\ref{xprimexbar}), (\ref{xhatxbar}), which, in turn, stems from the choice of the sets 
of reference diffeomorphisms ${\cal D}_{[{\bar c}_{(i)0}]}, {\cal D}_{[{\bar c}_{({}_{\hat{\;}})    0 }]}$. 
As we shall see in section 7, our final results in the 
continuum limit are independent of the choice of reference diffeomorphisms.

First consider the Jacobian 
$\frac{\partial x_0^{\lambda}}{\partial x_{\delta}^{\prime\nu^{\prime}}}|_{v_{I_{v_0},\delta}^{\prime}}$
where we have defined 
\begin{equation}
\{x_{\delta}^{\prime}\}:= \phi_{\delta}^* \{x_{\delta_0}^{\prime}\}
\label{5.16}
\end{equation}
Similar to section 4.2.2 we have that 
\begin{eqnarray}
\frac{\partial x_0^{\lambda}}{\partial x_{\delta}^{\prime\nu^{\prime}}}|_{v_{I_{v_0},\delta}^{\prime}}
&=&\frac{\partial x_0^{\mu}}{\partial (\phi_{\delta}^{*}x_0)^{\lambda}}|_{v_{I_{v_0},\delta}^{\prime}}
\frac{\partial x_0^{\lambda}}{\partial x_{\delta_0}^{\prime\nu^{\prime}}}|_{v_{I_{v_0},\delta_0}^{\prime}}\nonumber \\
&=& 
S^{\mu}_{\lambda} H^{\lambda}_{I_{v_0}\; \nu^{\prime}}\nonumber\\
&=&
G^{\mu}_{\;\tau^{\prime}} (H_{I_{v_0}}^{-1})^{\tau^{\prime}}_{\lambda}H^{\lambda}_{I_{v_0}\; \nu^{\prime}}\nonumber\\
&=& 
G^{\mu}_{\;\nu^{\prime}}
\label{jacobianref}
\end{eqnarray}
where we have used equations (\ref{phi0st}),(\ref{defS})- (\ref{defHi}), (\ref{phi0st}) and (\ref{defShatL}), 
(\ref{defHhat}),
(\ref{phi0hatst}). Note that equation (\ref{jacobianref}) implies that the  determinant of the  Jacobian, 
$\det\frac{\partial x_0^{\lambda}}{\partial x_{\delta}^{\prime\nu^{\prime}}}|_{v_{I_{v_0},\delta}^{\prime}}$ 
vanishes as $\delta\rightarrow 0$ by virtue of the behaviour of $G^{\mu}_{\;\nu^{\prime}}$ in (\ref{defGdelta}).
This is similar to what happens in equation (\ref{detjacobian}) in section 4.2.3 except that 
we no longer have  a dependence on the matrix $H_{I_{v_0}}^{-1})^{\tau^{\prime}}_{\lambda}$ by virtue
of the modification of the scrunching detailed in sections 5.3 and 5.4.

Next, similar to section 4.2.2, let $V(p)$ be a unit vector at 
$p=v_{I_{v_0},\delta_0}^{\prime}$ in the $\{x^{\prime}_{\delta_0} \}$ coordinates. 
so that $(\phi^*_{\delta}V)(p)=: V_{\delta}(p)$ at $p=v_{I_{v_0},\delta}^{\prime}$
is a unit vector in the $\{x^{\prime}_{\delta } \}$ coordinates.
The components $V^{\mu}_{\delta}(p)$ of $V_{\delta}(p)$  in the $\{x_0\}$ coordinates are related to
its components $V^{\mu^{\prime}}_{\delta}$ in the $\{x^{\prime}_{\delta} \}$ coordinates
at 
$p= v_{I_{v_0},\delta}^{\prime}$ through:
\begin{equation}
V^{\mu}_{\delta} (v_{I_{v_0},\delta}^{\prime} ) = 
V^{\nu^{\prime}}_{\delta}(v_{I_{v_0},\delta}^{\prime})
\frac{\partial x_0^{\mu}}{\partial x_{\delta}^{\prime\nu^{\prime}}}|_{v_{I_{v_0},\delta}^{\prime}} 
= V^{\nu^{\prime}}(v_{I_{v},\delta_0}^{\prime})
\frac{\partial x_0^{\mu}}{\partial x_{\delta}^{\prime\nu^{\prime}}}|_{v_{I_{v_0},\delta}^{\prime}}
\label{5.18}
\end{equation} 
where we have used, similar to (\ref{4.7}), that the components $V^{\nu^{\prime}}_{\delta}(v_{I_{v_0},\delta}^{\prime})$
in the $\{x^{\prime}_{\delta} \}$ are the same as the components
$V^{\nu^{\prime}}(v_{I_{v_0},\delta_0}^{\prime})$ in $\{x^{\prime}_{\delta_0} \}$ system by virtue of (\ref{5.16}).
From (\ref{jacobianref}) it follows that:
\begin{equation}
V^{\mu}_{\delta} (v_{I_{v_0},\delta}^{\prime} )
= V^{\nu^{\prime}}(v_{I_{v_0},\delta_0}^{\prime})G^{\mu}_{\;\nu^{\prime}}
\label{5.1}
\end{equation}

Equation (\ref{defGdelta}) then implies that:
\begin{eqnarray}
V^{\mu =3 }_{\delta} (v_{I_{v_0},\delta}^{\prime} ) &= & V^{\nu^{\prime}=3}(v_{I_{v_0},\delta_0}^{\prime})
\label{V3}
\\
V^{\mu \neq 3}_{\delta} (v_{I_{v_0},\delta}^{\prime} ) &= & O(\delta^{q-1} )
\label{V12}
\end{eqnarray}
Next, recall that $V$ is a {\em unit} vector at $v_{I_{v_0},\delta_0}^{\prime}$ in the 
$\{x^{\prime}_{\delta_0}\}$ coordinates. Hence we may set 
\begin{equation}
V^{\mu^{\prime}=3}( v_{I_{v_0},\delta_0}^{\prime} )= \cos \theta ,
\;\;\;\;  (V^{\mu^{\prime}=2}(v_{I_{v_0},\delta_0})^{\prime})^2 + (V^{\mu^{\prime}=3}(v_{I_{v_0},\delta_0}^{\prime}))^2 
=\sin^2 \theta
\label{deftheta}
\end{equation}
Using this in equations (\ref{V3}), (\ref{V12}) we have that:
\begin{eqnarray}
V_{\delta}^{\mu =3} (v_{I_{v_0},\delta}^{\prime} )& =&
\cos \theta 
\label{4.16},
\\
V_{\delta}^{\mu \neq 3} (v_{I_{v_0},\delta}^{\prime} ) &=& O(\delta^{q-1} ).
\label{4.17}
\end{eqnarray}
Using (\ref{tildeex3}), we may write  equations (\ref{4.16}) and (\ref{4.17}) as
\begin{equation}
V_{\delta}^{a} (v_{I_{v_0},\delta}^{\prime} ) =
\cos \theta {\hat {\tilde e}}_{I_{v_0}}^a|_{v_{I_{v_0},\delta}^{\prime}} +O(\delta^{q-1} )
\label{4.171},
\end{equation}
Finally, set
$V^a|_{v_{I_{v_0},\delta_0}^{\prime}} =:{\hat {\tilde e}}_{J_{v_0}}^{\prime a}|_{v_{I_{v_0},\delta_0}^{\prime}}$.
It follows  from (\ref{4.171})
that for $J_{v_0}\neq I_{v_0} $ we have that
\begin{eqnarray}
{\hat {\tilde e}}_{J_{v_0}}^{\prime a}|_{v_{I_{v_0},\delta }^{\prime}}
& = & -\cos\theta_{J_{v_0}}{\hat {\tilde e}}_{I_{v_0}}^{a}|_{v_{I_{v_0},\delta }^{\prime}} 
+ O(\delta^{q-1}),\label{Jtngntdelta}\\
\Rightarrow
\lim_{\delta\rightarrow 0}
{\hat {\tilde e}}_{J_{v_0}}^{\prime a}|_{v_{I_{v_0},\delta }^{\prime}}
&=& - \cos\theta_{J_{v_0}}{\hat { e}}_{I_{v_0}}^{a}|_{v_0},
\label{contJtngnt}
\end{eqnarray}
and that for $J_{v_0}= I_{v_0}$ that 
\begin{eqnarray}
{\hat {\tilde e}}_{I_{v_0}}^{\prime a}|_{v_{I_{v_0},\delta }^{\prime}}
= {\hat {\tilde e}}_{I_{v_0}}^{a}|_{v_{I_{v_0},\delta }^{\prime}},\label{Itngntdelta}\\
\Rightarrow \lim_{\delta \rightarrow 0}
{\hat {\tilde e}}_{I_{v_0}}^{\prime a}|_{v_{I_{v_0},\delta }^{\prime}}
=  {\hat {e}}_{I_{v_0}}^{a}|_{v_0},
\label{contItngnt}
\end{eqnarray}
where $\theta_{J_{v_0}}$ is the angle between the negative $x^{\prime 3}_{\delta_0}$ direction and the edge
tangent to ${\tilde e}_{J_{v_0}}$ at the vertex $v_{I_{v_0},\delta_0 }^{\prime}$ in 
$c_0(i,v_{I_{v_0},\delta_0}^{\prime})$
 and we have used (\ref{ei=etildei}) in obtaining (\ref{contJtngnt}), (\ref{contItngnt}) from (\ref{Jtngntdelta}),(\ref{Itngntdelta}). 
\\

Upto this point we have retained the edge dependence of the angular variable $\theta_{J_{v_0}}$ in order to display
what our results would have been had the deformation of the reference chargenetworks at $\delta =\delta_0$ not been 
conical.
Now we recall from sections 5.2 and 5.4 that the deformations are conical so that 
$\theta_{ J_{v_0}\neq I_{v_0} }=:\theta$ independent of $J_{v_0}$.  
Restoring the subscripts $(i),({}_{\hat{\;}})$, 
we have that the 
corresponding conical deformations are characterised by the angles $\theta_{(i)}, \theta_{({}_{\hat{\;}})}$.
Next, recall the following:\\
(i) The charge nets $c_0(i,v_{I_{v_0},\delta_0)}^{\prime})$,
$c_0(v_{I_{v_0},\delta_0)}^{\prime})$ are siblings (see the discussion at the beginning section 5.4).\\
(ii) The sibling restrictions ensure that the coordinates associated with each such sibling are 
images of a single reference coordinate patch (see equations (\ref{xprimexbar}), (\ref{xhatxbar})).\\
(iii) The results of section 3 imply that the Jacobians between the pair of coordinate patches
associated with any pair of siblings when evaluated at the deformed vertx $v_{I_{v_0},\delta_0}^{\prime}$
are proportional to rotations about an axis in the direction of the $I_{v_0}$th edge tangent at 
$v_{I_{v_0},\delta_0}^{\prime}$.

Clearly point (iii) implies that $\theta_{(1)}= \theta_{(2)}= \theta_{(3)}=\theta_{({}_{\hat{\;}})}:= \theta$.

\subsubsection{$c \in [c_0], c\neq c_0$}

Fix the reference structures, including the sets of reference diffeomorphisms, as in sections 2.2 and 5.1 and 
let the deformation of the reference chargenets and the definition of $\phi_{\delta}$ be as in sections 5.2 and 5.3.
Let $\beta \in {\cal D}_{[c_0]}$ be such that $c$ is the image of the reference chargenet $c_0$ by $\beta$.
We write this as 
\begin{equation}
c =\beta \;c_0. 
\label{c=betac0}
\end{equation}
By definition we have that 
\begin{equation}
c(i, v_{I_{v},\delta}^{\prime})= \beta \;c_0 (i, v_{I_{v},\delta}^{\prime})
:= (\beta\circ\phi_{\delta})\; c_0 (i, v_{I_{v},\delta_0}^{\prime}) 
\label{c=betaphic0}
\end{equation}

From section 5.2 we have that ${\bar \gamma}_i {\bar c}_{(i)0} = c_0(i, v_{I_{v_0},\delta_0}^{\prime})$ which, together
with (\ref{c=betaphic0}) implies that:
\begin{equation}
c(i, v_{I_{v},\delta}^{\prime})= 
(\beta\circ\phi_{\delta}\circ {\bar \gamma}_i)\; {\bar c}_{(i)0}
\label{betaphigamma}
\end{equation}
Accordingly, we define the coordinate patch $\{x^{\prime}_{(i),\beta,\delta}\}$ around 
 $v_{I_{v},\delta}^{\prime}$ in $c(i, v_{I_{v},\delta}^{\prime})$ as the image of reference coordinate
patch $\{ {\bar x} \}$ by this diffeomorphism so that:
\begin{equation}
\{x^{\prime}_{(i),\beta,\delta}\}
=(\beta\circ\phi_{\delta}\circ {\bar \gamma}_i)^*\{ {\bar x}\}
= (\beta\circ\phi_{\delta})^*\{x^{\prime}_{(i),\delta_0}\}= \beta^*\{x^{\prime}_{(i),\delta}\}.
\label{7.8}
\end{equation}
Here, we used equation (\ref{xprimexbar}) in the second last equality.
As suggested by the commutator calculation of section 4, the relevant Jacobian is 
$\frac{\partial x^{\mu}_{\beta}\;\;\; }{  \partial x^{\prime\nu^{\prime}}_{(i),\beta,\delta}}$
where, from equation (\ref{c=betac0}),  we have that 
\begin{equation}
\{x_{\beta}\}:= \beta^*\{x_0\}
\label{7.9}
\end{equation}
 are the coordinates associated with $c$
(recall from section 5.3 that $\{x_0\}$ is associated with $c_0$). Using equations (\ref{7.8}), (\ref{7.9}), we
have that:
\begin{eqnarray}
\frac{\partial x^{\mu}_{\beta}\;\;\;\;\;\;\;\; }{  \partial x^{\prime\nu^{\prime}}_{(i),\beta,\delta}}
|_{v_{I_{v},\delta}^{\prime}}&=&
\frac{\partial (\beta^* x_0)^{\mu}\;\;\; }
{  \partial ((\beta\circ\phi_{\delta}\;)^*x_{(i),\delta_0})^{\prime\nu^{\prime}}}
|_{v_{I_{v},\delta}^{\prime}}\\
&=&
\frac{\partial x_0^{\mu}\;\;\;\;\;\; }
{  \partial (\phi_{\delta}^*x_{(i),\delta_0})^{\prime\nu^{\prime}}}
|_{v_{I_{v_0},\delta}^{\prime}}\\
&=&
\frac{\partial x_0^{\mu}\;\;\;\;\; }{\partial(\phi_{\delta})^*x_0)^{\lambda} }|_{v_{I_{v_0},\delta}^{\prime}}
\frac{\partial(\phi_{\delta}^*x_0)^{\lambda} }
{  \partial (\phi_{\delta}^*x_{(i),\delta_0})^{\prime\nu^{\prime}}}
|_{v_{I_{v_0},\delta}^{\prime}}\\
&=&
\frac{\partial x_0^{\mu}\;\;\; \;\;}{\partial (\phi_{\delta}^*x_0)^{\lambda} }|_{v_{I_{v_0},\delta}^{\prime}}
\frac{\partial  x_0^{\lambda}\;\;\; }
{  \partial x^{\prime\nu^{\prime}}_{(i),\delta_0}}
|_{v_{I_{v_0},\delta_0}^{\prime}}\\
&=&
S^{\mu}_{(i)\lambda} H^{\lambda}_{(i)I_{v_0} \nu^{\prime}} = G^{\mu}_{\;\;\nu^{\prime}},
\label{7.14}
\end{eqnarray}
where we have used (\ref{phi0st}), (\ref{defS}) and (\ref{defH}) in the last equality.

Exactly the same analysis can be applied to $c(v_{I_{v},\delta}^{\prime})$. Using obvious notation, it is easy
to verify that such an analysis  
replaces $c(i, v_{I_{v},\delta}^{\prime})$ by $c(v_{I_{v},\delta}^{\prime})$ and the subscript $i$ by the 
subscript ${}_{\hat{\;}}$ so that we obtain
\begin{equation}
\frac{\partial x^{\mu}_{\beta}\;\;\;\;\;\;\;\; }{  \partial x^{\prime\nu^{\prime}}_{({}_{\hat{\;}}),\beta,\delta}}
|_{v_{I_{v},\delta}^{\prime}}
=G^{\mu}_{\;\;\nu^{\prime}}
\label{7.15}
\end{equation}
where 
\begin{equation}
\{ x^{\prime}_{({}_{\hat{\;}}) ,\beta,\delta}\} = (\beta\circ \phi_{\delta}\circ {\bar{\gamma}_{({}_{\hat{\;}})}})^*\{{\bar x}\}
=(\beta\circ \phi_{\delta})^*\{ x^{\prime}_{({}_{\hat{\;}}) \delta_0}\}
=\beta^*\{ x^{\prime}_{({}_{\hat{\;}}) \delta }\}
\label{7.8d}
\end{equation}
where we have used (\ref{xhatxbar}) in the second last equality.

The behaviour of the edge tangents at $v_{I_{v},\delta}^{\prime}$ may be ascertained by pushing forward
equations (\ref{Jtngntdelta}) - (\ref{contItngnt}) by $\beta$. More in detail, it follows from 
equation (\ref{7.9}) that 
\begin{eqnarray}
\beta^*({\hat {\tilde e}}_{I_{v_0}}^{a}|_{v_{I_{v_0},\delta }^{\prime}})&=&
{\hat {\tilde e}}_{I_{v}}^{a}|_{v_{I_{v},\delta }^{\prime}}
\label{pushhattilde} \\
\beta^*({\hat {e}}_{I_{v_0}}^{a}|_{v_0}) &=& {\hat {e}}_{I_{v}}^{a}|_{v}
\label{pushhat}
\end{eqnarray}
where ${\hat {\tilde e}}_{I_{v_0}}^{a}, {\hat {e}}_{I_{v_0}}^{a}$ are normalized with repect to the $\{x_0\}$
coordinates and 
${\hat {\tilde e}}_{I_{v}}^{a}, {\hat {e}}_{I_{v}}^{a}$ are normalized with respect to the $\{x_{\beta}\}$ coordinates.

Suppressing the $\beta, (i), ({}_{\hat{\;}})$ subscripts,
it follows from equations (\ref{7.8}) and (\ref{7.8d}) that 
\begin{equation}
\beta^*({\hat {\tilde e}}_{J_{v_0}}^{\prime a}|_{v_{I_{v_0},\delta }^{\prime}})=
{\hat {\tilde e}}_{J_{v}}^{\prime a}|_{v_{I_{v},\delta }^{\prime}}
\label{pushhattildeprime}
\end{equation}
where ${\hat {\tilde e}}_{J_{v_0}}^{\prime a}$ is normalized with respect to the 
$\{ x^{\prime}_{\delta}\}$ coordinates and
${\hat {\tilde e}}_{J_{v}}^{\prime a}$ is normalized with respect to the $\beta^*\{ x^{\prime}_{\delta}\}$
coordinates.

Using equations (\ref{pushhattilde}), (\ref{pushhat}), (\ref{pushhattildeprime}) to evaluate the 
pushforward, by $\beta$,
of equations (\ref{Jtngntdelta})- (\ref{contItngnt}), we obtain
for $J_{v}\neq I_{v} $,
\begin{eqnarray}
{\hat {\tilde e}}_{J_v}^{\prime a}|_{v_{I_{v},\delta }^{\prime}}
& = & - \cos\theta_{J_{v_0}}{\hat {\tilde e}}_{I_v}^{a}|_{v_{I_{v},\delta }^{\prime}} 
+ O(\delta^{q-1}),\label{Jtngntdeltac}\\
\Rightarrow
\lim_{\delta\rightarrow 0}
{\hat {\tilde e}}_{J_v}^{\prime a}|_{v_{I_{v},\delta }^{\prime}}
&=& - \cos\theta_{J_{v_0}}{\hat { e}}_{I_v}^{a}|_{v},
\label{contJtngntc}
\end{eqnarray}
and  for $J_v= I_v$, 
\begin{eqnarray}
{\hat {\tilde e}}_{I_v}^{\prime a}|_{v_{I_{v},\delta }^{\prime}}
=  {\hat {\tilde e}}_{I_v}^{a}|_{v_{I_{v},\delta }^{\prime}},\label{Itngntdeltac}\\
\Rightarrow \lim_{\delta \rightarrow 0}
{\hat {\tilde e}}_{I_v}^{\prime a}|_{v_{I_{v},\delta }^{\prime}}
=  {\hat {e}}_{I_v}^{a}|_{v}.
\label{contItngntc}
\end{eqnarray}
The vectors on the left hand sides of these equations are normalized with respect to the 
$\beta^*\{ x^{\prime}_{\delta}\}$     coordinates
and the ones on the right hand side, with  respect to the $\{x_{\beta}\}= \beta^*\{x_0\}$ coordinates.
In evaluating the $\delta \rightarrow 0$ limits in (\ref{contJtngntc}), (\ref{contItngntc}) we have 
used that $\beta$ is independent of $\delta$. 

The remarks at the end of section 5.5.1 apply so that  $\theta_{J_{v_0}}$ is independent of $J_{v_0}$ as well
as the (hitherto suppressed) subscripts $(i), ({}_{\hat{\;}})$ and (restoring the suppressed subscripts)
we have $\theta_{(i)J_{v_0}}=\theta_{({}_{\hat{\;}})J_{v_0}}=: \theta$.

\noindent{\em Note:} In the above analysis we fixed the value of $I_{v_0}$ and the deformations of $c_0$ were along the 
the $I_{v_0}$th edge of $c_0$. In section 7, we will have to combine the contributions of deformations
along all the different edges emanating from $v_0$ in $c_0$ (or, more generally, from $v$ in $c \in [c_0]$).
Therefore it is appropriate to set $\theta \equiv \theta_{I_{v_0}}$ so as to remind us that the cone angles
for deformations along different edges are, in general, different. Using this notation the continuum limit behaviour
of the edge tangents of edges deformed along the $I_v$th edge of the charge network being acted upon is given by the
following rewriting of equations (\ref{Jtngntdeltac}),(\ref{contJtngntc}),(\ref{Itngntdeltac}) and (\ref{contItngntc}):
\\
For $J_{v}\neq I_{v} $
we have that
\begin{eqnarray}
{\hat {\tilde e}}_{J_v}^{\prime a}|_{v_{I_{v},\delta }^{\prime}}
& = & - \cos\theta_{I_{v_0}}{\hat {\tilde e}}_{I_v}^{a}|_{v_{I_{v},\delta }^{\prime}} 
+ O(\delta^{q-1}),\label{IJtngntdeltac}\\
\Rightarrow
\lim_{\delta\rightarrow 0}
{\hat {\tilde e}}_{J_v}^{\prime a}|_{v_{I_{v},\delta }^{\prime}}
&=& - \cos\theta_{I_{v_0}}{\hat { e}}_{I_v}^{a}|_{v},
\label{IcontJtngntc}
\end{eqnarray}
and  for $J_v= I_v$ that 
\begin{eqnarray}
{\hat {\tilde e}}_{I_v}^{\prime a}|_{v_{I_{v},\delta }^{\prime}}
=  {\hat {\tilde e}}_{I_v}^{a}|_{v_{I_{v},\delta }^{\prime}},\label{IItngntdeltac}\\
\Rightarrow \lim_{\delta \rightarrow 0}
{\hat {\tilde e}}_{I_v}^{\prime a}|_{v_{I_{v},\delta }^{\prime}}
=  {\hat {e}}_{I_v}^{a}|_{v} .
\label{IcontItngntc}
\end{eqnarray}

\section{Modified VSA topology}
In this section we detail 
the modification of the VSA topology (relative to P1) with respect to which the continuum limit is defined. 
The bra set $B_{VSA}$ as well as the vertex smooth functions are modified.
While the modification of $B_{VSA}$ is minor, the new vertex functions are qualitatively different from
those of P1 in that their dependence extends to 
additional vertices of the bras in $B_{VSA}$.
These extended vertex functions are restricted to
have a precise short distance behaviour when certain vertices in their arguments 
approach coincidence. As we shall see in section 7, 
it is this short distance behaviour which cures the divergence of section 4.3.
While it is not hard to find a choice of additional vertices together with appropriate short distance
behaviour which cures this divergence, it is quite hard to find a such a choice for which the final
result is independent of the choice of reference diffeomorphisms.
Below, we display one such choice which involves a dependence on 
$M$ (where $M$ is the valence of the nontrivial 
vertex) 
additional vertices of the bras in $B_{VSA}$.

In section 6.1 we construct the modified $B_{VSA}$, in section 6.2 we define the new vertex functions
and in section 6.3 we detail their `short distance' behaviour.
\subsection{The modified bra set.}
The bra set is essentially the same as in P1. The only difference is that the deformations generated by the 
Hamiltonian constraint and the electric diffeomorphism constraint of P1 are replaced by the deformations
defined in  section 5. We provide a brief summary of the construction of the bra set with a change
in notation consistent with that used hitherto in this work.

Let $|c_{p}\rangle$ be a charge network with a single non-degenerate GR vertex
of valence $M$.  In the language of P1 $c_p$ is a primordial state. Let $|n,{\vec{\alpha}%
},c_{p}\rangle$ be the state obtained by $n$ successive finite-triangulation
Hamiltonian constraint-type  deformations 
applied to $|c_{p}\rangle$, these deformations having been defined in sections 5.2 and 5.3. Here,
$\vec{\alpha}:=\{\alpha_{i}~|~i=1,..n\},$ and each $\alpha_{i}$
is a
collection of labels corresponding to the internal charge, vertex, edge, and
finite triangulation  parameter value which go into the specification of the Hamiltonian
constraint-type deformations. Let the set of all distinct diffeomorphic images of $\langle n,{\vec{\alpha}}%
,c_{p}\vert$ be $B_{[n,\vec{\alpha},c_{p}]}$. Notwithstanding the differences in the deformations of section 5 
and P1, these two sets of  deformations share all the features relevant to defining the causal properties of the 
deformed chargenets in section 6.2, P1. Thus if $\langle c| \in B_{[n,\vec{\alpha},c_{p}]}$, the definitions of the
1- past $\gamma_{1-p}(c)$ of $\gamma (c)$, the future graph of $\gamma_{1-p}(c)$ in $c$, a future 
 graph of $\gamma_{1-p}(c)$ with respect to $c$ and a causal completion of the 1 past of $c$ go through
without modification. Note that the graph underlying a causal completion of the 1 past of $c$ need not
(and generically is not) obtained by deforming $\gamma_{1-p}(c)$ along the lines described in section 5.

Next, as in P1, we consider all possible causal completions of the 1 past of $c$ 
for every $c\in B_{[n,{\vec{\alpha}},c_{p}]}$ and call the resulting set of bras
as  $B_{\langle n,{\vec{\alpha}},c_{p}\rangle}$.
Consider all possible single
Hamiltonian constraint-type deformations (i.e. for all values of `$\alpha$')
of elements of $B_{\langle n,{\vec{\alpha}},c_{p}\rangle}$, where the deformations are as described in section 5,
 and take all
distinct diffeomorphic images of the resulting set of charge networks. Call
the resulting set $B_{[H\langle n,{\vec{\alpha}},c_{p}\rangle]}$. Repeat this
procedure again. That is, once again consider all Hamiltonian constraint-type
deformations of the elements of this set and then take distinct diffeomorphic
images of such deformed charge networks. Call this set $B_{[H[H\langle
n,{\vec{\alpha}},c_{p}\rangle]]}$.

Next, consider a pair of successive deformations of a charge net $\langle c| \in B_{<n,{\vec{\alpha}},c_{p}>}$, 
each deformation being of the type described in section 5.4.
As in P1, let $\beta$ be a
label which specifies the vertex at which the deformation takes place, the two
edge labels along which the deformations take place and the parameters
$\delta,\delta^{\prime}$ which quantify the amount of deformation. Denote the resulting state by 
$\hat{D}^{2}(\beta)|c\rangle$. Act by $\hat{D}^{2}(\beta)$ for all $\beta$ on elements of $B_{\langle
n,{\vec{\alpha}},c_{p}\rangle}$ and then take all distinct diffeomorphic
images thereof to form the set $B_{[D^{2}\langle n,{\vec{\alpha}},c_{0}%
\rangle]}$.
\footnote{More precisely, the sets $B_{<n,{\vec{\alpha}},c_{p}>}$,  $B_{[D^{2}\langle n,{\vec{\alpha}},c_{0}\rangle]}$
are sets of bras, so we mean to conctruct the latter out of diffeomorphic images of 
$\langle c |(\hat{D}^{2}(\beta))^{\dagger}$, $\langle c |\in B_{<n,{\vec{\alpha}},c_{p}>}$.}

Finally define $B_{\mathrm{VSA}}$ as:
\begin{equation}
B_{\mathrm{VSA}}:=B_{[H[H\langle n,{\vec{\alpha}},c_{0}\rangle]]}\cup
B_{[D^{2}\langle n,{\vec{\alpha}},c_{0}\rangle]} \label{defbvsa}%
\end{equation}

\subsection{Extended  vertex smooth functions.}
Let the set $B_{VSA}$ be constructed as in the previous section.
Upto semianalytic diffeomorphisms,
each bra in this set is obtained by successive deformations of its `parent' bra in 
$B_{\langle n,{\vec{\alpha}},c_{p}\rangle}$.
The successive deformations of such a parent bra are either both of the type generated
by the Hamiltonian constraint or both of the `singular diffeomorphism' type
generated by the diffeomorphism constraint smeared with an electric field dependent shift.
In both cases, each such deformation deforms the nontrivial parent  charge network
at its nontrivial vertex
(almost) along one of its edges in such a way that each deformed edge meets its undeformed counterpart
at a ($C^1$ or $C^0$) kink.
Thus in both cases, the last 2 deformations of the nontrivial parent charge net yield, upto $C^k$ semianlaytic
diffeomorphisms, 
a charge net in $B_{VSA}$ with a doubly deformed nontrivial vertex and an extra set of 2$M$ kinks.
From P1 and section 5 of this work, it follows that the following properties  hold with regard to 
these additional kink vertices 
The set of 2$M$ kinks can be divided into those kinks created by the first deformation and those created by the 
second. The smallest number of edges which connect each kink vertex of the former set 
to the final nontrivial vertex is exactly two and the two edges in question can be uniquely 
located in the doubly deformed graph (see the figures in P1). Similarly the
smallest number of edges which connect each kink vertex  which is produced by the second deformation, 
to the final nontrivial vertex,  is exactly one  and the edge in question is also uniquely 
located in the doubly deformed graph. In each set of these kinks, there is a single $C^1$ kink vertex
(where the deformed edge meets its undeformed counterpart in such a way that their edge tangents 
are collinear)  and $M-1$ $C^0$ kink vertices.

We require that the extended vertex smooth functions be functions of the final nontrivial vertex
and the set of kink vertices created by the {\em first} deformation. 
Note that in the `chronological' language  of P1, these kinks are the `second last' set of kinks
with the last set of kinks and the final nontrivial vertex to their `future'.
Accordingly, we refer to these kinks as the second last set of kinks.
%
%
Thus, every vertex smooth function $f$ is a smooth (or more precisely, semianalytic, $C^k$ in the context of the semianalyticity of
$\Sigma$) function from 
$M+1$  copies of the manifold to the complex numbers i.e. $f:\Sigma^{M+1}\rightarrow {\bf C}$,
$f= f(p_1,p_2,.., p_{M+1}), p_i\in \Sigma$.
and the (extended) VSA state $\Psi^f_{B_{VSA}}$ is the distribution corresponding to the weighted sum of 
bras in $B_{VSA}$ with weights $f$:
\begin{equation}
\Psi^f_{B_{VSA}}:=  \sum_{{\bar c}\in B_{VSA}}
\kappa_{\bar c} f(v^{\bar c}_{1},...,v^{\bar c}_{M-1},v^{\bar c}_{M},v^{\bar c}_{M+1}) \langle {\bar c}\vert,
\label{kappa}
\end{equation}
where $v^{\bar c}_{i}, i=1,..M-1$ are the second last $C^0$ kinks of ${\bar c}$, 
$v^{\bar c}_{M}$ is the second last $C^1$ kink of ${\bar c}$, 
$v^{\bar c}_{M+1}$ 
is the nontrivial vertex of ${\bar c}$ and $\kappa_{\bar c}$ is a number defined
in P1. 
In what follows, for simplicity we require that the  
functional dependence of the vertex smooth functions be symmetric with respect to the interchange of any two 
of its first $M-1$ arguments so that we need not concern ourself with the detailed nature of graph
symmetries of ${\bar c}$.

\subsection{Short distance behaviour}

In order to prescribe the short distance behaviour of the vertex smooth functions we fix a Riemmanian
metric $h_{ab}$ on $\Sigma$. 
We require that $f$ be of the form:
\begin{equation}
f (p_1,..,p_{M+1}) = f_1(p_1,..,p_{M}) g(p_{M+1})
\label{f=f1g}
\end{equation}
where $g:\Sigma \rightarrow {\bf C}$ is a semianalytic $C^k$ function and $f_1$ has the following short distance behaviour.
Let $B_{\epsilon}(p)$ be a convex normal neighbourhood of  the point $p$ such that the 
geodesic distance $d(q_1,q_2)$ between any points
$q_1,q_2\in B_{\epsilon}(p)$ is always less than $\epsilon$. Then for small enough $\epsilon$,
and for $p_i\neq p_M, i=1,..,M-1$, we require that  $f_1$ has the following behaviour
when $\{p_j, j=1,..,M\} \subset B_{\epsilon}(p)$:
\begin{equation}
f_1(p_1,..,p_{M})= f_2(p_1,..,p_{M})
\left( \frac{\sum_{i=1}^{M-1}\sum_{j=1}^{M-1}d(p_i,p_j) }{ \sum_{i=1}^{M-1}d(p_i,p_M) }\right)^{\frac{2}{3}},
\label{sd}
\end{equation}
where $f_2(p_1,..,p_{M})$ is a semianalytic $C^k$ function from $\Sigma^{M} \rightarrow {\bf C}$
which is invariant with respect the interchange of any two of its first $M-1$ arguments.

An example of $f_1$ is as follows. Let $U\subset \Sigma $ be a convex normal neighbourhood. Let $h (p)$
be  a semianlaytic $C^k$ function of compact support which vanishes outside $U$. A function $f_1$ which displays
the short distance behaviour (\ref{sd}) may be defined as:
\begin{equation}
f_1(p_1,..,p_{M})=(\prod_{i=1}^M h(p_i)) 
\left( \frac{\sum_{i=1}^{M-1}\sum_{j=1}^{M-1}d(p_i,p_j) }{ \sum_{i=1}^{M-1}d(p_i,p_M) }\right)^{\frac{2}{3}}.
\label{blowup}
\end{equation}
The function is ill defined at certain points in $\Sigma^M$, namely at those points 
where 
$p_{i}, i=1,..,M$ coincide. Nevertheless, as will be apparent in section 7, $f_1$ of the type in equation 
(\ref{blowup}) is an acceptable
choice for the construction of vertex smooth functions through equation (\ref{f=f1g}).

\section{The continuum limit of the Hamiltonian constraint and its commutator}

In section 7.1 we compute the continuum limit action 
of a single Hamiltonian constraint in the VSA topology of 
section 6. In section 7.2 we show that this action is diffeomorphism covariant. 
In section 7.3 we compute the continuum limit of the commutator and show that it is well defined 
and anomaly free.

\subsection{Continuum Hamiltonian constraint.}
The  continuum limit action of
the Hamiltonian constraint is $\lim_{\epsilon \rightarrow 0}\Psi^f_{B_{VSA}}({\hat C}_{\epsilon}(M)|c\rangle$
and a nontrivial limit is obtained as in P1 iff $|c\rangle$ is such that its deformations, produced by the 
action of ${\hat C}_{\epsilon}(M)$,  lie in $B_{VSA}$. 
%
Accordingly, consider the case where $c$ is such that at  non-zero $\epsilon$, 
$\Psi^f_{B_{VSA}}({\hat C}_{\epsilon}(M)|c\rangle \neq 0$. 
This is exactly case (c), section 6.3, P1.
It follows from an argumentation identical to that in 
section 6.3, P1 that $c$ is such that 
{\em all} the charge nets generated
by the action of {\em any} single finite triangulation Hamiltonian constraint on $c$ are in $B_{VSA}$
and that $c$ is, upto the action of semianalytic $C^k$ diffeomorphisms, 
itself produced by the single action of a finite triangulation Hamiltonian constraint on 
a state in $B_{<n,{\vec \alpha},c_p>}$ so that it has  a set of  last $M-1$ $C^0$ kinks and a last $C^1$ kink.
Let the positions of these kinks be $(v_1,.., v_{M-1})$ and $v_M$. 
The constraint operator ${\hat C}_{\epsilon}(M)$ deforms $c$ only in a very small vicinity of 
of its nontrivial vertex $v$.
The deformation moves the nontrivial vertex $v$ to $v_{I_{v},\epsilon}^{\prime} $ and creates a new set of $M$ kinks
in the vicinity of $v_{I_{v},\epsilon}^{\prime} $ while  leaving the graph structure in 
the vicinity of $(v_1,...,v_M)$ unaffected.
Consequently, the kinks $(v_1,...,v_M)$ are now the second last kinks of the deformed chargnet
$c(i,v_{I_{v},\epsilon}^{\prime})$.

It follows that at non-zero $\epsilon$, $\Psi^f_{B_{VSA}}({\hat C}_{\epsilon}|c\rangle$ leads to the evaluation 
of the function $g (p_{M+1})$ of equation (\ref{f=f1g})  at the point $p_{M+1}=v_{I_{v},\epsilon}^{\prime} $
and the evaluation of the 
function $f_1(p_1,..,p_M)$ of equation (\ref{f=f1g}) at the points $p_i=v_i, i=1,..,M$. Since the second last kinks 
of the deformed charge net  do not
move as $\epsilon \rightarrow 0$, we focus on the change in $g$ as the position of $v_{I_{v},\epsilon}^{\prime} $
varies, and suppress the dependence of $f_1$ on its arguments.

Recall that the modifications of section 5 do not alter the position of the displaced vertex 
$v_{I_{v},\epsilon}^{\prime} $ 
relative to P1. Hence it follows from the second equation of
section 4.4.1 of P1, that
in terms of the coordinates $\{x \}_c\equiv \{x\}$ at the nontrivial vertex $v$ of $c$ 
we have that:
\begin{equation}
x^{\mu}(v_{I_{v},\epsilon}^{\prime})  =   x^{\mu} (v) + \epsilon \;{\hat e}^a_{I_v}  +O(\epsilon^p )
\;\;, p>2
\label{xvepsilon}
\end{equation}

From (\ref{cdeltancfinal}) it follows that 
\begin{eqnarray}
\Psi^f_{B_{VSA}} ({\hat C}_{\epsilon}[N]|c\rangle
&=& 
\frac{\hbar}{2\mathrm{i}}\frac{3}%
{4\pi}N(v, \{x\}_c))\nu_{v}^{-2/3}\sum_{I_v, i}q_{I_{v}}^{i}\frac{1}{\epsilon}
\Psi^f_{B_{VSA}}(|c(i,v_{I_{v},\epsilon}^{\prime}) \rangle \nonumber\\
&=&\frac{\hbar}{2\mathrm{i}}\frac{3}%
{4\pi}N(v, \{x\}_c))\nu_{v}^{-2/3}\sum_{I_v, i}q_{I_{v}}^{i}\frac{1}{\epsilon}
f_1g( v_{I_{v},\epsilon}^{\prime})\label{f122.1}\nonumber \\
&=&\frac{\hbar}{2\mathrm{i}}\frac{3}%
{4\pi}N(v, \{x\}_c))\nu_{v}^{-2/3}\sum_{I_v, i}q_{I_{v}}^{i}
f_1\frac{g( v_{I_{v},\epsilon}^{\prime})- g(v)}
{\epsilon} ,
\label{f122.2}
\end{eqnarray}
where in the last line we have used gauge invariance to add the $g(v)$ term.
Note that from equation (\ref{xvepsilon}),  
we have that 
\begin{equation}
\lim_{\epsilon\rightarrow 0}\frac{g(v_{I_{v},\epsilon}^{\prime})- g( v)}
{\epsilon} = \left({\hat e}_{I_v}^{\mu} \frac{\partial g(p_{M+1})}{\partial x^{\mu}(p_{M+1})}\right)_{p_{M+1}=v}
\label{df123}
\end{equation}

Equations (\ref{f122.2}) and 
(\ref{df123}) together with equation (\ref{f=f1g})
imply that the continuum limit action of a single Hamiltonian 
constraint is given by:
\begin{equation}
\lim_{\epsilon \rightarrow 0} \Psi^f_{B_{VSA}} ({\hat C}_{\epsilon}[N]|c\rangle
= \frac{\hbar}{2\mathrm{i}}\frac{3}%
{4\pi}N(v, \{x\}_c))\nu_{v}^{-2/3}\sum_{I_v, i}q_{I_{v}}^{i}
{\hat e}_{I_v}^{a_v}\partial_{a_v}f(v_1,..,v_M, v),
\label{singlehamcont}
\end{equation}
where
the subscript $v$ on the index $a_v$ reminds us that the derivative acts on the $(M+1)$th argument 
of $f$.\\

\noindent {\em Remark:}
From the above calculation it is easy to see that  
the properties of the deformation other  than the positioning  (\ref{xvepsilon}) of the displaced vertex, 
such as the scrunching of deformed edge tangents, are {\em not}
relevant to the computation of the continuum limit of the single action of the Hamiltonian constraint.

\subsection{Diffeomorphism covariance.}

Since the key coordinate dependent elements which appear in the above continuum limit (\ref{singlehamcont}), 
namely, the coordinate 3- form used to evaluate the density weighted lapse
and the unit edge tangents which are normalized with respect to the coordinate metric, are the same as for the case
when the vertex smooth function depends only on the nontrival vertex, it is easy to see that the considerations
of sections 2 and 3 go though unchanged. In more detail, it is straightforward to verify that  the action (\ref{singlehamcont}) is diffeomorphism covariant 
provided:\\
\noindent
(i)  the reference coordinate patches for reference chargenets are chosen as in section 3 
\\
\noindent (ii) the coordinate 
patches associated with diffeomorphic  charge nets are diffeomorphic images of each other along the lines of section
2.2 
\\
\noindent 
(iii) the finite triangulation deformations at parameter $\delta$ of any charge network are related to the deformations
at the same parameter value, of the reference charge network along the lines of section 2.4.

\subsection{Well defined, anomaly free commutator}

The computation of the commutator between a pair of Hamiltonian constraint operators 
is described in section 7.3.2  and that of the
operator correspondent of the Poisson bracket between a pair of Hamiltonian constraints 
in section 7.3.3.
As in P1 we refer to the former as the LHS and the latter as the RHS.
Section 7.3.1 contains a note on the choice of coordinate
structures 
relevant to sections 7.3.2 and 7.3.3.

\subsubsection{Note on the choice of coordinates at $v_{I_{v},\delta}^{\prime}$.}

In the next section we shall evaluate the continuum limit of the action of the commutator between a pair of Hamiltonian
constraints. The continuum limit commutator is evaluated, as in equation (\ref{4.2}) by first taking the 
$\delta^{\prime}\rightarrow 0$ continuum limit of the action of ${\hat C}_{\delta^{\prime}}
(M)$ on the states $|c(i, v_{I_{v},\delta}^{\prime})\rangle$ generated by the action of  ${\hat C}_{\delta}(N)$ on
$|c\rangle$. Subsequent to this the $\delta\rightarrow 0$ contributions are evaluated. Their evaluations
depend on the behaviour of the  coordinate dependent edge tangents of section 5.5 as well the Jacobians
of the coordinates associated with the vertex  $v_{I_{v},\delta}^{\prime}$ of 
$c(i, v_{I_{v},\delta}^{\prime})$  with respect to the coordinates associated with the nontrivial vertex $v$ of $c$.

From section 5.3, the reference chargenets for $[c(i, v_{I_{v},\delta}^{\prime})]=[c_0 (i, v_{I_{v},\delta}^{\prime})]$
are ${\bar c}_{(i)0}$ amd the sibling restrictions imply that ${\bar c}_{(i)0}, i=1,2,3$ have the same 
nontrivial vertex ${\bar v}$ and same reference coordinate system ${\bar x}$.
%
Let  ${\bar \tau}_i \in {\cal D}_{[    {\bar c}_{(i)0}          ]}$
be such that ${\bar \tau}_i {\bar c}_{(i)0} = c(i, v_{I_{v},\delta}^{\prime})$  so that  the coordinates associated
with $v_{I_{v},\delta}^{\prime}$ in $c(i, v_{I_{v},\delta}^{\prime})$  are
${\bar \tau}_i^*\{   {\bar x}  \}$. It follows that the $\delta^{\prime}\rightarrow 0$ continuum limit of the 
single action of the Hamiltonian constraint referred to above is computed with respect to these coordinates.

Note, however that 
equations (\ref{c=betaphic0}), (\ref{betaphigamma}) of section 5.5.2 together with diffeomorphism covariance results of 
sections 7.2  and 3 (especially section 3.4), imply  that the $\delta^{\prime}\rightarrow 0$ continuum limit of the 
single action of the Hamiltonian constraint can equally well be 
evaluated with respect to the coordinates
$\{x^{\prime}_{(i),\beta,\delta}\}$ of equation (\ref{7.8}) of section 5.5.2.
Henceforth we shall use these coordinates to evaluate the $\delta^{\prime}\rightarrow 0$ contribution to the 
commutator (see equations (\ref{4.1}),(\ref{4.2})).

Finally we note that exactly the same argumentation  can be applied to the choice of coordinates pertinent to 
the evaluation of  the RHS. From section 5,P1,   the  RHS can be expressed as the 
commutator between a pair of electric diffeomorphisms. 
The continuum limit commutator is evaluated, as in equation (\ref{4.2}) by first taking the 
$\delta^{\prime}\rightarrow 0$ continuum limit of the action of ${\hat D}_{\delta^{\prime}}
({\vec M}_i)$ on the states $|c(v_{I_{v},\delta}^{\prime})\rangle$ generated by the action of  
${\hat D}_{\delta}({\vec N}_i)$ on
$|c\rangle$. Subsequent to this the $\delta\rightarrow 0$ contributions are evaluated. 
The first (i.e. the $\delta^{\prime}\rightarrow 0$) limit results in the  expression (\ref{7.17d}) which, by virtue
of the results of section 3, is independent of the choice of the set of reference diffeomorphisms,
${\cal D}_{[c(v_{I_{v},\delta}^{\prime})]}$. This, in turn, validates the use of the coordinates 
$\{x^{\prime  }_{({}_{\hat{\;}}),\beta,\delta}\}$ of section 5.5.2 (see equation (\ref{7.15})
around $v_{I_{v},\delta}^{\prime}$ in $c(v_{I_{v},\delta}^{\prime})$.\\

\noindent{\em Note}: The entire discussion above is phrased in language pertinent to the 
generic case when $c\in [c_0], c\neq c_0$.
For $c=c_0$ the discussion applies unchanged except that, above, we set $\beta$ to be the identity map.

\subsubsection{LHS}
The analysis closely parallels that of sections 4.5,4.6, P1 and section 4 of this paper.
Equation (\ref{cdeltancfinal}) implies that
\begin{equation}
\lim_{\delta^{\prime}\rightarrow 0}\Psi_{B_{VSA}}^f (
{\hat{C}}_{\delta^{\prime}}[M]{\hat{C}}_{\delta}[N]|c\rangle)
=
\frac{\hbar}{2\mathrm{i}}\frac{3}%
{4\pi}N(x(v))\nu_{v}^{-2/3}\sum_{I_v, i}q_{I_{v}}^{i}\frac{1}{\delta
} \lim_{\delta^{\prime}\rightarrow 0} \Psi_{B_{VSA}}^f (
{\hat{C}}_{\delta^{\prime}}[M] |c(i,v_{I_{v},\delta}^{\prime})\rangle ) 
\label{7.16}
\end{equation}
Noting that the charges at the edges emanating from $v_{I_{v},\delta}^{\prime}$ in $c(i,v_{I_{v},\delta}^{\prime})$
are flips of the original charges at the edges emanating from $v$ in $c$, Section 7.1 implies that, when nontrivial, the
limit below evaluates to:
\begin{equation}
\lim_{\delta^{\prime}\rightarrow0}\Psi^f_{B_{\mathrm{VSA}}}(\hat{C}%
_{\delta^{\prime}}[M]|c(i,v_{I_{v},\delta}^{\prime})\rangle)=\frac{\hbar}{2\mathrm{i}%
}\frac{3}{4\pi}M(x_{(i)\delta}^{\prime}(v_{I_{v},\delta}^{\prime}))\nu_{v_{I_{v}%
}}^{-2/3}\sum_{J_{v},i^{\prime}}\left.  ^{(i)}\!q_{J_{v}}^{i^{\prime}}\right.
f_1({\hat{\tilde{e}}}{}_{(i) J_{v}}^{\prime})^{a}\partial_{a}g(v_{I_{v},\delta
}^{\prime}).
\label{7.17}
\end{equation}
Here the subscript $(i)$ reminds us of the $(i)$- dependence of the primed coordinates (\ref{7.8}) 
and we have suppressed the subscript $\beta$ of equation (\ref{7.8}) to reduce notational clutter.
As in P1 $\nu_{v_{I_{v}}}^{-2/3}$ is the inverse volume eigen value at the vertex 
$v_{I_{v},\delta}^{\prime}$ of $c(i,v_{I_{v},\delta}^{\prime})$.
\footnote{From Appendix A, P1 this eigenvalue is invariant under charge flips and diffeomorphisms.
Consequently its notation is independent of $i, \delta$.}
The function
$f_1$ is evaluated at the second last kinks of the deformed chargenets obtained by the action of 
${\hat{C}}_{\delta^{\prime}}[M]$ on $|c(i,v_{I_{v},\delta}^{\prime})\rangle$. 
Similar to the discussion of 
section 7.1, these kinks are the last kinks of $|c(i,v_{I_{v},\delta}^{\prime})\rangle$. In the notation of P1,
these  kinks are placed on the edges $e_{J_v}$ emanating from $v$ in $c$ at the points ${\tilde v}_{J_v}$.
Since these points move (towards $v$) as $\delta$ decreases, we denote them by 
${\tilde v}^{\delta}_{J_v}\equiv {\tilde v}_{J_v}$. From P1, we have that  ${\tilde v}^{\delta}_{J_v\neq I_v}$ are 
$C^0$ kinks and ${\tilde v}^{\delta}_{ I_v}$ is the $C^1$ kink.
Thus, in equation (\ref{7.17}) we have that:
\begin{eqnarray}
&&f_1:= f_1(p_1,..,p_M),\nonumber \\
&&\{p_1,..,p_{M-1}\} = \{
{\tilde v}^{\delta}_{J_v\neq I_v}
\}, \;\;\;\;\; p_M= {\tilde v}^{\delta}_{I_v},
\label{f1dep}
\end{eqnarray}
where, by virtue of the symmetric dependence of $f_1$ on its first $M-1$ entries (see the last sentence of section 6.2),
 it does not matter which 
$p_{i\neq M}$ is identified with which ${\tilde v}_{J_v\neq I_v}$. Hence in the above equation we may write 
\begin{equation}
f_1\equiv f_1(\;  \{{\tilde v}^{\delta}_{J_v\neq I_v}\}, {\tilde v}^{\delta}_{I_v}\;)
\label{f1depsymm}
\end{equation}

Equations (\ref{7.16}), (\ref{7.17}) together with the density $-\frac{1}{3}$ property of $M$ imply that:
\begin{align}
\lim_{\delta^{\prime}\rightarrow0}\Psi^f_{B_{\mathrm{VSA}}}(\hat{C}%
_{\delta^{\prime}}[M]\hat{C}_{\delta}[N]c)  &  =\left(  \frac{\hbar
}{2\mathrm{i}}\frac{3}{4\pi}\right)  ^{2}\frac{1}{\delta}\nu
_{v}^{-2/3}N(x(v))\label{7.19}\\
&  \qquad\times\sum_{I_{v},i}M(x(v_{I_{v},\delta}^{\prime}))\left[  \det\left(
\frac{\partial x}{\partial x_{(i)\delta}^{\prime}}\right)  _{v_{I_{v},\delta}^{\prime}%
}\right]  ^{-1/3}f_1     \{\cdots\}_{(i),I_{v},\delta},\nonumber
\end{align}
with $f_1$ as in equation (\ref{f1dep}) and
where
\begin{equation}
\{\cdots\}_{(i),I_{v},\delta}:=
q_{I_{v}}^{i}\nu_{v_{I_{v}}}^{-2/3}%
\sum_{J_{v},i^{\prime}}\left.  ^{(i)}\!q_{J_{v}}^{i^{\prime}}\right.
({\hat{\tilde{e}}}{}_{(i) J_{v}}^{\prime})^{a}\partial_{a}g(v_{I_{v},\delta
}^{\prime}), \label{curly7}%
\end{equation}
As in section 4.6,P1 and section 4 of this work, we use equation (\ref{xvepsilon}) in equation (\ref{7.19}) 
to Taylor expand the 
lapse at $v_{I_{v},\delta}^{\prime}$ in terms of its derivative at $v$. This, 
 together with the antisymmetric dependence 
of the commutator on the lapses $M,N$ implies that:
\begin{align}
&  \lim_{\delta^{\prime}\rightarrow0}(\Psi^f_{B_{\mathrm{VSA}}}|(\hat{C}%
_{\delta^{\prime}}[M]\hat{C}_{\delta}[N]-\left(  N\leftrightarrow M\right)
)|c\rangle\nonumber\\
&  =\left(  \frac{\hbar}{2\mathrm{i}}\frac{3}{4\pi}\right)  ^{2}%
\nu_{v}^{-2/3}\nonumber\\
&  \times\sum_{I_{v}}\{N(x(v))\hat{e}_{I_{v}}^{a}\partial_{a}M(x(v))-\left(
N\leftrightarrow M\right)  +O(\delta)\}\left[  \det\left(  \frac{\partial
x}{\partial x_{(i)\delta}^{\prime}}\right)  _{v_{I_{v},\delta}^{\prime}}\right]
^{-1/3}f_1 \sum_i  \{\cdots\}_{(i),I_{v},\delta} \label{cmcn7}%
\end{align}
Using equatons (\ref{7.14}) and (\ref{defGdelta}) in the above equation yields:
\begin{align}
&  \lim_{\delta^{\prime}\rightarrow0}(\Psi_{B_{\mathrm{VSA}}}|(\hat{C}%
_{\delta^{\prime}}[M]\hat{C}_{\delta}[N]-\left(  N\leftrightarrow M\right)
)|c\rangle\nonumber\\
&  =\left(  \frac{\hbar}{2\mathrm{i}}\frac{3}{4\pi}\right)  ^{2}%
\nu_{v}^{-2/3}\nonumber\\
&  \times\sum_{I_{v}}\{N(x(v))\hat{e}_{I_{v}}^{a}\partial_{a}M(x(v))-\left(
N\leftrightarrow M\right)  +O(\delta)\}
\left[\delta^{-\frac{2}{3}(q-1)}f_1(\;\{{\tilde v}^{\delta}_{J_v\neq I_v}\}, {\tilde v}^{\delta}_{I_v}\;)\right]
\sum_i\{\cdots\}_{(i),I_{v},\delta}, 
\label{7.23}%
\end{align}
where we have used the notation (\ref{f1depsymm}).

We now evaluate the term in square brackets in equation (\ref{7.23}).
To do so recall from 6.,Appendix C.4, P1 that: \\
\noindent
(a)the $C^0$ kink vertices ${\tilde v}^{\delta}_{J_v\neq I_v}$
are placed at a coordinate distance $\delta^q$ from $v$ on $e_{J_v}$,
\\
\noindent
(b)the $C^1$ kink vertex ${\tilde v}^{\delta}_{I_v}$
is placed at a coordinate distance $2\delta$ from $v$ along $e_{I_v}$.
\\
Hence, to leading order in $\delta$ it follows from equation (\ref{sd})that:
\begin{equation}
\left( \frac{\sum_{J_v\neq I_v}\sum_{K_v\neq I_v}
d({\tilde v}^{\delta}_{K_v},{\tilde v}^{\delta}_{J_v}) }
{ \sum_{L_v\neq I_v}d( {\tilde v}^{\delta}_{L_v},{\tilde v}^{\delta}_{I_v}       ) }
\right)^{\frac{2}{3}}
= 
\left( 
\frac{
\sum_{J_v\neq I_v}\sum_{K_v\neq I_v} 
g_{ab}({\hat e}^a_{K_v}-{\hat e}^a_{J_v})({\hat e}^b_{K_v}-{\hat e}^b_{J_v})\delta^{2q}
}
{4 (M-1)
g_{ab}({\hat e}^a_{I_v}{\hat e}^b_{I_v}) \delta^2
}
\right)^{\frac{1}{3}}
\end{equation}
From this it follows that as $\delta\rightarrow 0$, the term in square brackets in equation (\ref{7.23}) behaves as:
\begin{equation}
\left[\delta^{-\frac{2}{3}(q-1)}f_1(\;\{{\tilde v}^{\delta}_{J_v\neq I_v}\}, {\tilde v}^{\delta}_{I_v}\;)\right]
= f_2(v,..,v)\left( 
\frac{
\sum_{J_v\neq I_v}\sum_{K_v\neq I_v} 
g_{ab}({\hat e}^a_{K_v}-{\hat e}^a_{J_v})({\hat e}^b_{K_v}-{\hat e}^b_{J_v})
}
{ 4(M-1)
g_{ab}({\hat e}^a_{I_v}{\hat e}^b_{I_v}) 
}
\right)^{\frac{1}{3}}
\label{sdf1delta}
\end{equation}
It is in this manner that the precise short distance behaviour of the 
vertex smooth functions detailed in  section 6.3 cures the divergence
of the commutator seen in section 4.3. Note that the resulting expression is invariant under a constant rescaling
of the unit tangents. This ensures the independence of our final result from the choice of reference diffeomorphisms.
Upto this point in the calculation, conicality of the deformations has not played a role, thus illustrating that
the short distance behaviour suffices to render the commutator well defined. As we shall see now, conicality
plays a key role in the evaluation of the term $\sum_i\{\cdots\}_{(i),I_{v},\delta}$ in 
equation (\ref{7.23}).
As will be apparent below and in the 
next section, it is the interplay between conicality and gauge invariance which renders the well defined commutator
anomaly free.

We now evaluate the term $\sum_i\{\cdots\}_{(i),I_{v},\delta}$ equation (\ref{7.23}) in the $\delta\rightarrow 0$
limit. Using equations (\ref{IJtngntdeltac}), (\ref{IItngntdeltac}) (recall that in these equations we suppressed
the subscript $(i)$ on their left hand sides) in (\ref{curly7}), we obtain
\begin{eqnarray}
&&\sum_i\{\cdots\}_{(i),I_{v},\delta}:= \nonumber \\
&&\sum_{i}q_{I_{v}}^{i}\nu_{v_{I_{v}}}^{-2/3}%
\sum_{i^{\prime}}\left( 
\left.  ^{(i)}\!q_{I_{v}}^{i^{\prime}}\right.
{\hat {\tilde e}}_{I_v}^{a}-  
(\sum_{J_{v}\neq I_v}\left.  ^{(i)}\!q_{J_{v}}^{i^{\prime}}\right. )
\cos\theta_{I_{v_0}}{\hat {\tilde e}}_{I_v}^{a}
\;+\;O(\delta^{q-1})
\right)
\partial_{a}g(v_{I_{v},\delta
}^{\prime}) \\
&&
=\sum_{i}q_{I_{v}}^{i}\nu_{v_{I_{v}}}^{-2/3}%
\sum_{i^{\prime}}
\left( 
\left.  ^{(i)}\!q_{I_{v}}^{i^{\prime}}\right.
{\hat {\tilde e}}_{I_v}^{a} 
+\left.  ^{(i)}\!q_{I_{v}}^{i^{\prime}}\right.
\cos\theta_{I_{v_0}}{\hat {\tilde e}}_{I_v}^{a}
\;+\;O(\delta^{q-1})
\right)
\partial_{a}g(v_{I_{v},\delta
}^{\prime}),
\label{curly7delta}
\end{eqnarray}
where in the last line we have used gauge invariance.
Using equation (\ref{IcontItngntc}) in (\ref{curly7delta}) we have, in the $\delta\rightarrow 0$ limit, that:
\begin{eqnarray}
\lim_{\delta\rightarrow 0}\sum_i\{\cdots\}_{(i),I_{v},\delta}
&=& 
2\nu_{v_{I_{v}}}^{-2/3}\cos^2\frac{\theta_{I_{v_0}}}{2}
\sum_{i,i^{\prime}}
q_{I_{v}}^{i}
\left.  ^{(i)}\!q_{I_{v}}^{i^{\prime}}\right.
{\hat {e}}_{I_v}^{a} \partial_{a}g(v)\nonumber\\
&=&2\nu_{v_{I_{v}}}^{-2/3}\cos^2\frac{\theta_{I_{v_0}}}{2}
\sum_{i}
(q_{I_{v}}^{i})^2
{\hat {e}}_{I_v}^{a} \partial_{a}g(v),
\label{curly7cont}
\end{eqnarray}
where we have used the definition of the flipped charges from section 4.4.3, P1.
Equations (\ref{curly7cont}), (\ref{sdf1delta}) and (\ref{7.23}) imply that  
\begin{align}
&  \lim_{\delta\rightarrow 0}\lim_{\delta^{\prime}\rightarrow0}(\Psi_{B_{\mathrm{VSA}}}|(\hat{C}%
_{\delta^{\prime}}[M]\hat{C}_{\delta}[N]-\left(  N\leftrightarrow M\right)
)|c\rangle\nonumber\\
&  =\left(  \frac{\hbar}{2\mathrm{i}}\frac{3}{4\pi}\right)  ^{2}%
\nu_{v}^{-2/3}
f_2(v,..,v)  \sum_{I_{v}}\{N(x(v))\hat{e}_{I_{v}}^{a}\partial_{a}M(x(v))-\left(
N\leftrightarrow M\right)  \}\nonumber\\
&
\left( 
\frac{
\sum_{J_v\neq I_v}\sum_{K_v\neq I_v} 
g_{ab}({\hat e}^a_{K_v}-{\hat e}^a_{J_v})({\hat e}^b_{K_v}-{\hat e}^b_{J_v})
}
{4 (M-1)
g_{ab}({\hat e}^a_{I_v}{\hat e}^b_{I_v}) 
}
\right)^{\frac{1}{3}}
2\nu_{v_{I_{v}}}^{-2/3}\cos^2\frac{\theta_{I_{v_0}}}{2}
\sum_{i}
(q_{I_{v}}^{i})^2
{\hat {e}}_{I_v}^{a} \partial_{a}g(v)
\label{lhsfinal}
\end{align}

\subsubsection{RHS}

The evaluation of the RHS proceeds exactly as for the LHS the only difference being that the deformed
chargenets $c(i,v_{I_{v},\delta}^{\prime})$ are replaced by their unflipped sibling
$c(v_{I_{v},\delta}^{\prime})$, the subscripts $(i)$ by the subscript $({}_{\hat {}})$
and the flipped charges $\left.  ^{(i)}\!q_{J_{v}}^{i^{\prime}}\right.$ by their unflipped 
correspondents $q^i_{J_v}$. In the interests of brevity we display the main steps of the 
calculation through equations (\ref{defRHS})- (\ref{rhsfinal}) below.
Our starting point is the definition of the RHS in section 5 of P1:
\begin{equation}
RHS = -3\lim_{\delta\rightarrow 0}\lim_{\delta^{\prime}\rightarrow0}\Psi^f_{B_{\mathrm{VSA}}}(
\sum_{i=1}^3(\hat{D}%
_{\delta^{\prime}}[{\vec M}_i]\hat{D}_{\delta}[{\vec N}_i]-\left(  N\leftrightarrow M\right)
)|c\rangle)
\label{defRHS}
\end{equation}
\begin{equation}
\lim_{\delta^{\prime}\rightarrow 0}\Psi_{B_{VSA}}^f (\sum_{i=1}^3
{\hat{D}}_{\delta^{\prime}}[{\vec M}_i]{\hat{D}}_{\delta}[{\vec N}_i]|c\rangle)
=
\frac{\hbar}{\mathrm{i}}\frac{3}%
{4\pi}N(x(v))\nu_{v}^{-2/3}\sum_{I_v, i}q_{I_{v}}^{i}\frac{1}{\delta
} \lim_{\delta^{\prime}\rightarrow 0} \Psi_{B_{VSA}}^f (
{\hat{D}}_{\delta^{\prime}}[{\vec M}_i] |c(v_{I_{v},\delta}^{\prime})\rangle ), 
\label{7.16d}
\end{equation}
where the factor of $2$ relative to equation (\ref{7.16}) can be traced to the finite triangulation 
definition of 
the electric diffeomorphism operator relative to that of the Hamiltonian constraint in P1.
\begin{equation}
\lim_{\delta^{\prime}\rightarrow0}\Psi^f_{B_{\mathrm{VSA}}}(\hat{D}%
_{\delta^{\prime}}[{\vec M}_i]c(v_{I_{v},\delta}^{\prime}))=-\frac{\hbar}{12\mathrm{i}%
}\frac{3}{4\pi}M(x_{    ({}_{\hat{}})       \delta}^{\prime}(v_{I_{v},\delta}^{\prime}))\nu_{v_{I_{v}%
}}^{-2/3}\sum_{J_{v}}  q_{J_{v}}^i
f_1({\hat{\tilde{e}}}{}_{ ({}_{\hat{}})  J_{v}}^{\prime})^{a}\partial_{a}g(v_{I_{v},\delta
}^{\prime}),
\label{7.17d}
\end{equation}
where the `extra' factor of $-\frac{1}{12}$ comes from the factor of $\kappa_{\bar c}$ in equation (\ref{kappa})
(see (2),section 5.4,P1). 
\begin{align}
\lim_{\delta^{\prime}\rightarrow0}\Psi^f_{B_{\mathrm{VSA}}}(\hat{D}%
_{\delta^{\prime}}[{\vec M}_i]\hat{D}_{\delta}[{\vec N}_i]c)  &  =-\frac{1}{12}\left(  \frac{\hbar
}{\mathrm{i}}\frac{3}{4\pi}\right)  ^{2}\frac{1}{\delta}\nu
_{v}^{-2/3}N(x(v))\label{7.19d}\\
&  \qquad\times\sum_{I_{v}}M(x(v_{I_{v},\delta}^{\prime}))\left[  \det\left(
\frac{\partial x}{\partial x_{ ({}_{\hat{}})    \delta}^{\prime}}\right)  _{v_{I_{v},\delta}^{\prime}%
}\right]  ^{-1/3}f_1     \{\cdots\}_{({}_{\hat{}}),I_{v},\delta},\nonumber
\end{align}
with $f_1$ as in equation (\ref{f1dep}) and
where
\begin{equation}
\{\cdots\}_{({}_{\hat{}}),I_{v},\delta}:=\sum_{i}q_{I_{v}}^{i}\nu_{v_{I_{v}}}^{-2/3}%
\sum_{J_{v}}q_{J_{v}}^{i}
({\hat{\tilde{e}}}{}_{ ({}_{\hat{}})   J_{v}}^{\prime})^{a}\partial_{a}g(v_{I_{v},\delta
}^{\prime}). \label{curly7d}%
\end{equation}
\begin{align}
&  \lim_{\delta^{\prime}\rightarrow0}(\Psi^f_{B_{\mathrm{VSA}}}|(\hat{D}%
_{\delta^{\prime}}[{\vec M}_i]\hat{D}_{\delta}[{\vec N}_i]-\left(  N\leftrightarrow M\right)
)|c\rangle\nonumber\\
&  =-\frac{1}{12}      \left(  \frac{\hbar}{\mathrm{i}}\frac{3}{4\pi}\right)  ^{2}%
\nu_{v}^{-2/3}\nonumber\\
&  \times\sum_{I_{v}}\{N(x(v))\hat{e}_{I_{v}}^{a}\partial_{a}M(x(v))-\left(
N\leftrightarrow M\right)  +O(\delta)\}\left[  \det\left(  \frac{\partial
x}{\partial x_{({}_{\hat{}})     \delta}^{\prime}}\right)  _{v_{I_{v},\delta}^{\prime}}\right]
^{-1/3}f_1\{\cdots\}_{({}_{\hat{}}),I_{v},\delta}, \label{cmcn7d}%
\end{align}
\begin{align}
&  \lim_{\delta^{\prime}\rightarrow0}(\Psi^f_{B_{\mathrm{VSA}}}|(\hat{D}%
_{\delta^{\prime}}[{\vec M}_i]\hat{D}_{\delta}[{\vec N}_i]-\left(  N\leftrightarrow M\right)
)|c\rangle\nonumber\\
&  =  -\frac{1}{12}        \left(  \frac{\hbar}{\mathrm{i}}\frac{3}{4\pi}\right)  ^{2}%
\nu_{v}^{-2/3}\nonumber\\
&  \times\sum_{I_{v}}\{N(x(v))\hat{e}_{I_{v}}^{a}\partial_{a}M(x(v))-\left(
N\leftrightarrow M\right)  +O(\delta)\}
\left[\delta^{-\frac{2}{3}(q-1)}f_1(\;\{{\tilde v}^{\delta}_{J_v\neq I_v}\}, {\tilde v}^{\delta}_{I_v}\;)\right]
\{\cdots\}_{({}_{\hat{}}),I_{v},\delta}, 
\label{7.23d}%
\end{align}
\begin{eqnarray}
\{\cdots\}_{  ({}_{\hat{}}) ,I_{v},\delta}&:=&
 q_{I_{v}}^{i}\nu_{v_{I_{v}}}^{-2/3}%
\left( 
q_{I_{v}}^{i}
{\hat {\tilde e}}_{I_v}^{a}-  
(\sum_{J_{v}\neq I_v}  q_{ J_{v}}^{i}  )
\cos\theta_{I_{v_0}}{\hat {\tilde e}}_{I_v}^{a}
\;+\;O(\delta^{q-1})
\right)
\partial_{a}g(v_{I_{v},\delta
}^{\prime}) \\
&=& q_{I_{v}}^{i}\nu_{v_{I_{v}}}^{-2/3}%
\left( 
q_{I_{v}}^{i}
{\hat {\tilde e}}_{I_v}^{a} 
+ q_{I_{v}}^{i}
\cos\theta_{I_{v_0}}{\hat {\tilde e}}_{I_v}^{a}
\;+\;O(\delta^{q-1})
\right)
\partial_{a}g(v_{I_{v},\delta
}^{\prime}),
\label{curly7deltad}
\end{eqnarray}
\begin{equation}
\lim_{\delta\rightarrow 0}\{\cdots\}_{(i),I_{v},\delta}
=
2\nu_{v_{I_{v}}}^{-2/3}\cos^2\frac{\theta_{I_{v_0}}}{2}
\sum_{i}
(q_{I_{v}}^{i})^2
{\hat {e}}_{I_v}^{a} \partial_{a}g(v),
\label{curly7contd}
\end{equation}
Equations (\ref{curly7contd}), (\ref{sdf1delta}) and (\ref{7.23d}) imply that  
\begin{align}
&  -3\lim_{\delta\rightarrow 0}\lim_{\delta^{\prime}\rightarrow0}(\Psi^f_{B_{\mathrm{VSA}}}|
\sum_{i=1}^3
(\hat{D}%
_{\delta^{\prime}}[{\vec M}_i]\hat{D}_{\delta}[{\vec N}_i]-\left(  N\leftrightarrow M\right)
)|c\rangle\nonumber\\
&  = \left(  \frac{\hbar}{2\mathrm{i}}\frac{3}{4\pi}\right)  ^{2}%
\nu_{v}^{-2/3}
f_2(v,..,v)  \sum_{I_{v}}\{N(x(v))\hat{e}_{I_{v}}^{a}\partial_{a}M(x(v))-\left(
N\leftrightarrow M\right)  \}\nonumber\\
&
\left( 
\frac{
\sum_{J_v\neq I_v}\sum_{K_v\neq I_v} 
g_{ab}({\hat e}^a_{K_v}-{\hat e}^a_{J_v})({\hat e}^b_{K_v}-{\hat e}^b_{J_v})
}
{4 (M-1)
g_{ab}({\hat e}^a_{I_v}{\hat e}^b_{I_v}) 
}
\right)^{\frac{1}{3}}
2\nu_{v_{I_{v}}}^{-2/3}\cos^2\frac{\theta_{I_{v_0}}}{2}
\sum_{i}
(q_{I_{v}}^{i})^2
{\hat {e}}_{I_v}^{a} \partial_{a}g(v),
\label{rhsfinal}
\end{align}
which agrees with equation (\ref{lhsfinal}).

\section{Concluding Remarks}
In section 8.1 we summarise our results. Section 8.2 is devoted to a technical aspect of
the property of nontriviality of a vertex. In section 8.3 we discuss our results and 
in section 8.4 we discuss open issues.

\subsection{Summary}

We briefly summarise our main results in terms of the following equations.
\\

\noindent (1)
The continuum limit action of the single Hamiltonian constraint, when non- trivial,
takes the form:
\begin{equation}
\lim_{\epsilon \rightarrow 0} \Psi^f_{B_{VSA}} ({\hat C}_{\epsilon}[N]|c\rangle
= \frac{\hbar}{2\mathrm{i}}\frac{3}%
{4\pi}N(v, \{x\}))\nu_{v}^{-2/3}\sum_{I_v, i}q_{I_{v}}^{i}
{\hat e}_{I_v}^{a}f(v_1,..,v_M) \partial_{a}g(v),
\label{singlehamcontfinal}
\end{equation}

\noindent (2)
The anomaly free continuum limit action of the commutator between a pair of Hamiltonian constraints,
when non- trivial, takes the form:
\begin{align}
&  \lim_{\delta\rightarrow 0}\lim_{\delta^{\prime}\rightarrow0}(\Psi_{B_{\mathrm{VSA}}}|(\hat{C}%
_{\delta^{\prime}}[M]\hat{C}_{\delta}[N]-\left(  N\leftrightarrow M\right)
)|c\rangle\nonumber\\
&  
=
\{N(x(v))\partial_{a}M(x(v))-\left(
N\leftrightarrow M\right)  \}
\left(  \frac{\hbar}{2\mathrm{i}}\frac{3}{4\pi}\right)  ^{2}%
\sum_{I_{v}}  
2\nu_{v}^{-2/3}\nu_{v_{I_{v}}}^{-2/3}
\hat{e}_{I_{v}}^{a}{\hat {e}}_{I_v}^{b}
\cos^2\frac{\theta_{I_{v_0}}}{2}
\sum_{i}
(q_{I_{v}}^{i})^2
\nonumber\\
&
\left( 
\frac{
\sum_{J_v\neq I_v}\sum_{K_v\neq I_v} 
g_{cd}({\hat e}^c_{K_v}-{\hat e}^c_{J_v})({\hat e}^d_{K_v}-{\hat e}^d_{J_v})
}
{4 (M-1)
g_{ef}({\hat e}^e_{I_v}{\hat e}^f_{I_v}) 
}
\right)^{\frac{1}{3}}
f_2(v,..,v)  
 \partial_{b}g(v)
\label{lhsfinalfinal}
\end{align}
which may be compared to the classical Poisson bracket:
\begin{equation}
\{C(N), C(M)\}= 
\int\mathrm{d}^{3}x~\left(  M\partial
_{a}N-N\partial_{a}M\right)  q^{-\frac{2}{3}}E_{i}^{c}E_{i}^{b}F_{bc}^{j}E_{j}^{c}
\end{equation}

\subsection{On the definition of nontriviality of a vertex}

An important element of our considerations hitherto is the property of nontriviality of a vertex
defined in section 2.1. This section is devoted to the discussion of a technical subtlety regarding this 
definition and may be skipped on a first reading of this paper.

Nontriviality of a vertex ensures a nontrivial continuum limit 
i.e it is assured that (for generic vertex smooth functions) there exist nontrivial 
chargenets  such that the continuum limit of the Hamiltonian constraint and its commutator are nontrivial.
More in detail, nontriviality of the continuum limit depends on (1) the nontriviality of the chargenet
bras which 
go into the construction of  the set $B_{VSA}$ which, in turn, goes into the specification of 
the VSA state  $\Psi^f_{B_{VSA}}$ in equations (\ref{singlehamcontfinal}) and (\ref{lhsfinalfinal})
  and (2) the nontriviality of the chargenet $|c\rangle$ being 
acted upon in equations (\ref{singlehamcontfinal}) and (\ref{lhsfinalfinal}).
Our remarks in this section pertain primarily to (1).

In the main body of this work we assumed that {\em non-degeneracy} (i.e. non-vanishing inverse 
metric eigen value) is a part of non- triviality (see section 2.1). However, while it is perhaps
reasonable to expect that the deformations of sections 5.2- 5.4 generically preserve a generic
property such as non- degeneracy, the expression for the inverse volume eigen value is so complicated
that we are, currently, unable to {\em prove} that nondegeneracy is {\em always}  preserved under these
deformations. Note however that the arguments of this work go through with a weaker definition of 
nontriviality  which {\em is} obviously preserved by these deformations. The weaker definition of a nontrivial
vertex substitutes (i) of section 2.1 by the  condition ${\rm (i)}^{\prime}$ below.
\\

\noindent{\em Definition}: A vertex of a chargenet is {\em weakly nontrivial} iff in addition to (ii), (iii) of 
section 2.1 it is such that:\\
\noindent ${\rm (i)}^{\prime}$(a) there exists no $i$ such that the $i$th charge vanishes on {\em all} edges
emanating from the vertex.
\\
\noindent ${\rm (i)}^{\prime}$(b) the valence of the vertex is greater than 3.
\\

We shall say that a chargenet is weakly nontrivial iff it has a single weakly nontrivial vertex.
\\

As mentioned above, weak nontriviality does not obtain any advantage with respect to the dependence of 
a nontrivial continuum limit on (2) above,  in that,  a nontrivial continuum limit can {\em only} ensue
if the chargenet being acted upon has a nondegenerate vertex. Rather, the virtue of weak nontriviality
is that it allows the construction of $B_{VSA}$  without
unwarranted assumptions in regard to the non- degeneracy of deformed vertices. Specifically, the deformations 
defined in section 5 not only preserve weak nontriviality but are also well defined for vertices which, prior to the 
deformations, are weakly nontrivial and degenerate. Moreover the `causal' definitions of section 6, P1 which play 
a crucial role in the demonstration of anomaly free- ness here, go through unchanged for weakly nontrivial 
chargenets. Since we have no control over the preservation of nondegeneracy by the deformations of section 5,
the construction of $B_{VSA}$ is at best logically incomplete and at worst invalid if we insist on the 
`strong nontriviality' of section 2.1.  For example, in the construction of $B_{VSA}$, the vertex of some $m$th 
generation chargenet arising from the primordial chargenet $c_p$ could in principle be weakly nontrivial and degenerate thus invalidating the strong nontriviality
property of section 2.1.

While weak nontriviality does yield an anomaly free commutator, it 
is no longer a guarantee of a nontrivial continuum limit. 
A nontrivial 
continuum limit (\ref{singlehamcontfinal}) (for generic $f$) 
is assured for the single Hamiltonian constraint if the final vertex of some  bra 
in $B_{[H<n,{\vec \alpha}, c_p>]}$ is nondegenerate, in addition to being weakly nontrivial.
A nontrivial anomaly free continuum limit 
(\ref{lhsfinalfinal}) (for generic $f$) is assured
for the commutator if, in addition to weak nontriviality, the final vertex of some bra in 
$B_{[H<n,{\vec \alpha}, c_p>]}$  {\em and} the final vertex of a  parent of this bra in 
$B_{<n,{\vec \alpha}, c_p>}$, 
are both nondegenerate. Note that in principle, the bras in $B_{[H<n,{\vec \alpha}, c_p>]}$ may be nondegenerate
without  those in $B_{<n,{\vec \alpha}, c_p>}$ being nondegenerate. Also note that nontriviality of the 
continuum limit is independent of the nondegeneracy or lack thereof, of the final vertex of the bras in 
$B_{VSA}$.

Finally, note that 
if  condition ${\rm (i)}^{\prime}$(b) is violated, gauge invariance implies that {\em all} vertices encountered 
in the construction of $B_{VSA}$ are {\em necessarily} degenerate thus trivialising the continuum limit of
the Hamiltonian constraint, the electric diffeomorphism operator, the LHS and the RHS.
This is the reason for demanding condition ${\rm (i)}^{\prime}$(b).

\subsection{Discussion of Results}
The consideration of higher than unity density weight Hamiltonian constraints 
requires regulating coordinate patches for their definition. This may be understood
as a consequence of the (to the best of our knowledge) fact that there is no 
backround independent density weighted object other than the unit density  Dirac delta function. 
It seems natural to expect that a diffeomorphism covariant construction of the higher density constraints
only requires a definition of their action on one representative in each diffeomorphism class of states.
This is exactly what happens. Thus, a single reference coordinate patch is chosen for each such 
representative and the construction is `spread' to other states through diffeomorphisms. Since the states
live on lower dimensional submanifolds of the spatial slice, there are many diffeomorphisms which map
the reference state to a particular element of its diffeomorphism class. Hence, to begin with, a 
set of reference diffeomorphisms need to be chosen as well. However, the final expressions for the
action of the Hamiltonian constraint (\ref{singlehamcont}) as well as its commutator (\ref{lhsfinal})
are independent of this choice. We have already discussed this independence in section 3.5 for the 
action of the constraint. A similar argument shows that the commutator (\ref{lhsfinal}) is also 
indpendent of this choice. This may be seen from the fact that a change in the choice  of reference diffeomorphisms
lead, by virtue of section 3, to a change in the choice of coordinates by a constant times a rotation. It is easily 
verified that equation  (\ref{lhsfinal}) is invariant under such a change.

This invariance with respect to, and the consequent independence of our results from, 
the choice of reference diffeomorphisms is a direct consequence of the restriction of the VSA states
to be of Grot- Rovelli type. Is this a natural restriction? As we argue now, {\em some} such restriction is 
to be expected and the GR restriction seems to be simple, elegant and extendible to the  $SU(2)$ case of
Euclidean gravity. The classical constraints are defined only if the `metric' $E^a_iE^{bi}$ is nondegenerate.
In contrast the quantum electric fields associated with a chargenet vanish almost everywhere. Hence it is 
not surprising, if at the microscopic quantum level, we want the constraint algebra to be anomaly free, that 
some version of non-degeneracy of the quantum geometry is required. We view the GR property as a reflection 
of the classical non-degeneracy at the quantum level.

To summarise: the only structures which are choice dependent are those of the reference coordinate patches.
For edge tangent sets with symmetries, even this choice is restricted by the requirement that the 
symmetries be proportional to rotations in the chosen coordinates. 

The final result for the commutator (\ref{lhsfinal}) depends on the choice of reference coordinates
through its dependence on objects which are defined with respect to the coordinates associated with the chargenet being acted upon, these coordinates, in turn, being diffeomorphic images of the reference coordinates. 
The coordinate dependent objects are the edge tangents which are normalised with respect to the coordinate metric
and the density weighted lapse which is evaluated with respect to the coordinate 3 form.
The result also depends on the angles $\theta_{I_{v_0}}$. These angles characterise the deformations generated
on the reference charge net by the action of the Hamiltonian constraint so that once again, it is only the 
deformation of the reference structures which play a role in the final expression.


\subsection{Open Issues}
The two key open issues are
the generalization of our considerations here to the context of a genuine habitat and 
the generalization of the techniques, ideas and results of this work, that of P1 and that of Reference \cite{aloku131}, 
from the 
case of internal group $U(1)^3$ to internal group $SU(2)$.
We have already discussed these issues in the final section of P1 so we refrain from repeating ourselves here.

Apart from these two issues, it is also important to know if our considerations here yield physically 
acceptable results for the kernel of the constraints. One possible 
`internal' check is to see if the commutator (\ref{lhsfinal})
is, like the single action of the Hamiltonian constraint (\ref{singlehamcont}), diffeomorphism covariant.
Our intuition is that the independence of (\ref{lhsfinal}) from the choice of reference diffeomorphisms
is indicative of an affirmative answer, this needs to be checked in detail. 
Of course the only conclusive way to settle the issue is to construct the kernel of the constraints, 
endow the kernel with an inner product and verify the existence of (semi)classical behaviour with respect to
a complete set of Dirac observables.
Now, it is easy to see that in the limited context of the 
VSA states of P1 and their generalization here, the kernel of the constraints is obtained by setting the vertex
functions to be constants. The resulting states are distributions and, from section 6.1 and from P1,
 each such state is generated from some `primordial'  charge net state.
A natural inner product  which suggests itself may be defined as follows: Define the inner product between
such distributions as the inner product between their associated primordial states. Whether or not this
inner product is the correct one can (only?) be known if we are able to check the self adjointness of a 
complete set of operators corresponding to classical Dirac observables. The availability of a large enough family
of classical observables is itself a challenging issue. However, recently, 
Barbero and
Villase$\rm{\tilde n}$or \cite{fereduardo} 
have made progress on this issue. We expect that their work  will provide
a starting point for the construction of quantum Dirac observables.

Assuming that we are indeed on the right track, it is also necessary to find a `large enough' kerenel of the 
constraints so as to be able to construct semiclassical states with respect to the putative Dirac observables.
In this regard it is important to extend the considerations of this work and of P1 to the context of 
primordial states with multiple nontrivial vertices. We believe that this should not be too hard, but it remains
to be worked out in detail.\\

\section*{Acknowledgements}

We thank Abhay Ashtekar,
Fernando Barbero, Alok Laddha, and Eduardo Villase$\rm{\tilde n}$or for their 
constant encouragement. We thank Pooja Singla for drawing our attention to Reference \cite{gl3r} and
to FB, AL and EV for going through a preliminary version of the manuscript.

\section*{Appendix}

\appendix

\section{Conical Alterations.}
\subsection{Conicality in $\{x_0\}$}
As in section 5.2.1 we abuse notation slightly and denote the chargenet obtained at the end of Step 3, Appendix C2,P1
by $c^{}_0(i, v_{I_{v_0},\delta_0}^{\prime})$ even though we have not ascertained the GR property of its 
vertex  $v_{I_{v_0},\delta_0}^{\prime}$.

Consider an edge ${\tilde e}^{}_{J_{v_0}\neq I_{v_0}}$ in $c_0(i, v_{I_{v_0},\delta_0}^{\prime})$ such that its tangent 
is not at the desired angle $\Theta_{x_0}$.
Our aim is to `rotate'  the  edge ${\tilde e}^{}_{J_{v_0}}$ so that its tangent points along this
direction. 
Since the rotation is {\em downward}, 
it follows that there exist 
$\alpha_{J_{v_0}}>0, 
\beta_{J_{v_0}}<0$ such that
\begin{equation}
{}^{\Theta_{x_0}}{\hat {\tilde e}}_{J_{v_0}}^{a}=
\alpha_{J_{v_0}} {\hat {\tilde e}}_{J_{v_0}}^{ a} + 
\beta_{J_{v_0}}{\hat {\tilde e}}_{I_{v_0}}^{a}.
\label{theta0}
\end{equation}
Next, recall from P1 
 that in a small enough vicinity of 
$v_{I_{v_0},\delta_0}^{\prime}$  we have that :\\
\noindent (i)
 the edges ${\tilde e}^{}_{K_{v_0}\neq I_{v_0}}$ are straight lines which 
lie below the plane $P$ where $P$ is the coordinate plane in the $\{x_0\}$ coordinates
which passes through $v_{I_{v_0},\delta_0}^{\prime}$ and is normal to ${\hat {\tilde e}}_{I_{v_0}}^{a}
|_{v_{I_{v_0},\delta_0}^{\prime}}$,
\\
\noindent (ii) the edge ${\tilde e}^{}_{I_{v_0}}$ lies above this plane,\\
\noindent (iii) the GR condition for $v$ (see 1., Appendix C2, P1) together with the second equation of 
3.,C2, P1,
implies that the coordinate plane $P_{J_{v_0}}$ 
which contains the point $v_{I_{v_0},\delta_0}^{\prime}$ and the `straight line' edge
${{\tilde e}}_{J_{v_0}}^{}$ and which is tangent to 
the vector ${\hat {\tilde e}}_{I_{v_0}}^{a}|_{v_{I_{v_0},\delta_0}^{\prime}}$, 
does not contain
any other straight line edge ${{\tilde e}}_{K_{v_0}}^{}, K_{v_0}\neq J_{v_0} \neq I_{v_0}$.

From (i)  and equation (\ref{theta0}) it follows that the putative rotated edge
${}^{\Theta_{x_0}}{ {\tilde e}}_{J_{v_0}}^{}$ must emanate below the plane $P$.
Consider the vector field which generates rotations around 
an axis passing through $v_{I_{v_0},\delta_0}^{\prime}$ in the direction normal to ${ P}_J$ in the coordinates
$\{x_0\}$. 
Multiplying this vector field by 
a semianalytic function of small enough compact support about $v_{I_{v_0},\delta_0}^{\prime}$
yields a
vector field of compact support which generates a diffeomorphism that rotates
the straight line edge 
${\tilde e}^{}_{J_{v_0}}$ at $v_{I_{v_0},\delta_0}^{\prime}$
into an edge ${}^{\Theta_{x_0}}{ {\tilde e}}_{J_{v_0}}^{}$ whose tangent
 is at the desired angle. We apply this diffeomorphism {\em only} to 
${\tilde e}^{}_{J_{v_0}}$. The desired conicality is obtained by applying this deformation procedure to
each edge (except the $I_{v_0}$th one)  in turn.
As in the main text, the resulting  chargenet is denoted by $c^{(0)}_0(i, v_{I_{v_0},\delta_0}^{\prime})$.

We now show that $v_{I_{v_0},\delta_0}^{\prime}$ in $c^{(0)}_0(i, v_{I_{v_0},\delta_0}^{\prime})$ is GR.
Note that the second equation of 3.,C2, P1 holds for the (unrotated) edge tangent set at  
$v_{I_{v_0},\delta_0}^{\prime}$  in $c^{}_0(i, v_{I_{v_0},\delta_0}^{\prime})$ which is obtained at the end of 
Step 3, C2, P1. This implies that for small enough $\delta_0$ the projections of the edge tangents 
of $c^{}_0(i, v_{I_{v_0},\delta_0}^{\prime})$
(other than the 
$I_{v_0}$th one) perpendicular (in the $\{x_0\}$ coordinates) to the $I_{v_0}$th one satisfy conditions 1.1, 1.2,
of C.2. i.e. no projection vanishes and no pair is linearly dependent. Projecting equation (\ref{theta0}) 
perpendicular to the $I_{v_0}$th edge tangent shows that these conditions hold for the edge tangents at
$v_{I_{v_0},\delta_0}^{\prime}$ in $c^{(0)}_0(i, v_{I_{v_0},\delta_0}^{\prime})$. From Appendix D, this, together
with downward conicality of the edge tangent set in $c^{(0)}_0(i, v_{I_{v_0},\delta_0}^{\prime})$ implies that 
$v_{I_{v_0},\delta_0}^{\prime}$ in $c^{(0)}_0(i, v_{I_{v_0},\delta_0}^{\prime})$ is GR.

\subsection{Conicality via an iterative procedure}.

Consider the chargenet $c^{(0)}_0(i, v_{I_{v_0},\delta_0}^{\prime})$ obtained at the end of 
previous section.
Note that since the diffeomorphisms  of the previous section correspond to  rigid rotations, 
(in the $\{x_0\}$ coordinates)
 very close to $v_{I_{v_0},\delta_0}^{\prime}$, 
there is a small 
enough neighbourhood of $v_{I_{v_0},\delta_0}^{\prime}$ such that (i)- (iii) continue to hold
with ${\tilde e}^{}_{J}$ replaced by 
${}^{\Theta_{x_0}}{ {\tilde e}}_{J_{v_0}}^{} \equiv {\tilde e}^{(0)}_{J_{v_0}}$.
Hence a procedure similar to the one in the previous section can be used to achieve
conicality in the $\{x^{\prime (0)}_{\delta_0}\}$ coordinates, and, as we shall see be iterated.

Consider ${\tilde e}^{(0)}_{J_{v_0}\neq I_{v_0}}$ such that its tangent 
is not at the desired angle. Let $T_{v_{I_{v_0},\delta_0}^{\prime}}$ be
the tangent space at $v_{I_{v_0},\delta_0}^{\prime}$ and let 
$P^J_{v_{I_{v_0},\delta_0}^{\prime}} \subset T_{v_{I_{v_0},\delta_0}^{\prime}}$ be the span of 
${\hat {\tilde e}}_{J_{v_0}}^{^{\prime}(0) a}, {\hat {\tilde e}}_{I_{v_0}}^{^{\prime} a}$. Clearly, 
since the rotation is  to be {\em downward}, there exists a unit  vector on the downward cone of angle $\Theta$
(as measured with respect to the $\{x^{\prime (0)}_{\delta_0}\}$ coordinates) about the third axis of 
the form:
\begin{equation}
{}^{\Theta}{\hat {\tilde e}}_{J_{v_0}}^{^{\prime}(0) a}=
\alpha^{\prime}_{J_{v_0}} {\hat {\tilde e}}_{J_{v_0}}^{^{\prime}(0) a} + 
\beta^{\prime}_{J_{v_0}}{\hat {\tilde e}}_{I_{v_0}}^{^{\prime} (0)a}, \;\;\alpha^{\prime}_{J_{v_0}}>0, 
\beta^{\prime}_{J_{v_0}}<0.
\label{A21}
\end{equation}
Our aim is to `rotate'  the  edge ${\tilde e}^{(0)}_{J_{v_0}}$ so that its tangent points along this
direction. Since the unit edge tangents in the $\{x_0\}$ coordinates  are positive rescalings of their
counterparts in the $\{x^{\prime (0)}_{\delta_0}\}$ coordinates it follows that there exist 
$\alpha_{J_{v_0}}>0, 
\beta_{J_{v_0}}<0$ such that
\begin{equation}
{}^{\Theta}{\hat {\tilde e}}_{J_{v_0}}^{(0) a}=
\alpha_{J_{v_0}} {\hat {\tilde e}}_{J_{v_0}}^{(0) a} + 
\beta_{J_{v_0}}{\hat {\tilde e}}_{I_{v_0}}^{a}.
\label{theta2}
\end{equation}

As mentioned at the beginning of this section, (i)- (iii) of the previous section continue to hold i.e. 
that in a small enough vicinity of 
$v_{I_{v_0},\delta_0}^{\prime}$  we have that :\\
\noindent (i)
 the edges ${\tilde e}^{(0)}_{K_{v_0}\neq I_{v_0}}$ are straight lines which 
lie below the plane $P$ where $P$ is the coordinate plane in the $\{x_0\}$ coordinates
which passes through $v_{I_{v_0},\delta_0}^{\prime}$ and is normal to ${\hat {\tilde e}}_{I_{v_0}}^{a}
|_{v_{I_{v_0},\delta_0}^{\prime}}$,
\\
\noindent (ii) the edge ${\tilde e}^{(0)}_{I_{v_0}}$ lies above this plane,\\
\noindent (iii) the GR condition (see 1., Appendix C2, P1) implies that the coordinate plane $P_{J_{v_0}}$ 
which contains the point $v_{I_{v_0},\delta_0}^{\prime}$ and the `straight line' edge
${{\tilde e}}_{J_{v_0}}^{(0)}$ and which is tangent to 
the vector ${\hat {\tilde e}}_{I_{v_0}}^{a}|_{v_{I_{v_0},\delta_0}^{\prime}}$, 
does not contain
any other straight line edge ${{\tilde e}}_{K_{v_0}}^{(0)}, K_{v_0}\neq J_{v_0} \neq I_{v_0}$.

From (i)  and equation (\ref{theta2}) it follows that the putative rotated edge
${}^{\Theta}{ {\tilde e}}_{J_{v_0}}^{(0)}$ must emanate below the plane $P$.
Consider the vector field which generates rotations around 
an axis passing through $v_{I_{v_0},\delta_0}^{\prime}$ in the direction normal to ${ P}_{J_{v_0}}$ in the coordinates
$\{x_0\}$. 
Multiplying this vector field by 
a semianalytic function of small enough compact support about $v_{I_{v_0},\delta_0}^{\prime}$
yields a
vector field of compact support which generates a diffeomorphism that rotates
the straight line edge 
${\tilde e}^{(0)}_{J_{v_0}}$ at $v_{I_{v_0},\delta_0}^{\prime}$
into an edge ${}^{\Theta}{ {\tilde e}}_{J_{v_0}}^{(0)}$ whose tangent
 is at the desired angle. We apply this diffeomorphism {\em only} to 
${\tilde e}^{0}_{J_{v_0}}$ and obtain the desired deformation.
Applying this procedure to each ${\tilde e}^{0}_{J_{v_0}}$ in turn results in the chargenet which we denote, as 
in the main text, by 
$c^{(1)}_0(i, v_{I_{v_0},\delta_0}^{\prime})$. This chargenet is downwardly conical with respect to 
the $\{x^{\prime (0)}_{\delta_0}\}$. We now argue that $v_{I_{v_0},\delta_0}^{\prime}$ is GR
in this chargenet. 
From Appendix A.1 we have that $v_{I_{v_0},\delta_0}^{\prime}$ is GR in $c^{(0)}_0(i, v_{I_{v_0},\delta_0}^{\prime})$.
As a result properties 1.1,1.2 of C2, P1 hold with respect to the $\{x^{\prime (0)}_{\delta_0}\}$
coordinates. The projection of equation (\ref{A21}) perpendicular (with respect to the 
$\{x^{\prime (0)}_{\delta_0}\}$ coordinate metric) to the $I_{v_0}$th edge tangent implies that these
properties hold for the rotated edge tangents. This in conjunction with conicality in $\{x^{\prime (0)}_{\delta_0}\}$ 
 of the edge tangent set
at $v_{I_{v_0},\delta_0}^{\prime}$ in $c^{(1)}_0(i, v_{I_{v_0},\delta_0}^{\prime})$, together with the considerations
of Appendix D imply that $v_{I_{v_0},\delta_0}^{\prime}$ is GR in $c^{(1)}_0(i, v_{I_{v_0},\delta_0}^{\prime})$.

Once again, note  that since the diffeomorphism which brings an edge tangent in  
$c^{(0)}_0(i, v_{I_{v_0},\delta_0}^{\prime})$ to its desired position in $c^{(1)}_0(i, v_{I_{v_0},\delta_0}^{\prime})$
corresponds to a rigid rotation (in the $\{x_0\}$ coordinates)
 very close to $v_{I_{v_0},\delta_0}^{\prime}$, 
there is a small 
enough neighbourhood of $v_{I_{v_0},\delta_0}^{\prime}$ such that (i)- (iii) continue to hold
with ${\tilde e}^{0}_{J_{v_0}}$ replaced by ${}^{\Theta}{ {\tilde e}}_{J_{v_0}}^{(0)}$ so that this procedure
can be iterated.

\section{Freedom in choice of reference coordinates.}

We work in the setting of the Claim of section 3.2. Let the set of vectors be characterised by some $a_{min}\geq 1$.
We are interested in characterising the freedom in the choice of  coordinate system such that:\\
\noindent (i) Equations (\ref{gmin}), (\ref{thetamin}) hold. 
\noindent (ii) The $x^3$ axis at $p$ is along $V_M$. 
\\

Let $\{x\}, \{x^{\prime}\}$ be  such that they satisfy  (i), (ii) above and let their Jacobian  matrix  at $p$ be 
\begin{equation}
J^{\mu^{\prime}}_{\; \nu}:= \frac{\partial x^{\mu^{\prime}}}{\partial x^{\nu}} .
\end{equation}
Condition (ii) above implies that
\begin{equation}
J^{\mu^{\prime}}_{\; 3} = C\delta^{\mu^{\prime}}_{\; 3}, \;\;C>0
\label{B1}
\end{equation}

We suppress the `$min$' subscript  of equation (\ref{gmin}) for notational convenience.
The components $G^{\mu^{\prime}}_{\nu^{\prime}}, G^{\mu}_{\;\nu}$ of $G^a_{min\;b}\equiv G^a_{\;b}$ in the 
coordinates $\{x\}, \{x^{\prime}\}$ are related by the matrix equation:
\begin{equation}
G^{\prime} = J G J^{-1}
\label{B2}
\end{equation}
where we have set $G^{\prime}\equiv G^{\mu^{\prime}}_{\nu^{\prime}},G \equiv G^{\lambda}_{\;\tau}, 
J\equiv J^{\mu^{\prime}}_{\; \lambda}, J^{-1}\equiv \frac{\partial x^{\tau}}{\partial x^{\nu^{\prime}}}$.

In this matrix notation, once again suppressing the `$min$' subscript, Condition (i) implies that:
\begin{equation}
G^{\prime}= C_G R (\theta ), \;\;\; G= C_G R (\theta )
\label{B3}
\end{equation}
where $R$ is given by:
\begin{equation}
R= 
\begin{pmatrix}
1& 0 &0 \\
0 & \cos\theta & -\sin\theta \\
0 & \sin \theta & \cos\theta
\end{pmatrix}
:= 
\begin{pmatrix}
1& 0  \\
0 & {}^{(2)}R
\end{pmatrix}
\label{B4}
\end{equation}
where in the last equality we have used obvious notation so that ${}^{(2)}R$ is  
the two dimensional  matrix corresponding to a rotation by the angle $\theta$.
Equations (\ref{B2}) and (\ref{B3}) imply that 
\begin{equation}
JR=RJ
\label{B5}
\end{equation}
Equation (\ref{B1}) implies that $J$ takes the form:
\begin{equation}
J= 
\begin{pmatrix}
C& j_1 & j_2 \\
0 & {}^{(2)}J_{11}&  {}^{(2)}J_{12}    \\
0 &  {}^{(2)}J_{21}   & {}^{(2)}J_{22}
\end{pmatrix}
=: 
\begin{pmatrix}
C& {\vec j}\\
0 & {}^{(2)}J   
\end{pmatrix}
\label{B6}
\end{equation}
where in the last equality, ${}^{(2)}J, {\vec j}$ are a 2 dimensional matrix and a  
2 vector respectively with $\det {}^{(2)}J>0$ by virtue of the right handedness of the coordinates under
consideration.
Equations (\ref{B4}), (\ref{B5}) and (\ref{B6}) imply that:
\begin{eqnarray}
j^a &= & R^{a}_{\;b}j^b \label{j=rj}
\\
{}^{(2)}J \;{}^{(2)}R&=& {}^{(2)}R \;{}^{(2)}J
\label{JR=RJ}.
\end{eqnarray}
Equation (\ref{j=rj}) implies that ${\vec j}=0$ since there is no nonvanishing vector which is invariant 
under a rotation in 2 dimensions. Equation (\ref{JR=RJ}) together with equations (\ref{B4}), (\ref{B6}) imply that:
\begin{equation}
\begin{pmatrix}
{}^{(2)}J_{11}\cos \theta + {}^{(2)}J_{12}\sin \theta & {}^{(2)}J_{12}\cos \theta-{}^{(2)}J_{11}\sin \theta \\
{}^{(2)}J_{21}\cos \theta + {}^{(2)}J_{22}\sin \theta & {}^{(2)}J_{22}\cos \theta-{}^{(2)}J_{21}\sin \theta 
\end{pmatrix}
=
\begin{pmatrix}
\cos \theta {}^{(2)}J_{11} - \sin \theta {}^{(2)}J_{21}     &  \cos \theta {}^{(2)}J_{12} - \sin \theta {}^{(2)}J_{22}\\
\sin \theta {}^{(2)}J_{11}+ \cos \theta {}^{(2)}J_{21}      & \sin \theta {}^{(2)}J_{12}+ \cos \theta {}^{(2)}J_{22}
\end{pmatrix}
\label{expjr=rj}
\end{equation}
Since the set of vectors is GR, we have that $\theta\neq 0, \theta \neq \pi$ (see 1.2, C.2,P1). It then follows from
(\ref{expjr=rj}) that ${}^{(2)}J_{11}= {}^{(2)}J_{22}$, ${}^{(2)}J_{12}= -{}^{(2)}J_{21}$, which means that 
${}^{(2)}J$ is  proportional to a rotation by the angle $\theta_J$ with 
$(\cos\theta_J, \sin\theta_J) =(\frac{ {}^{(2)}J_{11} }{ \sqrt{ \det{}^{(2)}J   } }, -\frac{{}^{(2)}J_{12}}
{\sqrt{\det{}^{(2)}J}})$
with proportionality constant $\sqrt{ \det{}^{(2)}J }  $, where $\det{}^{(2)}J= ({}^{(2)}J_{11})^2+ ({}^{(2)}J_{12})^2$.
Setting $\sqrt{ \det{}^{(2)}J }= C_J$ the above analysis together with equation (\ref{B6}) implies that 
$J$ takes the form
\begin{equation}
J=
\begin{pmatrix}
C& 0\\
0 & C_J \;{}^{(2)}R (\theta_J  ).
\end{pmatrix}
\label{Jfinal}
\end{equation}
It is easy to verify that transformations of the form equation (\ref{Jfinal}) 
preserve mainifest conicality (see the Definition, section 5.1). For example, if  (downward)
conicality is manifest in the 
$\{x\}$
coordinates with cone angle $\theta_x \in (\frac{\pi}{2}, \pi)$, then (downward) conicality is manifest in the 
$\{x^{\prime}\}$ coordinates with cone angle $\theta_{x^{\prime}} \in (\frac{\pi}{2}, \pi)$ where 
$\tan \theta_{x^{\prime}} = \frac{C_J}{C}\tan \theta_x$.

\section{Derivation of Equation (\ref{negativeprojprime}).}

Consider the Claim of Section 3.2 and its proof in Section 3.3. 
Let the set of vectors there be characterised by some $a_{min}\geq 1$.
Let the vectors $V_1,..,V_{M-1}$ be numbered in the anticlockwise order of their transverse projections. 
Since any map subject to the conditions of the Claim 
can at best translate by $la_{min}, l\geq 0, l\in {\bf Z}$ (see Step 4, 
section 3.3), it follows that the set of $M-1$  vectors $V_1, .., V_{M-1}$ splits 
into $a_{min}$ mutually exclusive sets such that no vector in one of these sets
can be mapped to a vector in a distinct set  and such that the vectors in a singe set are
mapped to each other by $G_{min}$ (where, by mapping, we mean mapping modulo rescalings).
This implies that $\frac{M-1}{a_{min}}:=n$  is a positive integer so that from 
equation (\ref{thetamin}) we have that:
\begin{equation}
\theta_{min} = \frac{2\pi}{n}
\label{2pibyn}
\end{equation}
The integer $n$ cannot be even, else $(G_{min})^{\frac{n}{2}}$ maps
each $V_i$ to a vector which is antiparallel to $V_i$ which cannot be the case due to the GR 
condition (see 1.1,Appendix C1,P1). Moreover $n\neq 1$ else $a_{min}=0$.
Thus, we have that 
\begin{equation}
n= 2p+1, \;\;\;p\geq 1, p\in {\bf Z}.
\label{nodd}
\end{equation}

Next, fix a coordinate system appropriate to $G_{min}$ so that $G_{min}$ is proportional to a rotation by 
$\theta_{min}$ about an axis in the direction of $V_M$. 
We now construct a triplet of vectors such that 
\begin{equation}
\alpha_1 V_{I_1\perp}+\alpha_2 V_{I_2\perp}+\alpha_3 V_{I_3\perp}= 0, \;\;\alpha_i>0.
\label{downtriple}
\end{equation}
where the subscript `$\perp$' refers to projections in the cooordinate plane $P_{\perp}$ perpendicular to $V_M$.
We choose $I_1=1, I_2= 1+a_{min}$. The vectors $V_{1\perp}$ and $V_{1+a_{min}\perp}$ are at angle $\theta_{min}$
in $P_{\perp}$. If there exists $K$ such that  $V_{K\perp}$  lies between the angles $\pi$ and $\theta_{min}+ \pi$
from $V_1$, it is easy to verify that the choice $I_1=1, I_2=2, I_3 =K$ satisfies equation (\ref{nodd}) above.
Accordingly, we set $K= ja_{min}+1, j<n$ and seek $j$ such that
\begin{equation}
\pi < j\theta_{min} < \pi + \theta_{min}.
\label{jseek}
\end{equation}
From equations (\ref{2pibyn}), (\ref{nodd}) it is easy to see that the above inequality  
is satisfied if we choose $j=p+1$. Thus, we have shown the existence of a triplet of vectors 
for which equation (\ref{downtriple}) holds. Clearly, this implies that the triplet satsifies the condition: 
\begin{equation}
\alpha_1 V_{I_1}+\alpha_2 V_{I_2}+\alpha_3 V_{I_3}= \gamma V_M,
\label{downtriple1}
\end{equation}
for some $\alpha_i>0$ and some real $\gamma$.

Next, consider the case where the edge tangent set at $v_{I_{v_0},\delta_0}^{\prime}$ in 
$c^{(0)}_0(i, v_{I_{v_0},\delta_0}^{\prime})$ is characterised by some $a^{(0)}_{min}\geq 1$. 
As in the main text, the coordinate system appropriate to $c^{(0)}_0(i, v_{I_{v_0},\delta_0}^{\prime})$
is denoted by $\{x^{\prime (0)}_{\delta_0}\}$. 
Equation (\ref{downtriple1}) implies that there exists a triple of edges
${\tilde e}^{(0)}_{J^i_{v_0}}, i=1,2,3, J^i_{v_0}\neq I_{v_0}$
such that:
\begin{equation}
\sum_i\alpha^{}_i{\hat {\tilde e}}_{J^i_{v_0}}^{^{\prime}(0) a} =  -\gamma {\hat {\tilde e}}_{I_{v_0}}^{^{\prime}(0)a},
\label{tedge}
\end{equation} 
for some $\alpha_i>0$ and some real $\gamma$. Here the edge tangents 
${\hat {\tilde e}}_{J^i_{v_0}}^{^{\prime}(0) a}$ are normalised with respect to the $\{x^{\prime (0)}_{\delta_0}\}$
coordinates. This equation together with the fact that these edge tangents when normalized with respect to the 
$\{x_0\}$ coordinates are {\em positively} rescaled implies that there exist some 
${\bar\alpha}_i>0$ and some real ${\bar \gamma}$ such that:
\begin{equation}
\sum_i{\bar \alpha}^{}_i{\hat {\tilde e}}_{J^i_{v_0}}^{(0) a} =  -{\bar \gamma} {\hat {\tilde e}}_{I_{v_0}}^{(0)a}.
\label{tedgebar}
\end{equation} 
where the unprimed hats refer to normalization in the $\{x_0\}$ coordinates.
The fact that {\em all} the edges except the $I_{v_0}$th  one point {\em downwards} in the  $\{x_0\}$
coordinates imply that ${\bar \gamma}>0$. The positive rescaling property then implies that
$\gamma >0$ in equation (\ref{tedge}).

\section{Conicality and the GR condition.}
In this appendix we prove the following Lemma.\\

\noindent{\em Lemma:} 
Let $\{x\}$
be a coordinate system at $p\in \Sigma$. 
Let $W_1,..,W_M$ be a set of {\em unit} vectors in the tangent space $T_p$ to the point $p\in \Sigma$ where unit norm
is with respect to the coordinate metric.
Let $W_M$ point along the positive $x^3$ direction and let $W_J, J\neq M$ point 
along the downward coordinate cone i.e. in the coordinate metric defined by $\{x\}$, let $W_J, J\neq M$
subtend the same $J$ independent angle $\theta$ with respect to the $x^3$ direction with $\theta >\frac{\pi}{2}$.
Further, let the projections $W_{i\perp}, i=1,..,M-1$ perpendicular to the $x^3$ direction be such that\\
\noindent (i) 
no such projection  vanishes\\
\noindent (ii) 
no two such projections are linearly dependent.\\

Then the set of vectors $W_1,..,W_M$ is GR i.e. no triplet of these vectors is coplanar.
\\

\noindent{\em Proof}: 
First consider  a triplet of distinct vectors $W_{J_1}, W_{J_2},W_M$. If they are coplanar, then there exist 
$\alpha_{i}, i=1,2$ such that 
\begin{equation}
\sum_{i=1}^2\alpha_{i}W_{J_i}= W_M
\end{equation}
The perpendicular projection 
of this equation is in contradiction with (i), (ii) of the Lemma. 

Next consider a triplet $\{ W_{J_i}, i=1,2,3\}, J_i\neq M$.
Coplanarity of this triplet implies that there exist $\alpha_{i}, i=1,2$  such that 
\begin{equation}
\sum_{i=1}^2\alpha_{i}W_{J_i}= W_{J_3}.
\end{equation}
The projection of this equation along the $x^3$ direction together with conicality (and $\theta \neq \frac{\pi}{2}$)
yields 
\begin{equation}
\alpha_1 + \alpha_2 =1.
\label{D1}
\end{equation}
The projection of this equation in the $x^1, x^2$ directions yields:
\begin{eqnarray}
\sum_{i=1}^2 \cos\phi_i\alpha_i &=& \cos \phi_3\\
\sum_{i=1}^2 \sin\phi_i\alpha_i &=& \sin\phi_3
\end{eqnarray}
where $\phi_i$ is the angle of the perpendicular components $W_{J_i\perp}$ with respect to the $x^1$ axis.
Squaring these equations and adding them
yields:
\begin{equation}
(\alpha_1 + \alpha_2)^2   - 2(1- \cos(\phi_1- \phi_2)) \alpha_1\alpha_2 = 1
\end{equation}
which together with equation (\ref{D1}) implies that $\phi_1= \phi_2$ or $\alpha_1\alpha_2=0$
or both, which, in turn is in contradiction with (i),(ii) of the Lemma.

\end{document}